\documentclass{aa}
\usepackage[dvips]{rotating}
\usepackage{amsmath,txfonts,graphicx,natbib,epsfig,amssymb,rotating}
\usepackage{times,textcomp,longtable,lscape}

\newcommand\Ltable{\onecolumn\clearpage\setlength{\LTcapwidth}{\textheight}}

\bibpunct{(}{)}{;}{a}{}{,}

\newcount\longrefs
\def\aap{\ifnum\longrefs=1 {Astron.\ Astrophys.}\else 
                           {A\hbox{\rm \&}A}\fi}
\def\aapr{\ifnum\longrefs=1 {Astron.\ Astrophys.\ Rev.}\else 
                            {A\hbox{\rm \&}AR}\fi}
\def\aaps{\ifnum\longrefs=1 {Astron.\ Astrophys.\ Suppl.}\else 
                            {A\hbox{\rm \&}AS}\fi}
\def\aj{\ifnum\longrefs=1 {Astron.\ J.}\else 
                          {AJ}\fi} 
\def\ao{\ifnum\longrefs=1 {Applied Optics}\else 
                           {Appl.\ Opt.}\fi} 
\def\aspcs{\ifnum\longrefs=1 {Astron.\ Soc.\ Pacific Conf. Series}\else 
                           {ASP Conf.\ Ser.}\fi} 
\def\apj{\ifnum\longrefs=1 {Astrophys.\ J.}\else 
                           {ApJ}\fi} 
\def\apjl{\ifnum\longrefs=1 {Astrophys.\ J. Lett.}\else 
                            {ApJ}\fi} 
\def\aplett{\ifnum\longrefs=1 {Astrophys.\ J. Lett.}\else 
                            {ApJ}\fi} 
\def\apjs{\ifnum\longrefs=1 {Astrophys.\ J. Suppl.}\else 
                            {ApJS}\fi}
\def\apss{\ifnum\longrefs=1 {Astrophys.\ and Space Science}\else 
                            {Ap\hbox{\rm \&}SS}\fi}
\def\araa{\ifnum\longrefs=1 {Ann.\ Rev.\ Astron.\ Astrophys.}\else 
                            {ARA\hbox{\rm \&}A}\fi}
\def\azh{\ifnum\longrefs=1 {Astronomicheskii Zhurnal}\else 
                            {Astron.\ Zhur.}\fi}
\def\baas{\ifnum\longrefs=1 {Bull.\ Am.\ Astron.\ Soc.}\else 
                            {BAAS}\fi}
\def\bain{\ifnum\longrefs=1 {Bull.\ Astronom.\ Institutes Netherlands}\else
                            {Bull.\ Astr.\ Inst.\ Neth.}\fi}
\def\gca{\ifnum\longrefs=1 {Geochim.\ Cosmochim.\ Acta}\else 
                           {Geochim.\ Cosmochim.\ Acta}\fi}
\def\grl{\ifnum\longrefs=1 {Geophys.\ Res.\ Lett.}\else 
                           {Geoph.\ Res.\ Lett.}\fi}
\def\iaucirc{\ifnum\longrefs=1 {IAU Circulars}\else 
                          {IAU Circ.}\fi}
\def\ip{\ifnum\longrefs=1 {in press}\else 
                          {in press}\fi}
\def\jgr{\ifnum\longrefs=1 {J.\ Geophys.\ Res.}\else 
                           {J.\ Geophys.\ Res.}\fi}  
\def\jrasc{\ifnum\longrefs=1 {J.\ Royal Astron.\ Soc.\ Canada}\else 
                           {JRAS Can.}\fi}  
\def\mnras{\ifnum\longrefs=1 {Mon.\ Not.\ Roy.\ Astron.\ Soc.}\else 
                             {MNRAS}\fi} 
\def\nat{\ifnum\longrefs=1 {Nature}\else 
                           {Nat}\fi}
\def\pasj{\ifnum\longrefs=1 {Pub.\ Astron.\ Soc.\ Japan}\else 
                            {PASJ}\fi} 
\def\pasp{\ifnum\longrefs=1 {Pub.\ Astron.\ Soc.\ Pacific}\else 
                            {PASP}\fi} 
\def\physscr{\ifnum\longrefs=1 {Physica Scripta}\else 
                            {Phys.\ Scrip.}\fi} 
\def\planss{\ifnum\longrefs=1 {Planetary \& Space Science}\else 
                            {Plan. \& Space Sci.}\fi} 
\def\procspie{\ifnum\longrefs=1 {Proc.\ SPIE}\else 
                            {Proc.\ SPIE}\fi} 
\def\qjras{\ifnum\longrefs=1 {Quarterly J.\ Royal Astron.\ Soc.}\else 
                            {QJRAS}\fi} 
\def\sa{\ifnum\longrefs=1 {Soviet Astron..}\else 
                               {Sov.\ Astron.}\fi}
\def\skytel{\ifnum\longrefs=1 {Sky \& Telescope}\else 
                            {Sky \& Tel.}\fi} 
\def\solphys{\ifnum\longrefs=1 {Solar Phys.}\else 
                               {Solar Phys.}\fi}
\def\ssr{\ifnum\longrefs=1 {Space Science Rev.}\else 
                               {Space\ Sci.\ Rev.}\fi}

\def\nl{,\ } 

\def\Leuven{Instituut voor Sterrenkunde\nl K.U.Leuven\nl Celestijnenlaan 200B\nl B--3001 Leuven\nl Belgium}

\def\Munich{Universit\"atssternwarte M\"unchen\nl Scheinerstr. 1\nl
D-81679 M\"unchen\nl Germany}

\def\Nijmegen{Departement Astrofysica\nl Radboud Universiteit van
Nijmegen\nl PO Box 9010\nl 6500 GL Nijmegen\nl the Netherlands}

\def\dutch{\def\refname{Referenties}\def\abstractname{Samenvatting}%
  \def\bibname{Bibliografie}\def\chaptername{Hoofdstuk}%
  \def\appendixname{Bijlage}\def\contentsname{Inhoudsopgave}%
  \def\listfigurename{Lijst van figuren}\def\listtablename{Lijst van tabellen}%
  \def\indexname{Index}\def\figurename{Figuur}\def\tablename{Tabel}%
  \def\partname{Deel}\def\enclname{Bijlage(n)}\def\ccname{Ter attentie van}%
  \def\headtoname{Aan}\def\headpagename{Pagina}%
  \def\today{\number\day\space\ifcase\month\or januari\or februari\or maart\or%
     april\or mei\or juni\or juli\or augustus\or september\or oktober\or%
     november\or december\fi \space\number\year}%
  \typeout{
              >>>>> use hlatex209 for Dutch hyphenation <<<<< 
         }}
\newcounter{onefig} \newcounter{fignumber}
\newcount\nocaptions \newcount\nofigures \newcount\figwidth
\newcount\viewgraphs
  \def\paper{}  \def\figlabel{} 
\long\def\nextfig#1{\setcounter{figure}{\value{fignumber}}
  \addtocounter{fignumber}{1}
  \ifnum \viewgraphs=1 \newpage \pagestyle{empty} \fi 
  \ifnum\value{onefig}=0 #1 \fi                 
  \ifnum\value{onefig}=\value{fignumber} #1 \fi}
\def\figwidths#1#2{\ifnum \nocaptions=1 #2mm \else #1mm \fi}  
\def\paper#1{}  
\long\def\plotfig#1#2{\ifnum \nofigures=1 \else #2 \fi}
\long\def\captiontext#1{\ifnum \nofigures=1 \raggedright \fi 
   \ifnum \nocaptions=1 \paper
     \ifnum \viewgraphs=0 
       \newline  \mbox{}\hrulefill\mbox{} \newline 
       \newline label:~\{\figlabel\} 
     \fi 
     \else \ifnum \nofigures=0 \fi 
   #1 \fi}

\newcount\panelwidth \newcount\panelheight 
\newcount\bxmin \newcount\bymin \newcount\bxmax \newcount\bymax
\newcount\tbxmin \newcount\tbymin
\newcount\tpanelwidth \newcount\tpanelheight \newcount\tpdif
\def\panelsize #1,#2;{\panelwidth=#1 \panelheight=#2}  
\def\setbb #1,#2;#3,#4;#5,#6;{
  \tbxmin=#1 \tbymin=#2    
  \bxmin=#3 \bymin=#4      
  \bxmax=#5 \bymax=#6}     
\def\barepanel #1{%
  \ifnum\panelheight=0 
    \tpdif=\bymax \advance\tpdif by -\bymin
    \multiply \tpdif by \panelwidth
    \tpanelheight=\tpdif
    \tpdif=\bxmax \advance\tpdif by -\bxmin
    \divide \tpanelheight by \tpdif
  \else \tpanelheight=\panelheight \fi
  \epsfig{file=#1,%
     bbllx=\bxmin bp,bblly=\bymin bp,bburx=\bxmax bp,bbury=\bymax bp,clip=,%
     width=\panelwidth mm,height=\tpanelheight mm}}
\def\labelypanel #1{
  \ifnum\panelheight=0 
    \tpdif=\bymax \advance\tpdif by -\bymin
    \multiply \tpdif by \panelwidth
    \tpanelheight=\tpdif
    \tpdif=\bxmax \advance\tpdif by -\bxmin
    \divide \tpanelheight by \tpdif
  \else \tpanelheight=\panelheight \fi
  \tpdif=\bxmax \advance\tpdif by -\tbxmin
  \tpanelwidth=\panelwidth \multiply \tpanelwidth by \tpdif
  \tpdif=\bxmax \advance\tpdif by -\bxmin
  \divide \tpanelwidth by \tpdif
  \epsfig{file=#1,%
    bbllx=\tbxmin bp,bblly=\bymin bp,bburx=\bxmax bp,bbury=\bymax bp,%
    clip=,width=\tpanelwidth mm,height=\tpanelheight mm}}
\def\labelxpanel #1{%
  \ifnum\panelheight=0 
    \tpdif=\bymax \advance\tpdif by -\bymin
    \multiply \tpdif by \panelwidth
    \tpanelheight=\tpdif
    \tpdif=\bxmax \advance\tpdif by -\bxmin
    \divide \tpanelheight by \tpdif
  \else \tpanelheight=\panelheight \fi
  \tpdif=\bymax \advance\tpdif by -\tbymin
  \multiply \tpanelheight by \tpdif
  \tpdif=\bymax \advance\tpdif by -\bymin
  \divide \tpanelheight by \tpdif
  \epsfig{file=#1,%
    bbllx=\bxmin bp,bblly=\tbymin bp,bburx=\bxmax bp,bbury=\bymax bp,%
    clip=,width=\panelwidth mm,height=\tpanelheight mm}}
\def\labelxypanel #1{%
  \ifnum\panelheight=0 
    \tpdif=\bymax \advance\tpdif by -\bymin
    \multiply \tpdif by \panelwidth
    \tpanelheight=\tpdif
    \tpdif=\bxmax \advance\tpdif by -\bxmin
    \divide \tpanelheight by \tpdif
  \else \tpanelheight=\panelheight \fi
  \tpdif=\bxmax \advance\tpdif by -\tbxmin
  \tpanelwidth=\panelwidth \multiply \tpanelwidth by \tpdif
  \tpdif=\bxmax \advance\tpdif by -\bxmin
  \divide \tpanelwidth by \tpdif 
  \tpdif=\bymax \advance\tpdif by -\tbymin 
  \multiply \tpanelheight by \tpdif
  \tpdif=\bymax \advance\tpdif by -\bymin
  \divide \tpanelheight by \tpdif
  \epsfig{file=#1,%
    bbllx=\tbxmin bp,bblly=\tbymin bp,bburx=\bxmax bp,bbury=\bymax bp,%
    clip=,width=\tpanelwidth mm,height=\tpanelheight mm}}

\def\CC{\par \vspace*{-2ex} \footnotesize \baselineskip=8pt \begin{verbatim}}

\long\def\startignore #1\stopignore{}   

\def\setlistparams{         
  \topsep=0.7ex                 
  \itemsep=0.7ex                
  \leftmargini=3ex}             
\setlistparams                  

\newcounter{alistindex}       

\newcounter{romenumnr}

\newlength{\minipagewidth}

\newsavebox{\boxcontent}
\newcommand{\ovalhead}[1]{
  \unitlength=1cm
  \sbox{\boxcontent}{\mbox{~~{#1}~~}}
  \begin{center}
    \ifdim\wd\boxcontent>6ex 
    \ifdim\wd\boxcontent<8cm 
    \begin{picture}(8,3) \thicklines     
      \put(4.0,0.8){\oval(8,1.6)} 
      \put(0.0,0.7){\parbox{8cm}{
         \begin{center} \usebox{\boxcontent} \end{center}}}
    \end{picture}
    \else \ifdim\wd\boxcontent<12cm 
    \begin{picture}(12,3) \thicklines     
        \put(6.0,0.8){\oval(12,1.6)} 
        \put(0.0,0.7){\parbox{12cm}{
           \begin{center} \usebox{\boxcontent} \end{center}}}
    \end{picture}
    \else
    \begin{picture}(16,3) \thicklines     
        \put(8.0,0.8){\oval(16,1.6)} 
        \put(0.0,0.7){\parbox{16cm}{
           \begin{center} \usebox{\boxcontent} \end{center}}}
    \end{picture}
    \fi \fi \fi
  \end{center}} 

\newcounter{headnr}            
\newcounter{subheadnr}[headnr]
\newcounter{subsubheadnr}[subheadnr]
\def\head #1\par{
  \stepcounter{headnr}                          
  \vspace{2ex} \noindent                        
  {\bf \theheadnr~~~~#1}\\[1ex] \noindent}      
\def\subhead #1\par{  
  \stepcounter{subheadnr}
  \vspace{1.3ex} \noindent
  {\bf \theheadnr.\arabic{subheadnr}~~~#1}\\[0.3ex] \noindent}
\def\subsubhead #1\par{
  \stepcounter{subsubheadnr}
  \vspace{1.0ex} \noindent
  {\bf \theheadnr.\arabic{subheadnr}.\arabic{subsubheadnr}~~~#1}\\ \noindent}

\font\dropfont= cmr12 scaled \magstep5
\def\dropcap#1#2{{\noindent
    \setbox0\hbox{\dropfont #1}\setbox1\hbox{#2}\setbox2\hbox{(}%
    \count0=\ht0\advance\count0 by\dp0\count1\baselineskip
    \advance\count0 by-\ht1\advance\count0by\ht2
    \dimen1=.5ex\advance\count0by\dimen1\divide\count0 by\count1
    \advance\count0 by1\dimen0\wd0
    \advance\dimen0 by.25em\dimen1=\ht0\advance\dimen1 by-\ht1
    \global\hangindent\dimen0\global\hangafter-\count0
    \hskip-\dimen0\setbox0\hbox to\dimen0{\raise-\dimen1\box0\hss}%
    \dp0=0in\ht0=0in\box0}#2}

\def\Ha{\mbox{H$\alpha$}}

\def\level #1 #2#3#4{$#1 \: ^{#2} \mbox{#3} ^{#4}$}   

\def\sun{\hbox{$\odot$}}                   

\def\mathstacksym#1#2#3#4#5{\def#1{\mathrel{\hbox to 0pt{\lower 
    #5\hbox{#3}\hss} \raise #4\hbox{#2}}}}

\mathstacksym\lta{$<$}{$\sim$}{1.5pt}{3.5pt} 
\mathstacksym\gta{$>$}{$\sim$}{1.5pt}{3.5pt} 
\mathstacksym\lrarrow{$\leftarrow$}{$\rightarrow$}{2pt}{1pt} 
\mathstacksym\lessgreat{$>$}{$<$}{3pt}{3pt} 

\begin{document}

\title{Statistical properties of a sample of periodically variable B-type
supergiants 
\thanks{Figures of the spectral line fits and discussion of the individual
objects are only available in electronic form via http://www.edpsciences.org}}
\subtitle{Evidence for opacity-driven gravity-mode oscillations}
 
\author{K.\ Lefever\inst{1} \and J.\ Puls\inst{2} \and C.\ Aerts\inst{1,3}
\institute{\Leuven \and \Munich \and \Nijmegen} }

\authorrunning{Lefever et al.} 
\titlerunning{Periodically variable B supergiants}

\offprints{K. Lefever, \email{karolien@ster.\\kuleuven.be}}

\date{Received 14 July 2006 / Accepted 3 November 2006}

\abstract{} 
  {We have studied a sample of 28 periodically variable B-type
  supergiants selected from the HIPPARCOS mission and 12 comparison stars
  covering the whole B-type spectral range. Our goal is to test if their
  variability is compatible with opacity-driven non-radial oscillations.}  
  {We have used the NLTE atmosphere code FASTWIND to derive the atmospheric
  and wind parameters of the complete sample through line profile fitting.
  We applied the method to selected H, He and Si line profiles, measured
  with the high resolution CES spectrograph attached to the ESO CAT
  telescope in La Silla, Chile.} 
  {From the location of the stars in the ($\log T_{\rm eff}$, log $g$) diagram,
  we suggest that variability of our sample supergiants is indeed due to the
  gravity modes resulting from the opacity mechanism. We find nine of the
  comparison stars to be periodically variable as well, and suggest them to
  be new $\alpha\,$Cyg variables. We find marginal evidence of a
  correlation between the amplitude of the photometric variability and the
  wind density. We investigate the Wind Momentum Luminosity Relation for the
  whole range of B spectral type supergiants, and find that the later types
  ($>$ B5) perfectly follow the relation for A supergiants. Additionally, we
  provide a new spectral type - $T_{\rm eff}$ calibration for B supergiants.}
  {Our results imply the possibility to probe internal structure models of
  massive stars of spectral type B through seismic tuning of gravity modes.}


\maketitle

\section{Introduction}

One of the remarkable by-products of the ESA HIPPARCOS mission was the
discovery of a large amount of new periodically variable B stars, including
almost a hundred new slowly pulsating B stars (SPBs hereafter) and 29
periodically variable B-type supergiants \citep{Waelkens98, Aerts99,
Mathias01}. The photometric behaviour of different kinds of evolved massive
stars were analysed in detail from HIPPARCOS data. VanLeeuwen et al. (1998) 
performed an analysis of 24 known B- to G-type variable supergiants and
found periods of tens of days for each of them, in agreement with previous
ground-based studies. \citet{Marchenko98}, on the other hand, focused on the
HIPPARCOS variability of 141 O-type and WR stars and noted the remarkable
variety of variability, with very diverse causes, within that sample. The
study of \citet{Waelkens98} is quite different in this respect, as they came
up with a sample of 29 new variable B supergiants exhibiting clear periodic
microvariations at relatively short periods of one to a few days. The
current paper contains a follow-up study of the latter sample. Our goal is
to evaluate the suggestion by \citet{Waelkens98} that these periodically
variable B-type supergiants experience oscillations excited by the opacity
mechanism, in analogy to main sequence B stars.

The suggestion of occurrence of non-radial oscillation modes in massive
supergiants is not new. Microvariations with amplitudes between a hundredth
and a tenth of a magnitude in the visual, and periods ranging from some 5 to
100\,d, have been found in numerous supergiants of spectral type OBA, termed
$\alpha\,$Cyg variables \citep{Sterken77, Sterken83, Burki_etal78,
VanGenderen89, Lamers98, VanGenderen01}.  \citet{Burki78} considered a
sample of 32 B- to G-type supergiants and derived an empirical
semi-period-luminosity-colour (PLC) relation (see his Eq.\,5), from which
he suggested the variability to be caused by oscillations rather than mass
loss. \citet{Lovy84} indeed managed to come up with a theoretical PLC
relation in agreement with the observed one for this type of stars. However,
only 40\% of the variable supergiants have periods compatible with the
radial fundamental mode, while the majority must exhibit a different kind of
oscillation mode. \citet{Kaufer97} made an extensive high-resolution
spectroscopic study of 6 BA-type supergiants which they monitored for
several years. They concluded that the variability patterns are extremely
complex and point towards cyclical variations in the radial velocities. From
CLEANing the periodograms of the radial-velocity curves, they derived
multiple periods and assigned them to non-radial oscillations because the
travelling features in the dynamical spectra turned out to be incompatible
with the rotational periods of the stars.

\citet{Glatzel93} interpreted the periodic variability of supergiants with
masses above $40\,M_\odot$ in terms of strange-mode instabilities and showed
the classical opacity mechanism to be unimportant in such objects.
\citet{Glatzel99} subsequently performed detailed calculations for stars
with $M=64\,M_\odot$ and predicted irregular velocity and luminosity
variations with time scales of 1 to 20\,d. They also proposed that such
massive stars undergo pulsationally-driven mass-loss.  It is not clear how
their result will change for stars in the mass range of 10 to 30$\,M_\odot$,
which is the transition region from low to large mass-loss rates due to line
driving. Therefore, the periodic variations of the B supergiants found by
\citet{Waelkens98} might still be due to the classical opacity mechanism,
since the instability strips of the $\beta\,$Cep stars and the SPBs were
found to extend to higher luminosities shortly after the discovery paper
\citep{Pamyatnykh99, Balona99}.

\citet{Waelkens98} positioned the new periodic B supergiants in the HR
diagram on the basis of multicolour photometric calibrations (accurate
parallaxes are not available) and found them to be situated between the SPBs
and previously known $\alpha\,$Cyg-type variables (see their Fig.\,2).
Oscillations were not predicted in that part of the HR diagram. Given the
uncertainty in the effective temperature and luminosity, \citet{Aerts00a}
tried to improve upon the fundamental parameter determination by
constructing the spectral energy distribution of the stars as a better
diagnostic to derive their effective temperature and gravity. This did not
improve the large uncertainties on the latter parameters, however.
Nevertheless, the sample selected by \citet{Waelkens98} remained the most
valuable one to investigate observationally the occurrence of gravity modes
in supergiant stars, because it is unbiased in the sense that the stars were
not at all selected to be observed with HIPPARCOS on the basis of
variability.  For this reason, we conducted an extensive spectroscopic
campaign for the sample, with the goal to perform line-profile fitting as a
way to estimate the fundamental parameters of the stars with unprecedented
precision. We report upon the analysis of these new data, and the position
of the stars with respect to the instability strip of gravity modes, in the
current work.

The questions we will address here are the following. We elaborate further
on the HIPPARCOS data in order to search for multiperiodic variability,
which is a natural property of non-radial oscillators.  From selected H, He
and Si line profiles, we derive the physical parameters (effective
temperature, gravity, luminosity, mass-loss, rotational velocity, etc.) of
the stars in the sample by \citet{Waelkens98} using high-quality
spectroscopic data.  From this, we derive their position in the HR diagram
with high precision and check if the stars lie within the instability strips
of gravity mode oscillations predicted by \citet{Pamyatnykh99}
and \citet{Saio06}.  During our
analysis we also look for asymmetries in the metallic line profiles.
Further, we search for correlations between the physical parameters and the
photometric variability. In particular, we investigate if there is any
connection between the observed peak-to-peak amplitude and frequency of the
light variability, the wind density and the rotation of the stars. Finally,
we investigate the Wind momentum Luminosity Relation (WLR) for the complete
B-type spectral range.

\section{Observations and Data Reduction}

We selected all southern stars of luminosity class I or II brighter than 9th
mag in the sample by \citet{Waelkens98} which fitted the observation window
of the assigned telescope time. This concerns 21 stars. To this we added 10
more {\it candidate\/} $\alpha$~Cyg variables from the Catalogue of Periodic
Variables of HIPPARCOS \citep{Hipparcos_volume11}, in such way that the
complete sample fully covers the B-type spectral range. \citet{Waelkens98}
were unable to assign them to one of the five considered classes ($\beta$
Cep stars, SPBs, chemically peculiar stars, B-type supergiants, Be stars),
see also \citet{Kestens1998}. Our period analysis and their spectral type
clearly point towards $\alpha$~Cyg variables, however.

These 31 targets (for spectral types, see Table\,\ref{finalparameters_SG} 
and discussion in Sect.\,\ref{calib}) were added to the long-term
spectroscopic monitoring programme of periodic B stars conducted at Leuven
University \citep{Aerts99}. The spectra of the stars were gathered with the
CES spectrograph attached to the CAT telescope at La Silla during numerous
observing runs spread over two years. For most targets, we obtained two
exposures of the H$\alpha$ line in different seasons (in order to check for
its variability), one of the H$\gamma$ line, one of the He~I\,6678\AA\ line
and one of the He~I\,4471\AA\ line. Besides these, we observed one silicon
line for each star with the goal to obtain an accurate temperature estimate.
Depending on spectral type, this is either the Si~II\,4130\AA\ doublet (late
B-type) or the Si~III\,4560\AA\ triplet (early B-type). 

The spectra were reduced in the standard way, i.e., by means of
flatfielding, wavelength calibration through Th-Ar exposures and
rectification of the continuum through a cubic spline fit. The resolution
amounts to 70\,000 and the exposure times range from 3 to 50 minutes,
resulting in a high signal-to-noise ratio of about 250.

It became immediately evident that the three stars \object{HD\,71913},
\object{HD\,157485} and \object{HD\,165812} were misclassified in the Bright
Star Catalogue (BSC) as supergiants. They turned out to be new $\beta\,$Cep
stars. These have been studied in detail by \citet{Aerts00b} and are not
included here. This finally leads to 28 sample supergiant stars.

In order to assess the importance of having {\it periodic\/} light
variability in our sample, we have additionally selected 12 bright B
supergiants from the BSC, again chosen to cover the complete B-type spectral
range.  These variables were not classified as {\it periodic\/} variable by
the HIPPARCOS classification algorithm.  While for some of these bright
objects stellar parameters are available in the literature, we have invested
in collecting their spectra as well, in order to treat these stars in the
same way as the sample stars.

\section{Photometric Variability \label{photometry_section}}

The HIPPARCOS data of the 40 targets were subjected to detailed period
search by means of the \citet{Scargle82} and Phase Dispersion
Minimisation \citep{Stellingwerf78} method. In Fig.\,\ref{vb}, we show the phase
diagrams for the dominant frequency for two representative cases. The
detailed results of the frequency analyses are provided in
Tables\,\ref{freq} and \ref{freq_np}.

For most targets we recovered the main period found by \citet{Waelkens98},
but not for all of them. 
For six stars, the first harmonic of the main
frequency was also needed to obtain an accurate fit to the HIPPARCOS data
(see, e.g., Fig.\,\ref{vb}).  We found evidence for multiperiodicity for eleven
stars. The detected periods range from 1.15 to 25\,d, with only four stars
having a period longer than 10\,d. Thus we confirm that most of the 28
sample stars have periods which are significantly shorter than 
the ones of other $\alpha\,$Cyg variables.  

We do also find short periods (less than 10\,d) for some {\it comparison}
stars, though they were not classified as periodic by the HIPPARCOS team.
When comparing the observed periods and peak-to-peak amplitudes between the
sample and comparison stars, seven out of twelve comparison stars have lower
peak-to-peak variations, whereas the other five seem to have periods and
amplitudes comparable to the ones detected in our target sample.

\section{Spectroscopic Analysis: Results from Line Profile Fitting}

\subsection{$v \sin i$ from Metallic Lines}

The projected rotational velocity, $v \sin i$, was found from the automated
tool developed by \citet{SimonDiaz2006}, which is based on a Fourier method
to separate the effects of rotational broadening and macro-turbulence (which
can be significant in B-type supergiants, cf. \citealt{Ryans2002}). This
method was first described by \citet{Gray1973} and reviewed in
\citet{Gray1978}. It is discussed into detail more recently by
\citet{Piters1996}.

Weak metallic lines are the best diagnostic to derive $v \sin i$, since
they are (by definition) free from saturation effects and least affected by
collisional broadening.  We have the following lines at our disposal: either
Si~II 4128-4130\AA\ or Si~III~4552-4567-4574\AA\ (depending on spectral
type), Mg~II~4481\AA\ (in the same spectral order as He~I~4471\AA) and, for
slow rotators, also C~II~6578-6582\AA\ near H$\alpha$. Besides these primary
diagnostics, often also other, even weaker metallic lines can be found.

Table\,\ref{vsini} lists our results for $v \sin i$, together with its
standard deviation and the number of lines we used. For the 12 comparison
objects, not all selected orders were measured, and hence only few lines
could be used. For five stars we even lack useful metallic lines. Three out
of these five (\object{HD\,64760}, \object{HD\,157246} and
\object{HD\,165024}) have blended metal lines due to their fast rotation, and
we adopt a mean value for $v \sin i$ as provided by SIMBAD. For the other two
objects (\object{HD\,51110} and \object{HD\,147670}), no value is given in
SIMBAD, and it was not possible to use other lines to measure this quantity: for
HD\,51110 the He~I lines were too weak and peculiar, whereas for HD\,147670 only
H$\alpha$ had been secured.

The occurrence of {\it asymmetries\/} in the line profiles may reveal
the presence of time-dependent pulsational broadening (see, e.g.,
\citealt{Aerts03}).  In the current sample, clear asymmetries in the
Si lines were detected only for \object{HD\,54764}, \object{HD\,92964}
and HD\,109867.  Most probably, they are related to binarity
(HD\,54764) or to a large mass-loss rate (HD\,92964), which in both
cases affect the photospheric lines in an asymmetric way. Only for
HD~109867, we can speculate about a relation between line asymmetry and
pulsational variability.

\begin{figure}
\centering
\vspace*{-1.3cm}
\rotatebox{0}{\resizebox{6.5cm}{!}{\includegraphics{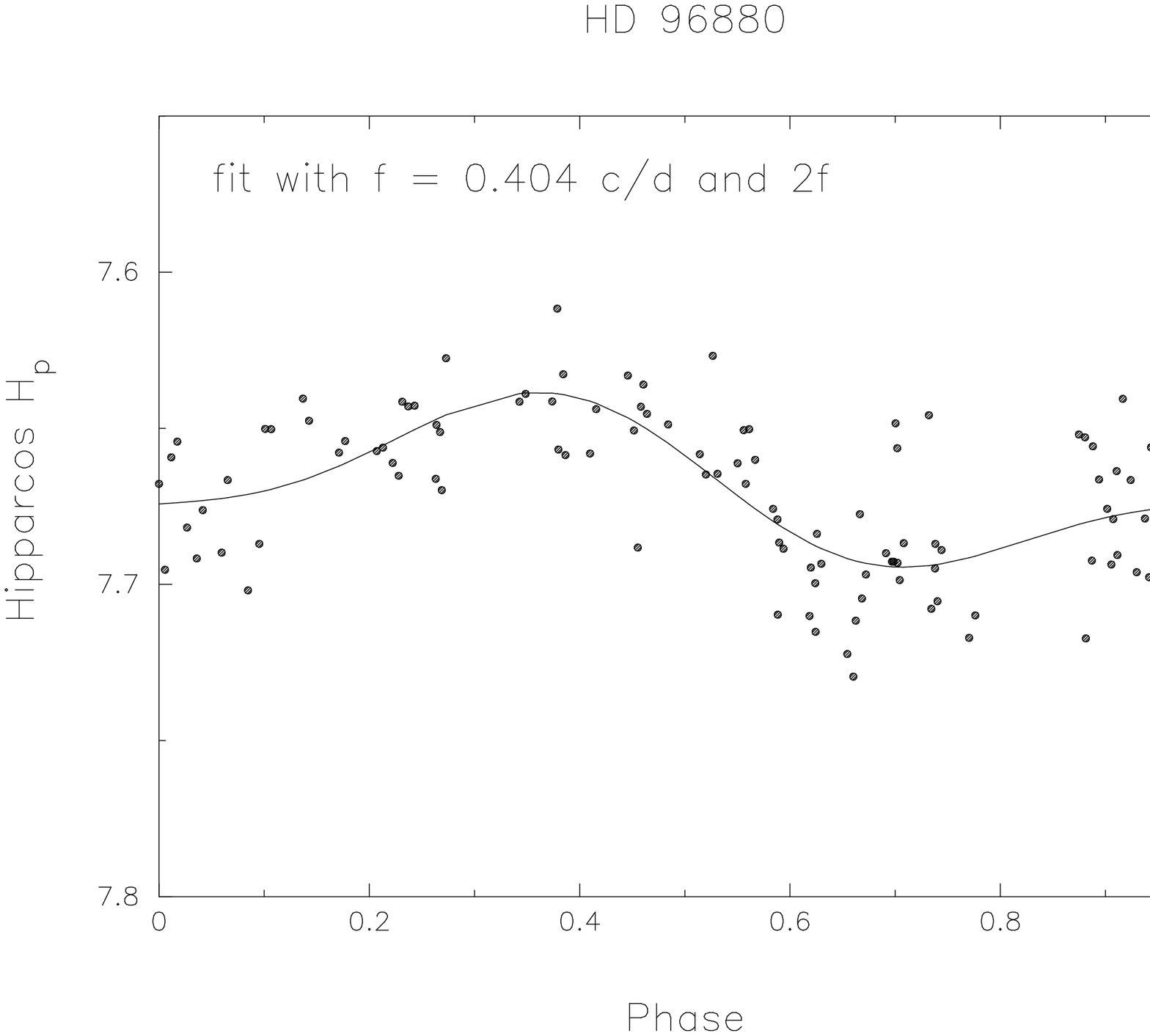}}}\\[-1cm]
\rotatebox{0}{\resizebox{6.5cm}{!}{\includegraphics{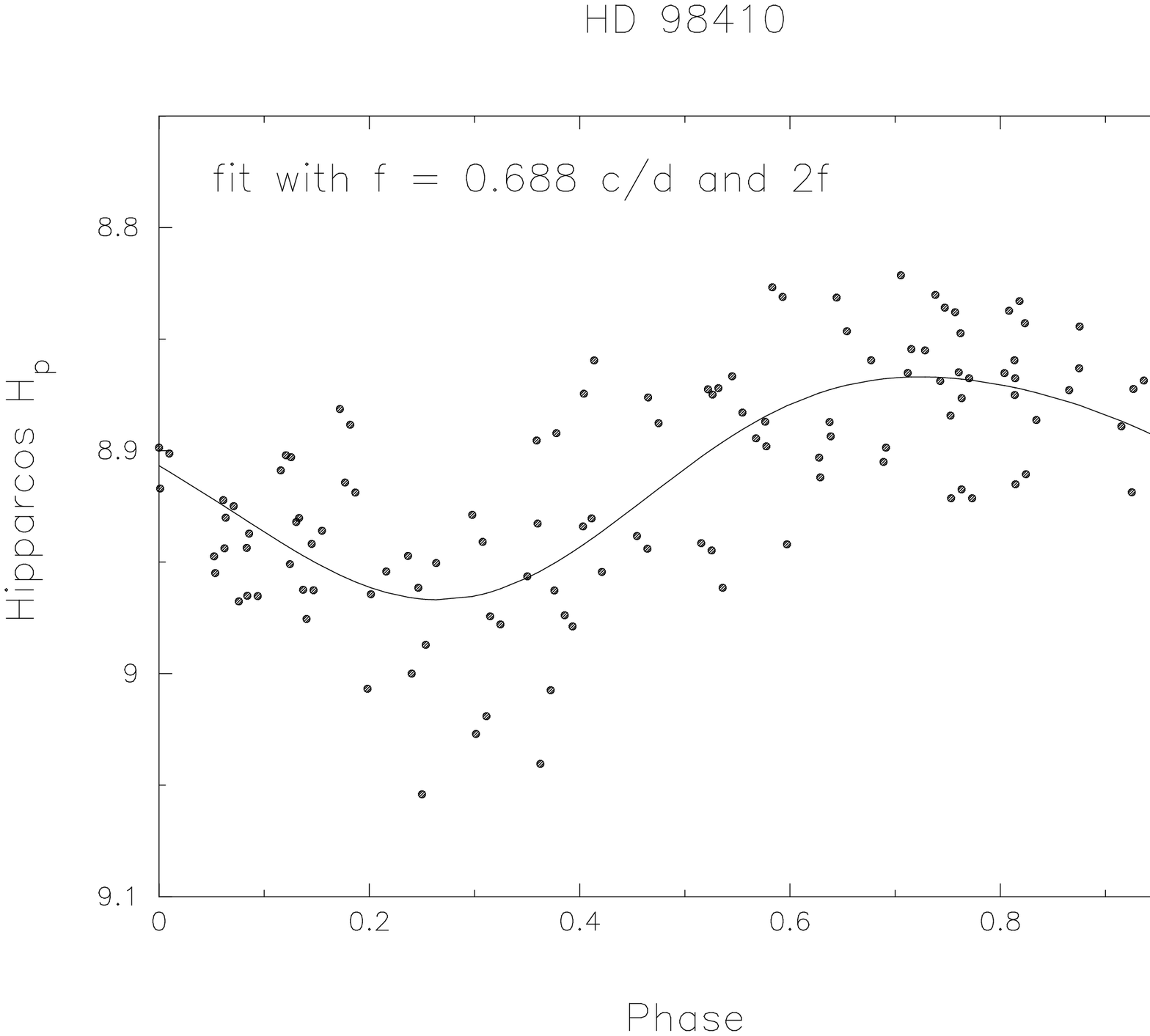}}}\\[0.5cm]
\caption{The HIPPARCOS lightcurve folded according to the dominant frequency
for the stars \object{HD\,96880} (B1\,Ia) and \object{HD\,98410} (B2.5\,Ib/II).
The dots are the observations and the full line is a least-squares fit for
the indicated frequency and its first harmonic.}
\label{vb}
\end{figure}

\begin{table*}
\tabcolsep=4pt
\caption{Results of the period analyses for the 28 sample stars.  The periods
P$_i$ are expressed in d and have an accuracy better than 0.001\,d. The
amplitudes A$_i$ and their $1\sigma$ error, $\sigma_{A_i}$, are given in
mmag. When the frequency's first harmonic is present in the lightcurve, the
label ``yes'' occurs in column ``H''. Column ``v.r.'' indicates the variance
reduction of the harmonic fit, in \%. $P_2$ is derived from the residuals, after
prewhitening with $P_1$.  The total variance reduction is obtained from a
harmonic fit to the data with both periods. In the last column the observed
peak-to-peak value (i.e., the difference between the largest and the smallest
observed magnitude) is given in mmag.}
\label{freq}
\centering
\begin{tabular}{rrcccccrcccrrr}
\hline \hline
HD & P$_1$ & A$_1 \pm \sigma_{A_1}$ & H & v.r. & P$_2$ & A$_2 \pm \sigma_{A_2}$
& H & v.r. &
(total v.r.) & $\triangle$H$_{P,obs}$  \\
\hline
47240 &  1.730 & 29 $\pm$ 4  &  no & 60\% &&&&&& 70 \\ 
51110 &  2.315 & 61 $\pm$ 6  &  no & 53\% &&&&&& 150 \\
53138 & 24.390 & 45 $\pm$ 5  &  no & 49\% & 3.690 & 35 $\pm$ 4 & no &
 45\% & (73\%) & 100 \\
54764 &  2.695 & 17 $\pm$ 4  &  no & 33\% &&&&&& 50 \\
68161 & 16.949 & 25 $\pm$ 2  &  no & 68\% &&&&&& 50 \\
80558 &  1.695 & 38 $\pm$ 6  &  no & 42\% & 5.814 & 26 $\pm$ 5 & no
 & 25\% & (57\%) & 90  \\
89767 &  1.153 & 28 $\pm$ 4  &  no & 34\% &&&&&& 100 \\
91024 &  2.398 & 36 $\pm$ 6  &  no & 42\% &&&&&& 100 \\
91943 &  6.452 & 24 $\pm$ 4  &  no & 40\% &&&&&& 70 \\
92964 & 14.706 & 43 $\pm$ 5  &  no & 45\% & 2.119 & 36 $\pm$ 5 & no & 43\% &
(71\%) & 110 \\
93619 &  4.310 & 27 $\pm$ 6  &  no & 24\% &&&&&& 100 \\
94367 &  7.937 & 48 $\pm$ 5  & yes & 58\% & 4.329 & 27 $\pm$ 5 & no & 34\% &
(73\%) & 100\\
94909 & 16.949 & 37 $\pm$ 6  &  no & 36\% & 1.256 & 25 $\pm$ 4 & no & 28\% &
(54\%) & 110\\ 
96880 &  2.475 & 46 $\pm$ 6  & yes & 47\% &&&&&& 120\\
98410 &  1.453 & 97 $\pm$ 8  & yes & 53\% & 8.696 & 40 $\pm$ 8 & yes & 45\% &
(75\%) & 230\\
102997 &  2.688 & 29 $\pm$ 6  & yes & 33\% & 2.976 & 23 $\pm$ 4 & yes & 37\% &
(58\%) & 110\\
105056 &  2.899 & 41 $\pm$ 8  &  no & 36\% & 7.299 & 51 $\pm$ 8 & yes & 46\% &
(67\%) & 130\\
106343 &  3.650 & 27 $\pm$ 6  & yes & 38\% & 3.906 & 23 $\pm$ 4 & no & 31\% &
(58\%) & 110 \\
108659 &  5.076 & 22 $\pm$ 4  &  no & 30\% &&&&&& 80 \\
109867 &  4.484 & 32 $\pm$ 3  &  no & 46\% & 4.785 & 20 $\pm$ 3 & no & 30\% &
(63\%) & 80 \\
111990 &  2.890 & 30 $\pm$ 4  &  no & 26\% &&&&&& 120 \\
115363 &  3.077 & 38 $\pm$ 8  &  no & 22\% &&&&&& 120 \\
141318 &  1.466 & 16 $\pm$ 2  &  no & 35\% &&&&&& 50 \\
147670 &  5.435 & 66 $\pm$ 4  &  no & 68\% &&&&&& 110 \\
148688 &  6.329 & 46 $\pm$ 6  &  no & 61\% & 1.845 & 24 $\pm$ 4 & no & 42\% &
(79\%) & 90 \\
154043 &  2.874 & 30 $\pm$ 7  &  no & 34\% &&&&&& 80 \\
168183 &  2.105 & 60 $\pm$ 5  &  no & 49\% &&&&&& 150 \\
170938 &  5.618 & 94 $\pm$ 18 & yes & 61\% &&&&&& 110 \\
\hline
\end{tabular}
\end{table*}

\begin{table*}
\tabcolsep=3pt
\caption{Results of the period analyses for the 12 comparison stars. 
Notations and units are the same as in Table\,\ref{freq}.} 
\label{freq_np}
\centering
\begin{tabular}{rrrccrrccrr}
\hline \hline
HD & P$_1$ & A$_1 \pm \sigma_{A_1}$ & H & v.r. & P$_2$ & A$_2 \pm \sigma_{A_2}$
& H & v.r. & (total
v.r.) &
$\triangle$H$_{P,obs}$  \\
\hline
 46769 & 0.1122 &  9 $\pm$ 2 &  no & 40\% & & & & & & 22  \\
 58350 & 6.6313 & 47 $\pm$ 6 &  no & 39\% & & & & & &145 \\
 64760 & 2.8090 & 11 $\pm$ 2 &  no & 34\% &  1.8447 &  9 $\pm$ 2 & no &  33\% &
(56\%) & 28\\
 75149 & 1.2151 & 33 $\pm$ 5 &  no & 43\% &  2.2143 & 20 $\pm$ 4 & no &  34\% &
(63\%) & 100\\
 86440 & 6.1996 &  9 $\pm$ 2 &  yes& 51\% &  0.2371 &  5 $\pm$ 2 & no &  22\% &
(62\%) & 36\\
106068 & 4.2644 & 42 $\pm$ 5 &  no & 46\% & 25.1889 & 24 $\pm$ 4 & no &  31\% &
(64\%) & 110\\
111904 & 3.3389 & 30 $\pm$ 4 &  no & 38\% & 19.1205 & 19 $\pm$ 3 & no &  32\% &
(58\%) & 110\\
125288 & 8.0906 &  9 $\pm$ 2 &  no & 21\% & & & & & & 38\\
149038 & 0.6390 & 19 $\pm$ 3 &  no & 61\% & & & & & & 37\\
157038 & 3.6430 & 48 $\pm$ 7 &  no & 68\% & 1.5432 & 27  $\pm$ 3 & no  & 50\% &
(84\%) & 100\\
157246 & 1.1811 & 10 $\pm$ 2 &  no & 40\% & 0.1281 &  9  $\pm$ 2 & no  & 40\% &
(64\%) & 20\\
165024 & 2.7693 &  6 $\pm$ 1 &  no & 28\% & 0.8455 &  6  $\pm$ 1 & no  & 24\% &
(46\%) & 24\\
\hline
\end{tabular}
\end{table*}

\begin{table}
\tabcolsep=4pt
\caption
{Projected rotational velocity, $v \sin i$, and its standard deviation (in
km/s) for all 40 sample stars, determined via the Fourier transform of
metallic lines.  When no metallic lines or only blended lines are available,
the corresponding values from SIMBAD are given in italic.  The column
indicated as ``n.o.l.'' gives the number of lines used to determine the mean
value of $v \sin i$ and its standard deviation.}
\label{vsini}
\centering
\begin{tabular}{rrrr|rrrr}
\hline \hline
HD       & $v \sin i$ 	& s.d.  & n.o.l. & HD  & $v \sin i$ & s.d. & n.o.l.  \\
\hline
46769 &  68 &  5 &  2 		& 102997 &  39 &  8 & 13 \\
47240 &  94 &  9 &  6 		& 105056 &  61 & 21 & 10 \\
51110 & \textit{n/a} &    &  	& 106068 &  26 &  5 &  6 \\
53138 &  38 &  4 & 14 		& 106343 &  44 &  7 & 14 \\
54764 & 108 & 15 & 10 		& 108659 &  29 &  5 & 12 \\
58350 &  37 &  5 &  4 		& 109867 &  50 & 14 & 16 \\
64760 & \textit{220} &    &  	& 111904 &  32 &  9 &  4 \\
68161 &  17 &  2 & 20 		& 111990 &  36 &  6 &  6 \\
75149 &  30 &  8 &  4 		& 115363 &  55 & 12 &  8 \\
80558 &  28 &  5 & 17 		& 125288 &  25 &  4 &  4 \\
86440 &  20 &  6 &  5 		& 141318 &  32 &  3 &  8 \\
89767 &  47 &  6 & 12 		& 147670 &  \textit{n/a} &    &    \\
91024 &  25 &  6 & 23 		& 148688 &  50 & 11 & 13 \\
91943 &  48 &  7 & 12 		& 149038 &  57 &  4 &  2 \\
92964 &  31 &  6 & 17 		& 154043 &  37 &  9 &  8 \\
93619 &  47 & 13 & 11 		& 157038 &  41 &  5 &  4 \\
94367 &  31 &  4 & 15 		& 157246 &  \textit{302}&    &    \\
94909 &  64 & 10 & 11 		& 165024 &  \textit{95} &    &    \\
96880 &  44 &  9 & 14 		& 168183 & 124 & 16 &  7 \\
98410 &  31 &  3 & 10 		& 170938 &  51 &  6 &  10 \\   
\hline
\end{tabular} 
\end{table}

\subsection{Determination of Physical Parameters \label{fittingprocedure}}

To investigate the position of our sample stars in the HR diagram on a solid
basis, we determine the fundamental parameters of the stars from our
high-resolution spectroscopic follow-up data.  For this purpose, we use the
non-LTE, spherically symmetric model atmosphere code FASTWIND (Fast Analysis
of STellar atmospheres with WINDs) which enables us to derive atmospheric
and wind parameters. The code was first introduced by \citet{Santolaya97}
and has meanwhile been updated for a number of improvements. The most
important ones concern the inclusion of line blocking/blanketing effects and
the calculation of a consistent temperature structure by exploiting the
condition of flux-conservation in the inner and the electron thermal balance
(e.g., \citealt{Kubat99}) in the outer part of the atmosphere. A detailed
description of the code was given by \citet{Puls05}, where also
representative results have been compared with those from alternative NLTE
codes, CMFGEN \citep{Hillier98} and WM-basic \citep{Pauldrach01}. Meanwhile,
a number of spectroscopic investigations of early type stars were performed
by means of FASTWIND, both in the optical (e.g., \citealt{Herrero2002,
Trundle2004, Repolust2004, Massey2004, Massey2005, Mokiem2005, Mokiem2006}) and in the
NIR \citep{Repolust2005}.

As stated earlier, for most stars two hydrogen lines (H$\alpha$ and
H$\gamma$), two He~I lines (the triplet line 4471\AA\ and the singlet line
6678\AA) and one silicon multiplet (Si~III~4552-4567-4574 for the early
types, up to B2, and Si~II~4128-4130 for the later spectral types) have been
observed.  The choice to measure these specific lines has not been made
randomly, but is based on their well-known specific dependency on one or
more of the basic parameters we want to unravel.  The model atom for silicon
used in this investigation is the same as used and described by
\citet{Trundle2004} in their analysis of SMC B supergiants.

\begin{figure}
\includegraphics[width=8.6cm]{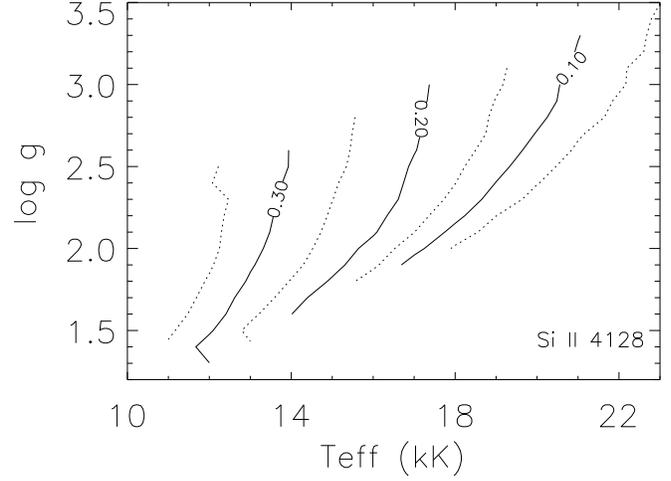}
\caption{Isocontour levels of equivalent line width (expressed in Angstrom)
for Si~II~4128 (solar Si abundance, negligible mass-loss), demonstrating that
this line is a good temperature indicator up to 20\,000~K.}
\label{isocontour_siII}
\end{figure}
\begin{figure}
\includegraphics[width=8.6cm]{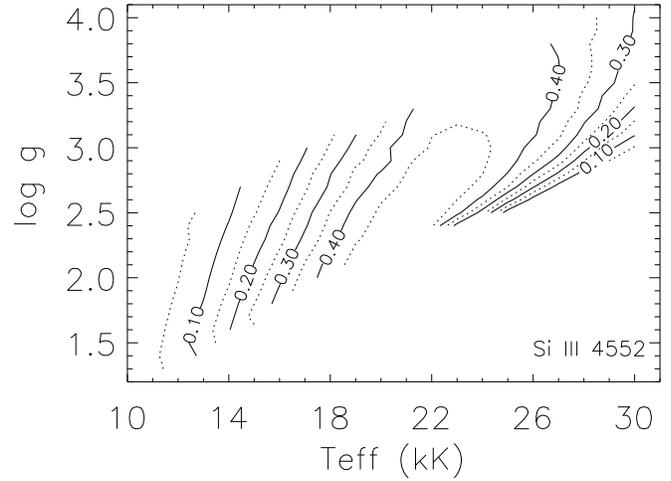}
\caption{As Fig.\,\ref{isocontour_siII}, but for
Si~III~4552. At temperatures around 23\,000K,
Si~III~4552 reaches a maximum in equivalent line width. The isocontours show
that from this maximum, theoretical line profiles can behave similarly towards
lower {\it and} higher temperatures, causing a dichotomy in the
determination of the effective temperature, when we lack Si~II or Si~IV.}
\label{isocontour_siIII}
\end{figure}
\begin{figure}
\includegraphics[width=8.6cm]{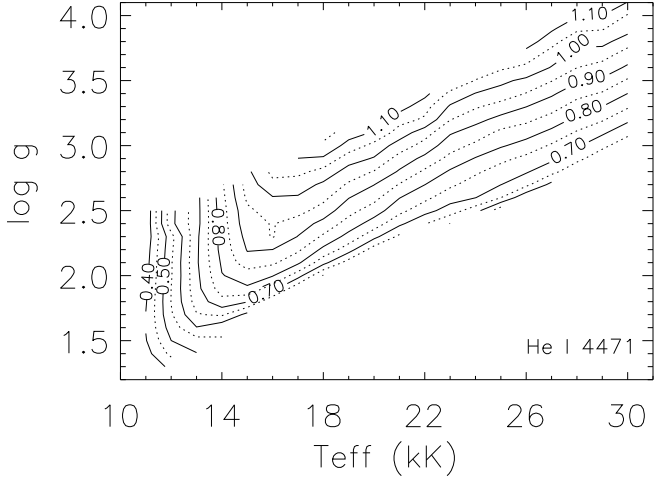}
\caption{Isocontour levels of equivalent line width (expressed in Angstrom)
for He~I~4471 (n(He)/N(H) = 0.1, negligible mass-loss). In the cool B-type
regime, He~I~4471 is a
perfect temperature indicator as the isocontours are almost vertical. At
higher temperatures, this line changes into a good diagnostic for the
gravity.}
\label{isocontour_heI}
\end{figure}

In Figs.\,\ref{isocontour_siII}, \ref{isocontour_siIII} and
\ref{isocontour_heI}, we show the isocontour levels of equivalent line width
of Si~II~4128, Si~III~4552 and He~I~4471 respectively, and their dependence
on the effective temperature and surface gravity based on an extensive grid
of synthetic models (see below). These figures show that Si~II is a very good
temperature indicator for B-type stars with an effective temperature below
20\,000~K (Fig.\,\ref{isocontour_siII}).  From then on, Si~III takes over as
a temperature diagnostic (Fig.\,\ref{isocontour_siIII}). Used in parallel,
both silicon multiplets could be used to infer information about the silicon
abundance. Since, however, we always have only one of both at our disposal,
we adopt a solar silicon abundance for our study ($\log$~(Si/H) = -4.45 by
number, cf.\ \citealt{Grevesse1998} and references therein), which has
changed only marginally (to $\log$~(Si/H) = -4.49) in the recent update of
the solar composition \citep{Asplund2005}.

The silicon abundance in B stars is heavily disputed. Depending on sample and
method, values range from roughly solar \citep{Gies1992, Gummersbach1998,
Rolleston2000} to a depletion by typically 0.3 dex \citep{Kilian1992,
Killian1994, McErlean1999, Daflon2004}, in both cases with variations by $\pm$
0.2 dex. Analyses of individual objects by \citet{Urbaneja2004} and
\citet{Przybilla2006} indicate a rather large scatter, again in the same range
of values. In view of this uncertainty, and the fact that \citet{Crowther2006}
in his analysis of Galactic B supergiants found no obvious problems in using
solar values, we also adopted this value. We will report on the influence of
this assumption on the final outcome later on.

He~I~4471 serves several purposes: for early B-types it is a good gravity
indicator (with a certain sensitivity to temperature as well), whereas for
the later B-types ($T_{\rm eff} <$ 15\,000~K) it becomes progressively
independent of gravity, but can be used to constrain the temperatures
perfectly (Fig.\,\ref{isocontour_heI}).  In those cases where the effective
temperatures (from Si) and the gravity (from H$\gamma$, see below) are well
defined, both He~I lines (He~I~4471 and He~I~6678) are useful to constrain
the helium content, as well as to check for the overall consistency of the
results, which in any case is the primary purpose of the second He~I
(singlet) line. The recent debate on the difficulty to use He~I {\it
singlet} lines as diagnostic tools (due to subtle line overlap effects of
the He~I resonance line in the FUV, \citealt{Najarro2006}) is (almost)
irrelevant in the present context, since it concerns only spectral types of
B0 and hotter.

Of course, He~I~4471 is not our primary gravity indicator. As usual, we
employ the Balmer line wings for this purpose, in our case particularly the
wings of H$\gamma$. Since the \Ha\ line is formed further out in the
atmosphere, it is affected by the stellar wind, and, for larger wind
densities, displays the typical emission or P Cygni type profile.  Depending
on the mass-loss rate, $\dot{M}$, the velocity law exponent, $\beta$, and
the terminal wind velocity, $v_{\infty}$ (with $v(r) \approx v_{\infty}
(1-R_{\ast}/r)^\beta$), the profile will have a more pronounced emission
profile, a steeper red wing or a broader absorption. Note that for lower
wind densities, only the core of \Ha\ is refilled by wind emission, and
the errors regarding the derived mass-loss rates become larger.

We used a ``by eye'' line profile fitting procedure to find the best fitting
synthetic FASTWIND spectrum for the observed line profiles of each object,
in order to derive their fundamental parameters. The synthetic profiles are
convolved with a rotational profile with appropriate $v \sin i$ (cf.
Table\,\ref{vsini}), whereas the macro-turbulence $v_{\rm macro}$ (well
visible in the wings of the Si-lines) is considered as an additional free
parameter, determined in parallel with the fit and accounting for a typical
radial-tangential distribution \citep{Gray1978}.

In a first step, we derive coarse parameters by using an extensive grid of
synthetic models, with effective temperatures ranging from 10\,000~K up to
30\,000~K (taking steps of 1\,000~K) and appropriate gravities in steps of
0.1 dex in $\log g$ (systematically shifting the gravity range to higher
values for higher effective temperatures). We consider the grid steps in
$T_{\rm eff}$ and $\log g$ as a {\it rough} estimate (see
Sect.\,\ref{erroranalysis}) for their uncertainty. For each $T_{\rm
eff}$/$\log g$ grid point, six different values (equidistant in the $\log$,
with appropriate boundaries) for the wind strength parameter $Q$ have been
calculated, with $Q$ = $\dot{M}/ (v_{\infty}\, R_{\ast})^{1.5}$, thus
combining the effect of mass-loss, terminal wind velocity and radius into
one single parameter. The reason for doing so is that different combinations
of these individual parameters, leading to the same $Q$-value, lead to
almost identical \Ha\ profiles \citep{Puls96}, whereas the other lines
remain also conserved \citep{Puls05}. In so far, the ``real'' fit quantity
is $Q$ and not the mass-loss rate itself, which is determined only after
$v_{\infty}$ and $R_{\ast}$ have been adopted and/or derived.

As a first guess, we adopted a ``typical'' radius for each grid point, as
calculated from evolutionary models and followed the observed terminal
velocities of massive hot supergiants from \citet{Prinja1998} to initialise
$v_{\infty}$. In combination with the predescribed $Q$-values, this leads to a
wide spread in mass-loss rates. As mentioned above, all models were calculated
for the ``old'' solar silicon abundance.  For the grid, we also considered 
solar values for the helium content (n(He)/n(H) = 0.10).  
However these values were adapted whenever required. 
Finally, all profiles have been calculated for three different values
of the micro-turbulent velocity, $v_{\rm micro}$, namely 5, 10 and 15 km/s.

After having derived a coarse fit by comparison with the synthetic spectra
from our model grid, we further refined the stellar and wind parameters (in
particular, $\dot M$ and $\beta$) in order to obtain the best possible fit.
Note that also the micro-turbulent velocity was adapted when necessary.  We
consider this quantity to be spatially constant and identical for {\it
all\/} investigated lines, i.e., we assume that $v_{\rm micro}$ does not
follow any kind of stratification throughout the atmosphere (see
Section\,\ref{erroranalysis}).

In a last step, the actual radius (contrasted to the adopted one) was
estimated in an iterative process, by comparing the V-band integrated
theoretical fluxes with corresponding absolute visual magnitudes M$_V$,
which in turn were taken from the calibrations by \citet{SchmidtKaler1982}
(except for HD\,168183 and HD\,111904, which are known cluster
members and hence we could derive M$_V$ from their distance). This procedure
(including corresponding updates of $\dot M$) was repeated until the
difference in input and output radius became negligible (usually one or two
iterations were sufficient).  

Note that the derived gravities are contaminated by centrifugal effects.
In order to obtain the ``true'' gravities needed to calculate, e.g., the
masses and to find the appropriate positions in the $\log
T_{\rm eff} - \log g$ diagram (Sect.\,\ref{position}), one has to apply a
``centrifugal correction'' with respect to the derived rotational velocities 
\citep[ and references therein]{Repolust2004}. These corrections have been
applied, and the corresponding values can be found in
Table\,\ref{finalparameters_SG} as entry $g_{\rm corr}$. For our further
analysis, we will use exclusively these data. 

\subsection{Results from the Line Fits \label{linefitting}}

Due to the restricted number of available spectral lines and because
different regimes of the parameter space lead to different accuracy, given
the available diagnostic, we subdivide our sample in three groups of stars,
depending on the reliability of the derived stellar parameters (mainly
$T_{\rm eff}$ and $\log g$). The first group (hereafter ``group~I'')
comprises sample stars for which we consider the results as very reliable.

The second group constitutes objects which suffer from the following
``defect''. From Fig.\,\ref{isocontour_siIII}, it is obvious that there will
be models at {\it each} side of the ``peak'' around 23\,000~K which produce
similar Si~III~4552 (and 4567-4574) profiles. Since these lines are our
major temperature indicator for the early type stars, 
this leaves us with two possibilities for
$T_{\rm eff}$, and only the strengths of additional Si~II or Si~IV lines
would allow for a conclusive answer. For sample stars of spectral type
B1-B2, only the Si~III multiplet is available to us.  In this case, we make
the appropriate choice between the high and the low temperature solution
relying either on He~I 4471 (which still has a certain sensitivity on the
temperature, but requires an assumption of the Helium abundance) or on its
spectral subtype (which we infer from SIMBAD or recent literature). Due to
the restricted reliability of this approach, all corresponding objects are
collected in ``group~II''. We discuss this approach in detail in the
appendix, for the prototypic example of HD\,54764 which is a B1 Ib/II
star. The effective temperatures and gravities derived for group I and II
objects will be finally used to obtain a new calibration of $T_{\rm eff}$ 
as a function of spectral subtype. 

``Group~III'' contains stars for which we have no means to derive accurate
stellar parameters, either because the objects are rather extreme or suffer from
additional effects not included in our atmospheric models, either because of
their peculiar spectrum that complicates a reliable fit, or a combination of
both. Therefore we classify these stars as ``unreliable'', and their parameters
should be considered with caution.

Apart from these three groups, we define a fourth group (hereafter
``group~IV'') consisting of the twelve comparison stars.  For these objects,
at most three lines have been observed (He~I~4471, H$\gamma$ and H$\alpha$),
which, in combination with our new $T_{\rm eff}$-calibration (see above),
will be used to estimate effective temperatures, surface gravities and wind
parameters.

For most of the sample stars we observed two H$\alpha$ profiles about one year
apart, in order to obtain an impression of the wind variability (for a detailed
investigation, see \citealt{Markova2005}).  We model each H$\alpha$ profile
separately by fixing all parameters except for the mass-loss rate.
The resulting two values for $\dot{M}$ are finally averaged to compute the
wind strength parameter $\log Q$ and the mean wind density which are
required for our further investigations in Sect.\,\ref{Mdot_variability}. 

In the appendix (only available in the electronic edition) we will
display and discuss, where necessary, the individual fits one by one, for
all four ``reliability groups''. The derived stellar and wind parameters are
presented in Table\,\ref{finalparameters_SG}. In the following, we give
some comments on general problems encountered during our fitting procedure.

\subsubsection*{Comments on General Problems \label{generalremarks}}

We have noted a certain discrepancy between the two lines of the
Si~II~4128-4130 doublet.  Theory predicts the Si~II 4130 line to be somewhat
stronger than the Si~II~4128 line, since the $gf$ value of the second
component is roughly a factor 1.5 larger than of the first component
(different atomic databases give very similar results). However, in most 
(but not all) cases, we {\it observe} an equal line strength for both lines. While
further investigation is needed, we approached this problem by finding a
compromise solution in the final fit.  The related errors are discussed in
Sect.\,\ref{erroranalysis}.

Thanks to the high spectroscopic resolution, a problem with the forbidden
component in the blue wing of He~I~4471 could be identified.  It appears that
this (density dependent) component is often (however not always) predicted too
weak for the early type stars ($<$ B1) and too strong for the late type stars
($>$ B2). For the cooler stars, this might be explained by a fairly
strong O~II blend at this position.

For some stars, He~I~6678 would need a higher macro-turbulence, $v_{\rm macro}$,
than the other lines. The clearest example of this situation is given
by HD\,92964, but, to a lesser extent, it also occurs in HD\,89767, 
HD\,94909, HD\,93619, HD\,96880 and HD\,106343, all of them being early type stars.
Their He\,I~6678 'fits' show that there must be an additional broadening mechanism
which we do not understand at present. 

In a number of cases we were not able to reproduce the shape of the
H$\alpha$ profile, mainly because of two reasons. On the one hand, the
assumption of spherical symmetry adopted in FASTWIND (and almost all other
comparable line-blanketed NLTE codes\footnote{but see \citet{Zsargo2006} for
recent progress regarding a 2-D modelling}) prohibits the simulation of
disks or wind compressed zones in case of large rotational speeds. On the
other hand, we neglected the effects of wind clumping (small scale density
inhomogeneities redistributing the matter into dense clumps and an almost
void interclump medium, see, e.g., \citealt[and references
therein]{Puls2006}), which can have a certain effect on the shape of the
\Ha\ profile and on the absolute value of the mass-loss rate. Recent
findings by \citet{Crowther2006} have indicated that this effect is rather
small in B-type supergiants though. Even if the detailed shape of \Ha\
is not matched, the error in the derived (1-D) mass-loss rate remains
acceptable, due to the strong reaction of the profile on this parameter, at
least if the wind densities are not too low. For such low wind densities then,
the discussed processes do not lead to any discrepancy with the observed
profile shape, since only the core of \Ha\ becomes refilled.

We stress that, in this kind of analysis, a reliable normalisation of the
spectra is of crucial importance. An incorrect normalisation of the silicon
lines, e.g., leads to errors in the derived effective temperatures, which
will propagate into the derived surface gravities.\footnote{To preserve the
H$\gamma$ profile, changing the temperature by 1\,000~K requires a
simultaneous change in gravity by roughly 0.1 dex.} Errors occurring in the
normalisation of H$\gamma$ additionally enlarge the error in $\log g$,
whereas an erroneous rectification of \Ha\ affects the $Q$-value and
thus the mass-loss rate. Though we were restricted to few selected orders
(thus cutting out the largest part of the available continuum), the
remaining spectral windows were generally sufficient to obtain a correct
normalisation thanks to the high S/N ratio which was obtained. For pure
emission H$\alpha$ profiles, on the other hand, this was more difficult, due
to the large width of the profiles.

A reliable {\it derivation} of terminal velocities, $v_{\infty}$, turned out
to be possible only for a restricted number of stars. As already explained,
we adopted the values determined by \citet{Prinja1998} from the UV as a
first estimate, with an extrapolation towards later B-types by using the
corresponding A supergiant data provided by \citet{Lamers1995}.  Both data
sets have been collected in Table~1 of \citet{Kudritzki00}.  By means of
these values, for most of the objects such a good fit in \Ha\ (and other
lines) had been obtained at first instance that further alterations seemed
to be unnecessary. Only for HD\,92964, the adopted $v_{\infty}$-value might
be too low, which could explain the mismatch of He~I~6678. However, the
first \Ha\ profile of this object is in complete emission and also the
second one displays only a tiny absorption dip, which makes it difficult to
derive a reliable value. Thus, also for this star, we adopted the UV-value.

On the other hand, by alternatively {\it deriving} $v_{\infty}$ from fitting 
the shape of \Ha, we generally found values which are either similar or
lower than the ones from the UV. Though the agreement was extremely good for
some objects, for others a discrepancy by more than a factor of two was
found, resulting in a mean difference of \Ha\ and UV terminal velocity
of 45\%. Note that the fits generally improved when adopting the UV values
finally.

Thus we conclude that it is not possible to precisely estimate $v_{\infty}$
from \Ha\ alone, at least in a large fraction of cases, where the typical
error by such an approach is given by a factor of two.

\section{Error Estimates \label{erroranalysis}}

Thanks to the high quality of our spectra, fitting errors due to resolution
limitations or instrumental noise do not play a role in our analysis. The
major problem encountered here is the very restricted number of available
lines, and the involved assumptions we are forced to apply (particularly
regarding the Si abundance and the micro-turbulent velocity, see below).

Apart from this principal problem, the major source of errors is due to our
``eye-fit'' procedure (contrasted to automated methods, e.g.,
\citealt{Mokiem2005}), which is initiated by manually scanning our
pre-calculated grid (see Section\,\ref{fittingprocedure}).  The effective
temperatures and gravities did not need refinement once a satisfactory
solution had been found from the grid, after tuning the mass-loss rates and
the velocity exponents. Therefore, the grid steps reflect the errors on
those parameters. In practice, this means that {\it typical\/} errors are of
the order of 1\,000K in $T_{\rm eff}$ and 0.1 in $\log g$, which is -- for
later spectral types -- somewhat larger than possible under optimal
conditions, i.e., if much more lines were available.  Finer step sizes or
further fine-tuning of the models with respect to effective temperature, on
the other hand, was regarded as irrelevant, due to the consequences of our
assumptions regarding abundance and micro-turbulence.

\subsection{Error Estimates for $T_{\rm eff}$}

Whenever many lines from one ion are present, the micro-turbulence can be
specified with a high precision given a ``known'' abundance. A few lines
from different ionisation stages, on the other hand, allow for a precise
temperature estimate, since in this case the ratios of lines from different
stages are (almost) independent of abundance (which is the reason that
spectral classification schemes use these ratios).  Missing or incomplete
knowledge becomes a major source of uncertainty if only few lines from
one ionisation stage are available. We assess the different effects
one by one.

\subsubsection{Influence of Si Abundance \label{error_abundance}}

Concentrating first on Si, a star with depleted abundance will display, at a
given temperature, weaker Si lines in {\it all} ionisation stages, and
vice versa, if Si is enhanced. Thus, if the line strengths decrease with
increasing temperature (as for Si~II, see Fig.\,\ref{isocontour_siII}), the
effective temperature would be overestimated if the actual abundance is
lower than the assumed solar value, and underestimated for increasing line
strength with temperature (e.g., for the low temperature region of Si~III,
cf. Fig.\,\ref{isocontour_siIII}).

To check the quantitative consequences of this uncertainty, we calculated,
for three different temperatures (15\,000, 20\,000 and 25\,000~K), various
models which are depleted and enhanced in Si by a factor of two (thus
comprising the lower values discussed in the literature, see
Sect.\,\ref{fittingprocedure}), and investigated how much the derived
temperature would change. When changing the effective temperature of the
model, one has to change the surface gravity in parallel, in order to
preserve the H$\gamma$ profile.  

For late-type stars at 15\,000~K (where only Si~II is available), such a
depletion/enhancement of Si corresponds to a decrease/increase of $T_{\rm
eff}$ by 2\,000K (and $\log g$ by 0.2). At 25\,000~K, which is a
representative temperature for the early-type objects for which we only have
the Si~III triplet, the effect was found to be identical.  At 20\,000~K, the
effect depends on whether we have Si~II or Si~III at our disposal, and the
overall effect is a bit smaller. If we have Si~II, we again find an
overestimation, now by 1\,500~K, if the star is depleted in Si, but assumed
to be of solar composition.  If we have Si~III, which is still gaining in
strength in this temperature regime, the effective temperature would be
underestimated by 1\,500~K.

In conclusion, due to the uncertainties in the Si abundance, we expect that
our temperature scale might systematically overestimate the actual one
(except for those group II objects which rely on Si~III, where an
underestimation is possible), by 1\,500~K to 2\,000~K {\it if\/} the average
abundance were actually 0.3 dex lower than solar.

\subsubsection{Influence of $v_{\rm micro}$ \label{error_vmic}}

One might argue that a Si depletion is not present in our sample, since
in almost all cases our secondary temperature diagnostic, He~I, was fitted
in parallel with Si without further problems. However, we have no
independent check of the He content (only one ion available).  In most
cases, the He~I line profiles were consistent with solar abundance, but
strongly evolved objects have processed material and should have a larger He
content (see, e.g., the corresponding discussion in \citealt{Crowther2006}).

Even if one would regard the consistency between Si and He as conclusive, it
depends on one of our additional assumptions, namely that the micro-turbulent
velocities are constant with height, i.e., identical for He and Si
lines.\footnote{For most of our objects, the analysed He lines are stronger
than the Si lines, i.e., they are formed above the Si lines.}. Though there
is no clear indication in the present literature that this hypothesis is
wrong, a stratified micro-turbulent velocity seems plausible. In such a case
(i.e., different $v_{\rm micro}$ for He and Si), we would no longer have a
clear handle on this quantity, and due to the well-known dependence of line
strength on this quantity (for Si lines, see, e.g., \citealt{Trundle2004},
\citealt{Urbaneja2004}, for He lines \citealt{McErlean1998}), an additional
source of error would be present.  From test calculations, it turned out
that a change of 1\,000~K (which is our nominal error in $T_{\rm eff}$)
corresponds to a change of $v_{\rm micro}$ by roughly 4 to 5~km/s in the Si
lines.

For the few objects with low macro-turbulent velocity, $v_{\rm macro}$,
and low rotational speed, we were able to directly
``measure'' $v_{\rm micro}$, thanks to the high resolution of our spectra.
In these cases, the profiles become too narrow in the wings and too strong
in the core, when decreasing $v_{\rm micro}$ (and vice versa when increasing
$v_{\rm micro}$). This behaviour cannot be compensated by changing $T_{\rm
eff}$.  For most of the objects, however, such precise information is washed
out by $v_{\rm macro}$. In the majority of cases we were able to obtain
satisfactory fits (He in parallel with Si) by keeping typical values
available in our grid, which are 15~km/s for early type objects and 10~km/s
for late type objects.  Changing $v_{\rm micro}$ by more than 2 to 3 km/s
would destroy the fit quality of either Si or He.

In conclusion, we are confident about the derived values of $v_{\rm micro}$
(and thus of the temperatures), provided that the He and Si lines are
affected by a similar micro-turbulent broadening, i.e., that stratification
effects are negligible.

\subsubsection{Influence of the Si~II Problem}

As mentioned in Section\,\ref{generalremarks}, there is a discrepancy
between the two lines of the Si~II~4128-4130 doublet for most of our
late-type objects. By allowing for a compromise solution (in which
Si~II~4128 is predicted as too weak and Si~II~4130 as too strong), we
minimise the error.  Indeed, to fit {\it either\/} of both lines
perfectly, we would have to change the effective temperature by roughly 500~K,
which is well below our nominal error.

\subsection{Error Estimates for Other Quantities\label{error_others}}

Although the fit-/modelling-error in $\log g$ is $\pm$0.1 for a {\it given\/}
effective temperature, $\log g$ itself varies with $T_{\rm eff}$ (as already
mentioned, typically by 0.1 for $\Delta T$ = 1\,000~K), so that for a
potentially larger error in $T_{\rm eff}$ (due to under-/overabundances of
Si) also the gravity has to be adapted.

The errors for the other parameters follow the usual error propagation (for
a detailed discussion, see \citealt{Markova2004} and \citealt{Repolust2004}),
being mainly dependent on the uncertainty in the stellar radius, which in turn
depends on the validity of the used calibration of the absolute visual
magnitude from \citet{SchmidtKaler1982}. 
A precise error estimate of the latter quantity is difficult to
obtain, but at least for early and mid Ia supergiants (until B3 Ia) we can
compare this calibration with the results from \citet[ Sect.~2]{Crowther2006},
who derived $M_V$ values either directly from cluster membership arguments
or from average subtype magnitudes of Magellanic Cloud stars. Comparing their
results with ours, we find similar average values, with a 1-$\sigma$
scatter of \[\Delta M_V \approx \pm0.43\;{\rm mag},\] which will be adopted in
the following, also for the later spectral types and the other luminosity
classes. 
From this number and the error in $T_{\rm eff}$, the error in the radius
becomes \[\Delta \log R_{\ast}/R_{\sun} \approx \pm0.088,\]
which corresponds to 22\%, and is consistent with the differences in
the radii derived by us and by \citet{Crowther2006}, see also
Table\,\ref{comp_Crowther}. We subsequently find a typical
uncertainty in the luminosity of 
\[\Delta \log L/L_{\sun} \approx 0.22{\ldots}0.19\] 
for $T_{\rm eff}$ = 12\,000~K{\ldots}25\,000~K, respectively. The
wind-strength parameter, $\log Q$, can be determined with rather high
precision. Adopting a combined contribution of fit error and uncertainty 
in $T_{\rm eff}$ of $\pm0.05$ (\citealt{Repolust2004}, their Sect.~6.2),
and an additional contribution of $\pm0.1$ accounting for the temporal
variability (Table\,\ref{finalparameters_SG} and
Sect.\ref{Mdot_variability}), we find \[\Delta \log Q \approx \pm0.11.\]
The precision in $\Delta \log \dot{M}$ amounts to 
\[\Delta \log \dot{M} \approx \pm0.24,\] 
if we estimate the error in $v_{\infty}$ as 30\%. Finally, the error in the 
derived wind-momentum rate is \[\Delta \log D_{\rm mom}\approx \pm0.34,\] 
i.e., somewhat larger than the error in log $L$.

\section{Comparison with Other Investigations\label{Comparison}}

We compare the results of our analysis with corresponding ones from similar
investigations, in particular those by \citet{Crowther2006, McErlean1999} and
\citet{Kudritzki1999} for five stars in common.

\subsection{Comparison with the Analysis by
\citet{Crowther2006}\label{comparison_crowther}}

\begin{figure*}
\begin{tabular}{cc}
\rotatebox[origin=l]{90}{\hspace{3.cm} Relative intensity} \hspace{0.2cm}
\includegraphics[angle=90,width=8.6cm,height=8.cm]{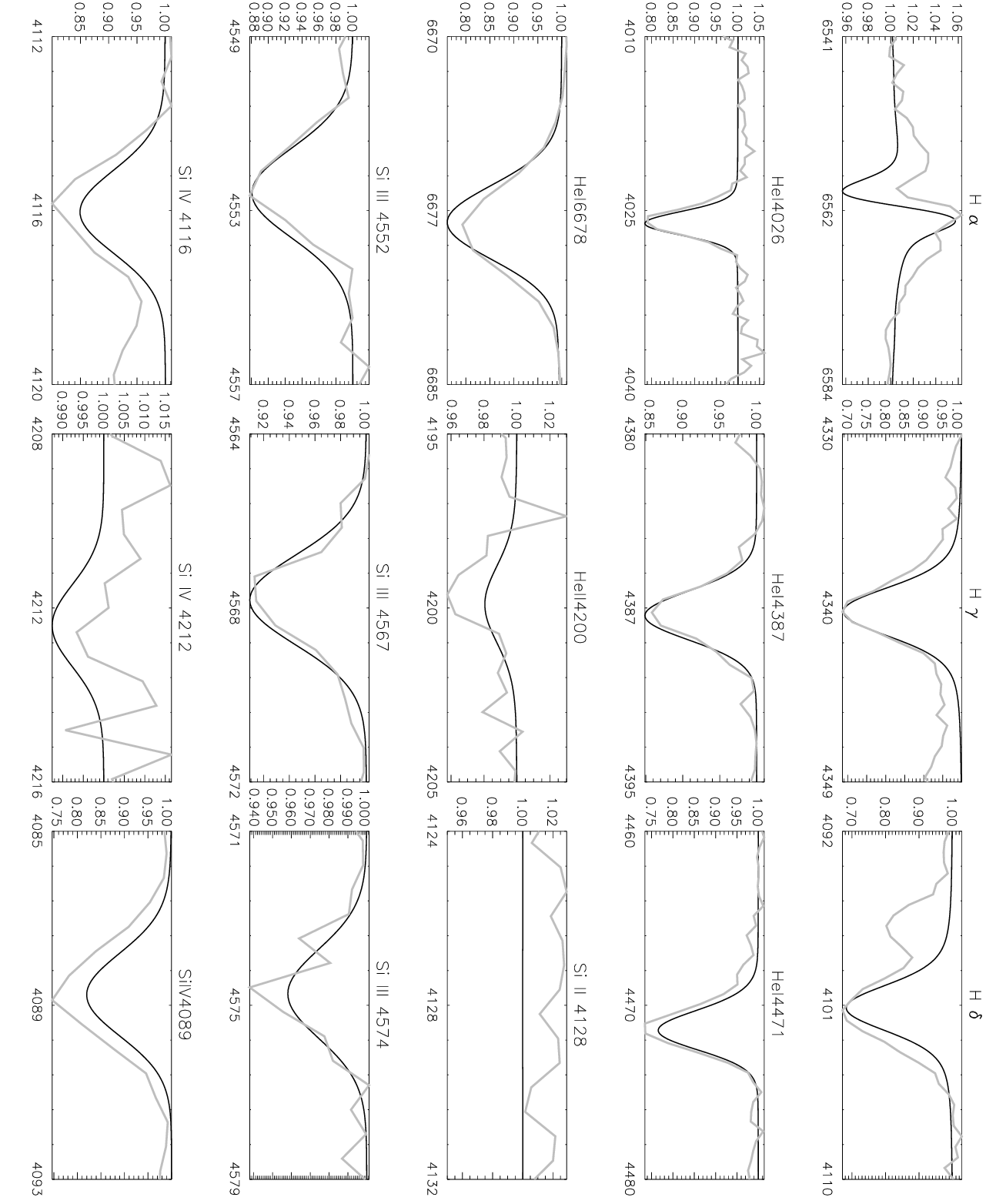} &
\rotatebox[origin=l]{90}{\hspace{3.cm} Relative intensity} \hspace{0.2cm}
\includegraphics[angle=90,width=8.6cm,height=8.cm]{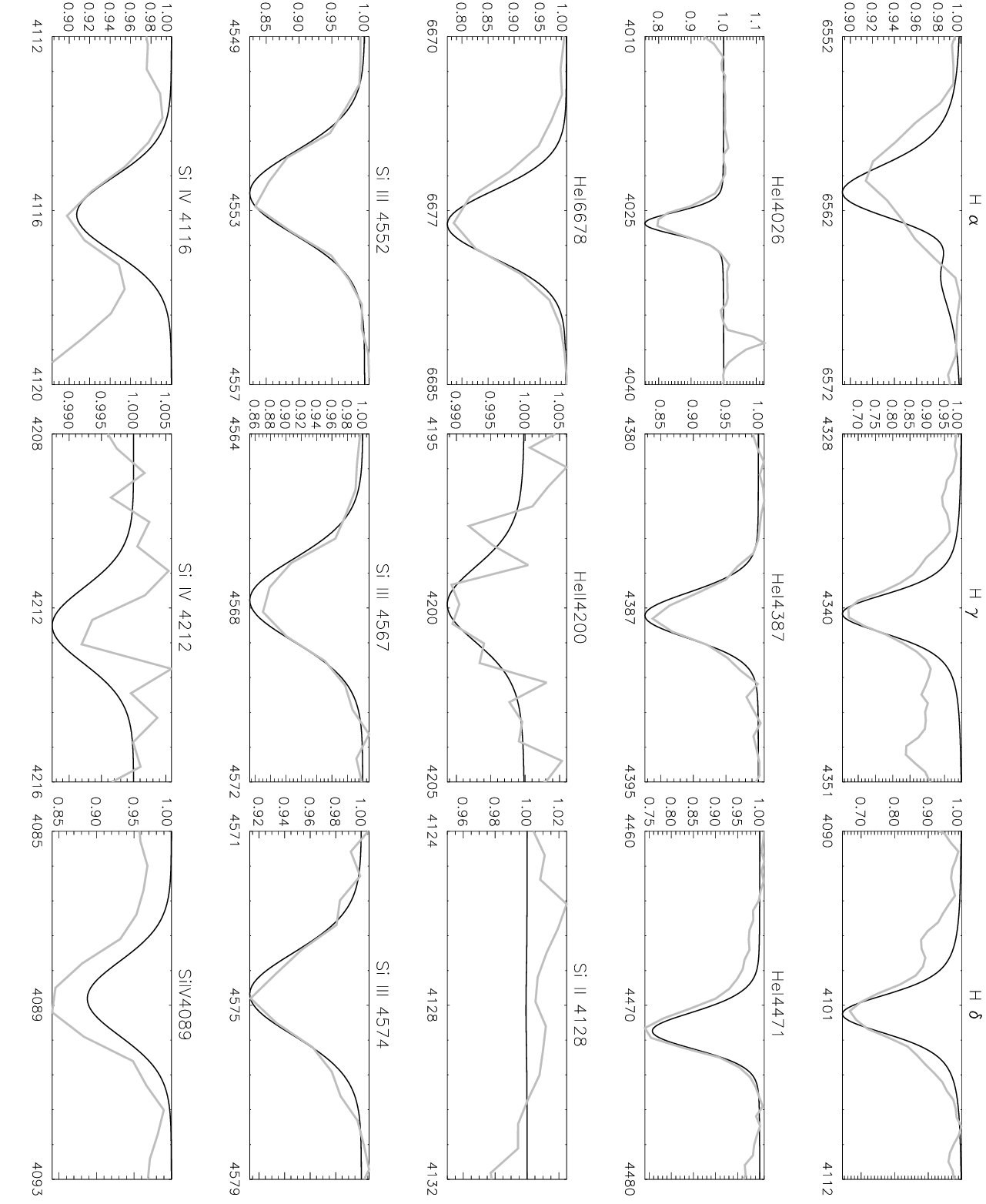}\\
\hspace{1.cm}(a) & \hspace{1.cm}(b) \\
& \\
\rotatebox[origin=l]{90}{\hspace{3.cm} Relative intensity} \hspace{0.2cm}
\includegraphics[angle=90,width=8.6cm,height=8.cm]{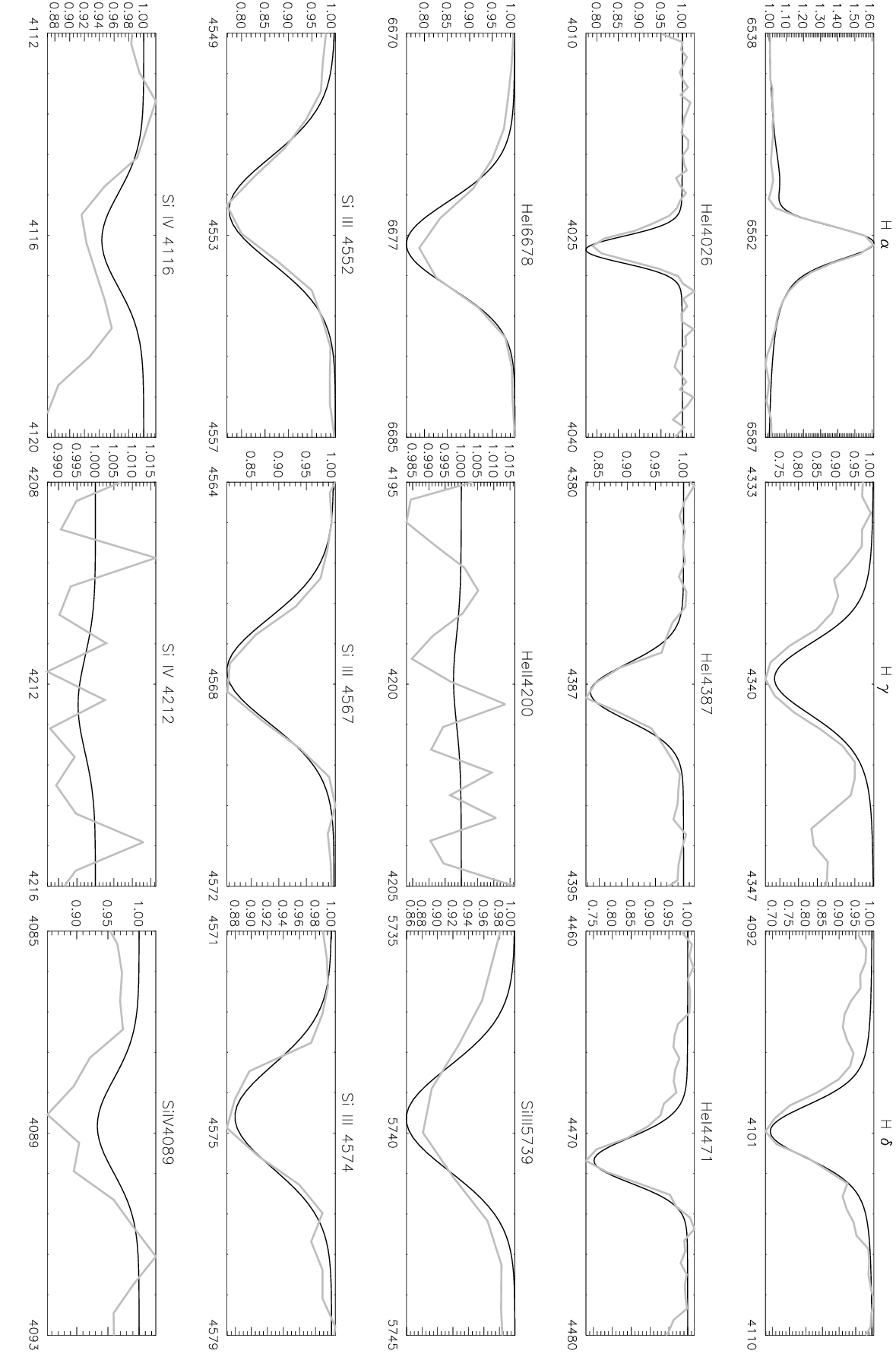} &
\rotatebox[origin=l]{90}{\hspace{3.cm} Relative intensity} \hspace{0.2cm}
\includegraphics[angle=90,width=8.6cm,height=8.cm]{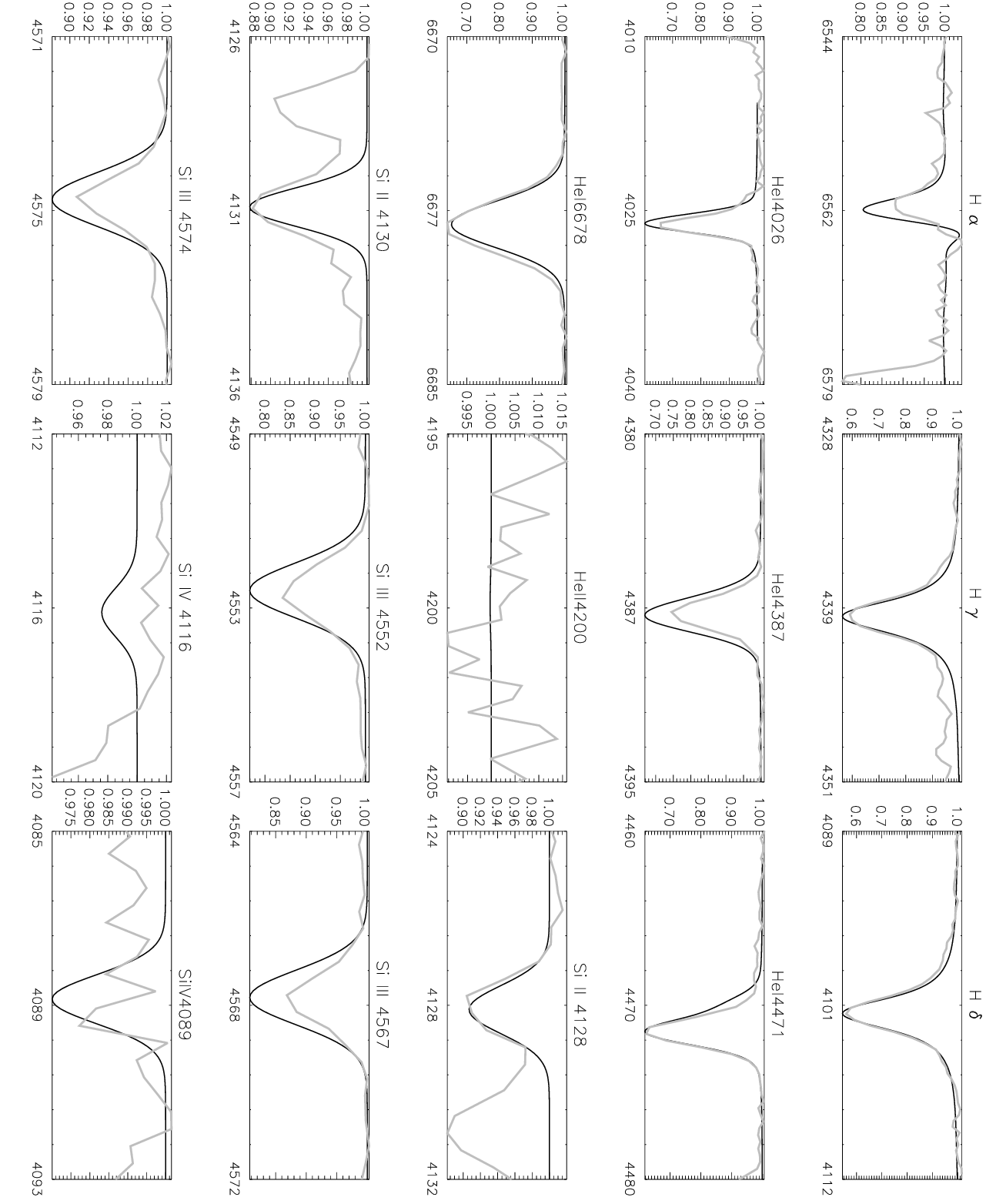}\\
\hspace{1.cm}(c) & \hspace{1.cm}(d) \\
\end{tabular}
\caption{Comparison between synthetic line profiles (downgraded) from our
best fitting models (black) and the low/intermediate dispersion
spectra of \citealt{Crowther2006} (grey), for four early
B-type supergiants in common: (a) HD\,94909; (b) HD\,91943; (c)HD\,148688; (d)
HD\,53138.}
\label{comparisonplots}
\end{figure*}

\begin{table*}
\caption{Comparison between the fundamental parameters derived in this study
and by \citet{Crowther2006, McErlean1999} and \citet{Kudritzki1999} for the
objects in common. The mass-loss rates were re-determined from the
low/intermediate-resolution spectra of Crowther et al., but do not differ
significantly from the values derived from our data. $Q$-values refer
to mass-loss rates in units of $M_{\odot}$/yr, terminal
velocities in km/s and stellar radii in $R_{\odot.}$}
\label{comp_Crowther}
\centering
\begin{tabular}{ccccccccccl}
\hline \hline
HD & $T_{\rm eff}$ & $\log g$ & M$_V$ & $R_*$  & $\log L/L_{\odot}$ 
& $v_{\infty}$ & $\beta$  & $\log Q$ & $v_{\rm micro}$ & reference\\
  & (kK)  & (cgs) &   &  (R$_{\odot}$) &  & (km/s) &  & (see capt.) & (km/s) &
\\
\hline
94909 & 25.0 & 2.7 & -6.9 & 36   & 5.65 & 1450 & 1.8 & -12.96 & 20 & this
study\\
      & 27.0 & 2.9 & -6.4 & 25.5 & 5.49 & 1050 & 1.5 & -12.34 & 10 &
\citet{Crowther2006} \\
\hline
91943 & 24.0 & 2.7 & -5.95& 23   & 5.19 & 1400 & 2.5 & -13.36 & 15 & this
study\\
      & 24.5 & 2.8 & -6.3 & 26.8 & 5.35 & 1470 & 1.2 & -13.01 & 10 &
\citet{Crowther2006} \\
\hline
148688& 21.0 & 2.5 & -6.9 & 42   & 5.49 & 1200 & 3.0 & -12.90 & 15 & this
study\\
      & 22.0 & 2.6 & -6.8 & 36.7 & 5.45 &  725 & 2.0 & -12.39 & 15 &
\citet{Crowther2006} \\
\hline
 53138& 17.0 & 2.15& -7.0 & 50   & 5.27 &  490 & 2.5 & -13.20 & 10 & this
study\\
      & 15.5 & 2.05& -7.3 & 65   & 5.34 &  865 & 2.0 & -13.57 & 20 &
\citet{Crowther2006} \\
      & 18.5 & 2.35&      &      & 5.04 &      &     &        & 10 &
\citet{McErlean1999} \\
      & 18.5 & 2.30&      & 39.6 & 5.22 &  620 & 2.5 & -13.61 & 40 &
\citet{Kudritzki1999} \\
\hline
 58350& 13.5 & 1.75& -7.0 & 65   & 5.10 &  250 & 2.5 & -13.17 & 12 & this study
(group IV)\\
      & 16.0 & 2.10&      &      & 5.36 &      &     &        & 15 &
\citet{McErlean1999} \\
\hline
\end{tabular}
\end{table*}

Our sample has four targets in common with the sample of Galactic early B
supergiants studied by \citet{Crowther2006}: HD\,94909 (B0 Ia), HD\,91943
(B0.7 Ib), HD\,148688 (B1 Ia) and HD\,53138 (B3 Ia). They used the
alternative NLTE model atmosphere code CMFGEN \citep{Hillier98} to derive
the physical parameters and wind properties of these stars. Compared to
their low dispersion CTIO and intermediate dispersion JKT/INT spectra, we
have the advantage of the very high resolution CES data. On the other hand,
Crowther et al. have complete spectra at their disposal (kindly
provided to us by P.\ Crowther). In order to compare the spectra with each
other, we first downgraded our synthetic spectral lines by convolving with an
appropriate Gaussian. Subsequently, we verified whether the best model we
found from a limited number of lines also provides a good fit to the
additional H, He and Si lines in the complete spectrum. In this procedure,
we only adapted $\dot{M}$ when necessary to fit the H$\alpha$ profile.
Crowther et al. used a solar Si abundance as well and 0.20 by number
for the helium abundance. The complete comparison is summarised in
Table\,\ref{comp_Crowther}.\\

\noindent \textbf{\object{HD\,94909} (B0 Ia)} From the CTIO spectrum of this
star, Crowther et al. estimated the effective temperature to be 27\,000~K and
$\log g$ 2.9.  It is impossible to fit our Si~III lines with such a high
temperature, because they become far too weak. Instead, we derive an effective
temperature of 25\,000~K, in combination with a $\log g$ of 2.7. When we compare
our (degraded) best fitting model with the CTIO spectrum
(Fig.\,\ref{comparisonplots}(a)), we see that Si~IV is predicted too weak, which
explains the higher temperature found by Crowther et al.  By exploring the
neighbouring parameter space, it turned out that we cannot simultaneously fit
Si~III and Si~IV, and we suggest that this star is {\it over}abundant in Si.\\

\noindent \textbf{\object{HD\,91943} (B0.7 Ib)} For this star, only a few
additional lines (besides those measured by us) are available due to the high
noise level. In Fig.\,\ref{comparisonplots}(b) we show
Si~III~4552-4567-4574 for consistency, together with lines
of an additional ionisation stage of Si (Si~II~4128 and Si~IV~4089/4116/4212) 
and helium (He~I~4026, He~I~4387 and He~II~4200). There might be a problem with the
normalisation of the Si~IV~4212 and Si~IV~4089 profiles, but still it is
clear that the strength of the observed and theoretical profiles agree
satisfactorily. In view of the low dispersion, we obtain a reasonable fit,
which gives us confidence in our results.\\

\noindent \textbf{\object{HD\,148688} (B1 Ia)} is often used as a comparison
star in UV studies as a galactic counterpart for early B-type supergiants in
M31, M33 or the SMC \citep[respectively]{Bresolin2002, Urbaneja2002,
Evans2004}. This star is one out of a few that show no radial-velocity changes.
We have only a few additional lines available in the CTIO spectrum. It is
very encouraging that they all nicely confirm our results (see
Fig.\,\ref{comparisonplots} (c)).  In order to fit the Si~III~triplet, we
need an effective temperature of 1\,000~K lower than the one suggested by
\citet{Crowther2006}.\\

\noindent \textbf{\object{HD\,53138} (B3 Ia)} can surely be named one of the
most ``popular'' B-type supergiants studied until now. Let us first
concentrate on our high resolution spectrum (Fig. A.3. in the appendix). 
Clear variations in the wind outflow are registered. In
the first measurement of H$\alpha$, the P Cygni profile has only a tiny
absorption trough and a considerable emission, whereas the second profile
indicates a much lower wind density.  This star is one of the objects to
show the discrepancy between predicted and actual line strength of the Si~II
doublet components (Sect.\,\ref{generalremarks}) for which we adopted a
compromise solution. In this way, our best fitting model gives an effective
temperature of 18\,000~K and a $\log g$ of 2.25.

When degrading the resolution of this best fitting model to the resolution
of the full spectrum provided by Crowther et al. (originating from the LDF
Atlas\footnote{\tiny
http://www.ast.cam.ac.uk/STELLARPOPS/hot\_stars/spectra\_lib/mw\_library/mw\_library\_index.html}
), we find some discrepancy in the Si~III triplet (see
Fig.\,\ref{comparisonplots} (d)). This discrepancy can be resolved by
decreasing the temperature in combination with either a lower micro-turbulent
velocity or a depletion in Si, since also Si~IV~4089 seems to be a bit too
strong. To cure this problem, we can go down as far as 17\,000~K (with log
$g$ = 2.15 and $v_{\rm micro}$ = 10~km/s), which is still 1\,500~K higher
than the value derived by \citet{Crowther2006} ($T_{\rm eff}$ = 15\,500 K,
log $g$ = 2.05). To find an explanation for this difference, we had a closer
look at their spectral line fits. Though the overall fit is good, the Si
lines are not matched perfectly. In particular, at their value of $T_{\rm
eff}$ = 15\,500 K, the Si~II~4128-4130 profiles are predicted too strong,
whereas Si~III~4552-4567-4574 is predicted too weak. This disagreement in
the ionisation balance suggests that the effective temperature should be
somewhat higher. We use the 17\,000~K model, which is at the lower bound of
the quoted error range of our original analysis, as a solution to all
discussed problems.

A summary of the main parameters resulting from both studies is given in
Table\,\ref{comp_Crowther}. Interestingly, in all cases, the low/intermediate
resolution H$\alpha$ profile observed by Crowther et al.\ lies amidst our two
H$\alpha$ profiles, so that the inferred mass-loss rates are very similar and
the variability is not large (cf.  Sect.\,\ref{Mdot_variability}).  As already
pointed out, the differences in radii obtained by us and Crowther et al.\ are of
similar order.  On the other hand, the $\beta$-values implied by our fits are
generally larger than the ones from \citet{Crowther2006}. Accounting
additionally for the moderate differences in $T_{\rm eff}$ for the first three
objects (in these cases, our values are lower), this explains the lower $\log Q$
values found by us.  For HD\,53138, on the other hand, the $\beta$ values are
similar, whereas our effective temperature is {\it larger}, explaining the
{\it higher} $\log Q$ value.

\subsection{More Comparisons\label{compmcerlean}}

By means of the plane-parallel, hydrostatic, NLTE atmosphere code TLUSTY
(\citet{Hubeny2000}, \citet{McErlean1999} deduced the photospheric parameters
and CNO abundances of 46 Galactic B supergiants.  Effective temperatures were
mostly obtained from the ionisation balance of Si. Two of their objects are in
common with our sample (HD\,53138 and HD\,58350, a B5Ia Group IV object, see
Appendix \ref{comments_groupIV}), and Table\,\ref{comp_Crowther} displays the
comparison. The effective temperature for HD\,53138 from McErlean et al.
compares well with our {\it high temperature solution\/} for this object, i.e.,
differs significantly from the much lower value derived by Crowther et al.
(see above), and also the gravities are consistent. Regarding HD\,58350, we have
{\it adopted\/} a temperature consistent with the calibrations provided in the
next section, which is significantly lower than the value found by
McErlean et al.  For details, we refer to Appendix \ref{comments_groupIV},
but we point out that the complete JKT spectrum allows for an increase in
temperature by roughly 1\,000~K.

Finally, on the basis of the temperature scale derived by \citet{McErlean1999},
\citet{Kudritzki1999} analysed the wind properties of a sample of early/mid B
and A supergiants, by means of a previous, unblanketed version of
FASTWIND. Their value for the wind-strength parameter of HD\,53138 coincides with
the value provided by Crowther et al., i.e., is lower than our result.  We
regard this agreement/disagreement as not conclusive, since (i) the \Ha\ spectra
used by Kudritzki et al.\ are different (less emission), and (ii) the analysis
was based on unblanketed models, which in certain cases can lead to differences
in the derived mass-loss rate (cf.\ \citealt{Repolust2004}, their Fig.~21). A
further, more comprehensive comparison will be given in
Sect.\,\ref{WLR_section}, with respect to modified wind-momenta.

Summarising the results from this comparative study, we conclude that, despite
of the multidimensional parameter space we are dealing with and the
interdependence of the parameters, we are able to derive rather accurate values
from only a few selected lines (of course, within our assumptions, in particular
regarding the Si abundance). This enables us to provide (and to use) new
calibrations for the effective temperatures, which will be discussed in the next
section.

\section{$T_{\rm eff}$ Calibration and Group IV Objects \label{calib}}

Thanks to the (almost) complete coverage of the B star range and the large
number of objects available, we are able to derive a $T_{\rm eff}$ calibration
as a function of spectral type, which we subsequently use to derive the
temperatures of our group IV objects.  The spectral types were taken from the
literature. In case different assignments were given, we have
provided arguments why we prefer one above the other, in the Appendix. Ideally,
one would want to re-assign spectral types from our high-quality data as in,
e.g., \citet{Lennon1993}.  However, one needs a fair number of spectral lines to
do this in a safe way.  Since we have only a few H, He and Si lines, we
preferred to use the spectral classifications from the literature, keeping in
mind that some of them may not be very refined. We confirmed or adopted the
luminosity class on the basis of the strength of H$\alpha$.

\begin{figure}
\centering
\includegraphics[angle=90,width=8.6cm]{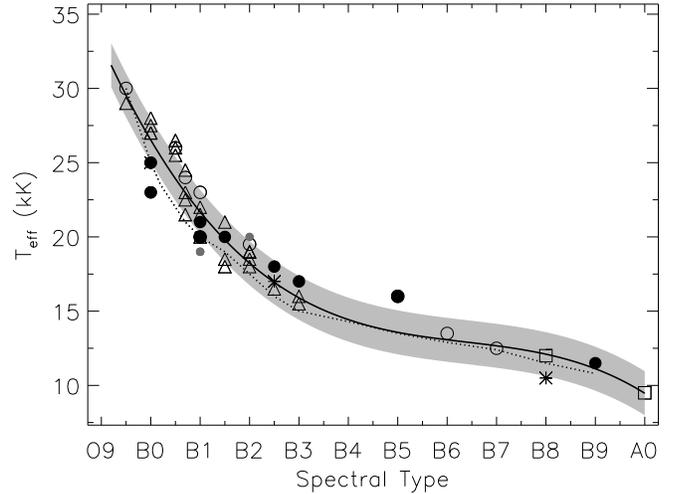}
\caption{$T_{\rm eff}$ as a function of spectral type for the sample B-type
supergiants: group I (black circles), group II (small grey filled circles)
and group III (asterisks). For group I, we subdivided according to luminosity class: Ia
(filled symbols) and Ib (open symbols). Triangles denote the early B Ia
supergiants from \citet{Crowther2006}, and rectangles two late-type Ia
supergiants with very precise parameters from \citet{Przybilla2006}. The dotted
line represents the effective temperature scale of \citet{Lennon1993} and the
full line shows our newly derived effective temperature scale based on group
I/II stars in addition to the objects from \citet{Crowther2006} and
\citet{Przybilla2006}. The grey area denotes the standard deviation of our
regression fit.}
\label{teff_spt}
\end{figure}

In Fig.\,\ref{teff_spt} we see that the effective temperature follows an
systematic decrease with spectral type. To derive the temperature
calibration, we joined our results for group I/II objects with those from
\citet{Crowther2006} (who used assumptions similar to ours in their
analysis) and added two more objects from \citet{Przybilla2006} at the low
temperature edge. In contrast to the errors inherent to our
analysis\footnote{generally, $\Delta T$ = 1\,000~K, except for the two B5
objects with $\Delta T$ = 2\,000~K (see below).}, which are identical with
those from \citet{Crowther2006}, these two objects (HD\,34085 = $\beta$ Ori
(B8 Iae:), HD\,92207 (A0 Iae)) could be analysed in a very precise way by
exploiting the complete spectrum, with resulting errors of only $\pm$ 200~K. 

By performing a polynomial fit to these data (including the quoted errors), 
we derive the following effective temperature scale for B-type
supergiants,
\[T_{\rm eff} = 26522.63 - 5611.21 x + 817.99 x^2 - 42.74 x^3,\]
with $x$ the spectral subclass. Note that we have included both luminosity
subclasses (Ia and Ib) to obtain this fit, and that the inclusion of the
objects by \citet{Crowther2006} changed the results only marginally compared
to a regression using our and Przybilla's data alone. The obtained standard
error for this regression is $\pm$1\,500~K.

When we consider the three group~III objects in this figure (the asterisks),
they match our derived calibration perfectly although we considered their
parameters as unreliable. The B0 star HD\,105056 lies at exactly the same
position as HD\,94909, just at the lower edge of our error bar. HD\,68161 (B8
Ib/II?) will be excluded from our further analysis, due to problems regarding
its classification (see Appendix\,\ref{comments_groupIII}).

As the Ia supergiants have, because of their larger luminosity, more mass
loss than Ib objects, they suffer more from line blanketing and mass-loss
effects. This is why supergiants of luminosity class Ia are expected to
appear cooler than the ones with luminosity class Ib.
Comparing the filled and open circles in Fig.\,\ref{teff_spt}, this
is indeed exactly what we observe.

\begin{figure*}
\begin{minipage}{8.8cm}
\resizebox{\hsize}{!}
   {\includegraphics[angle=90]{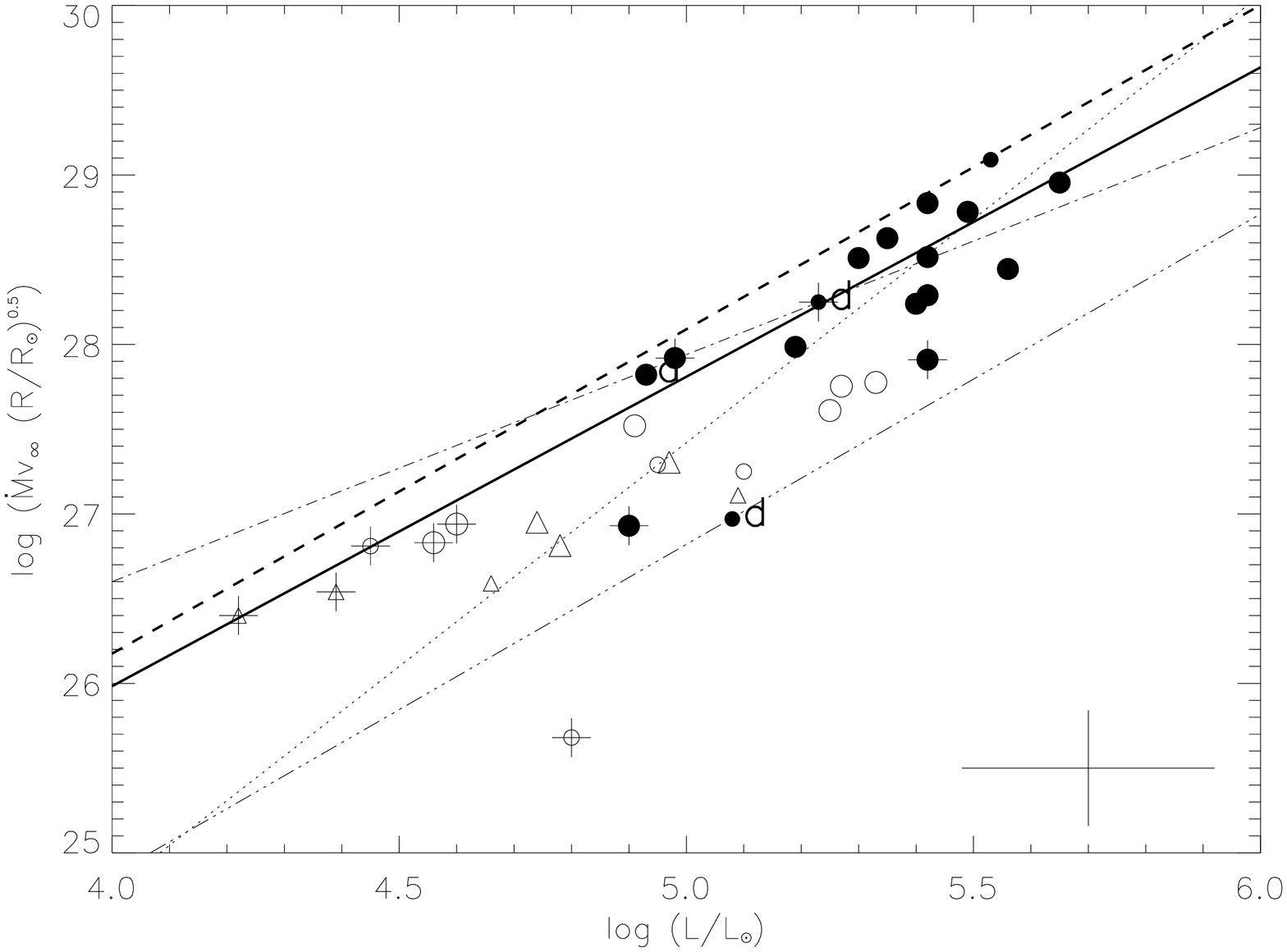}}
\end{minipage}
\hfill
\begin{minipage}{8.8cm}
   \resizebox{\hsize}{!}
   {\includegraphics[angle=90]{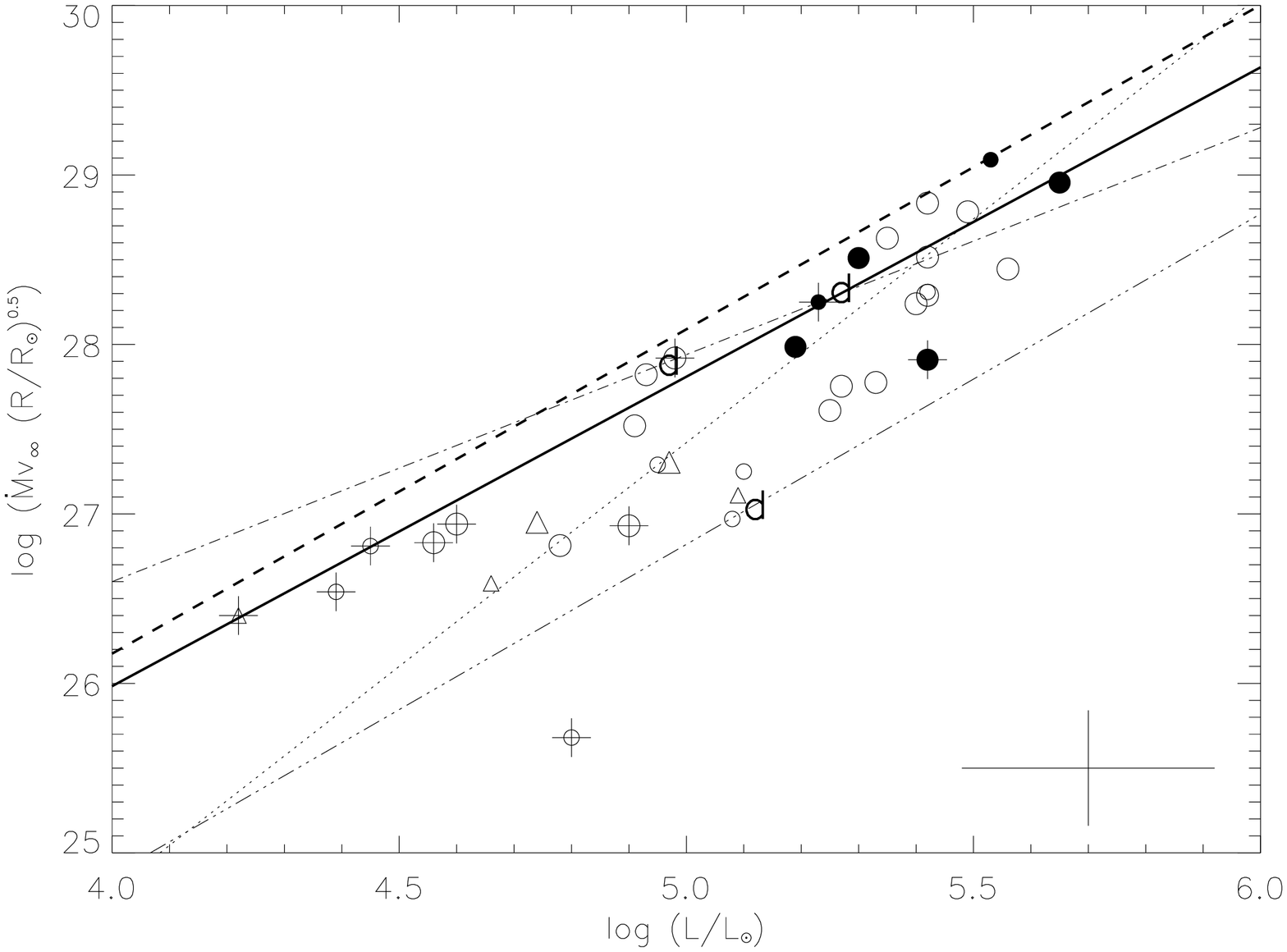}}
\end{minipage}
\caption{Wind momentum Luminosity Relation for our sample of B supergiants.
Only class I,II (large symbols) and IV objects (small symbols) have been
included, and the mass-loss rates
from both \Ha\ profiles were averaged. Bold solid/dashed:
Theoretical predictions from \citet{Vink2000} for objects with $T_{\rm eff} >$
23\,000~K and 12\,500~K $< T_{\rm eff} <$ 23\,0000~K, respectively.
Dashed-dotted, dashed-dotted-dotted and dotted are
the ``observed'' relations from \citet{Kudritzki1999}, for early, mid B
supergiants and A supergiants, respectively.\newline
Left: Results from this study, objects denoted as a function of
spectral type. Filled circles: B0/1; open circles: B2{\ldots}B5; triangles: B6
and
later.\newline
Right: as left, but objects denoted as a function of $T_{\rm eff}$.
Filled circles: $T_{\rm eff} > $23\,000~K; open circles: objects with
12\,500~K $< T_{\rm eff} <$23\,000~K; triangles: $T_{\rm eff} <$12\,500~K.
Note that the theoretical bi-stability jump is predicted at 23\,000~K.\newline
Typical error bars are indicated in the lower right. Overplotted crosses 
denote objects with \Ha\ in absorption, which have a larger error in
$\log Q$ than the other objects, due to the uncertainty regarding $\beta$.
``d'' denotes three objects with disk-like features (HD\,47240, HD\,64760
and HD\,157246).}
\label{WLR_figure}
\end{figure*}

Our temperature scale agrees well with the one provided by
\citet{Lennon1993}, which is indicated with the dotted line. The largest
differences occur between B0 and B2, where our scale lies roughly 1\,500~K
higher, mostly due to the objects analysed by \citet{Crowther2006}.

Let us point out one problem concerning the two B5 objects (with identical
$T_{\rm eff}$). We see in Fig.\,\ref{teff_spt} that both temperatures lie
clearly above the errors of the calibration. We can still fit the Si~II
lines well by decreasing the temperature (from 16\,000~K to 14\,500~K), when
we reduce $v_{\rm micro}$ by 4~km/s (see Sect.\,\ref{error_vmic}), which
would still be acceptable for these late-type objects. However, with such a
low value, a simultaneous fit to the helium {\it and\/} Si lines becomes
impossible, violating our general fitting strategy. Thus, if the lower
temperature would be the actual one, this might be due to two reasons,
outlined already in Sect.\,\ref{erroranalysis}: either He and Si have different
micro-turbulent velocities, or Si is underabundant in both B5 targets.  Note
that the calibration itself remains rather unaffected by these objects,
since we used larger error estimates of $\Delta T = 2\,000~K$ here.

Finally, by means of this calibration (for particular values, see
Table\,\ref{calibration_table}), we are able to derive the fundamental
parameters for the group IV comparison objects. Using the He~I~4471 line as
a double check for a consistent effective temperature, the surface gravity
is obtained from H$\gamma$ and the wind properties from H$\alpha$.
Corresponding comments are given in Appendix\,\ref{comments_groupIV}, and
all results are summarised in Table\,\ref{finalparameters_SG}.

\begin{table}
\tabcolsep=3pt 
\caption {Effective temperature calibration for B-supergiants, based on the
results from Sect.~\ref{calib}, used to derive the effective temperatures
for the group IV objects.} 
\label{calibration_table}
\centering
\begin{tabular}{lc|lc}
\hline \hline
SpT & $T_{\rm eff}$ & SpT & $T_{\rm eff}$\\
\hline 
    O 9.5  &     29\,500  &   B 3  &     15\,800\\
    B 0    &     26\,500  &   B 4  &     14\,400\\
    B 0.5  &     23\,900  &   B 5  &     13\,500\\
    B 1    &     21\,600  &   B 6  &     13\,000\\
    B 1.5  &     19\,800  &   B 7  &     12\,600\\
    B 2    &     18\,200  &   B 8  &     12\,100\\
    B 2.5  &     16\,900  &   B 9  &     11\,100\\
\hline
\end{tabular}
\end{table}

\section{Wind-momentum Luminosity Relation \label{WLR_section}}

In Figure\,\ref{WLR_figure} we present the position of our galactic B
supergiants in the wind momentum - luminosity diagram, where the
modified wind momentum rate is defined as $D_{\rm mom}~=~\dot M
v_{\infty}~(R_{\ast}/R_{\odot})^{0.5}$. The presence of such a relation (with
wind-momenta being a power law of luminosity, and exponents depending on
spectral type and metalicity) is a major prediction of the theory of
line-driven winds (for details, see, e.g. \citealt{Kudritzki00}), and has
been used in recent years to check our understanding of these winds.

To compare our results with earlier findings, we provide results from
different investigations relevant in the present context. In particular, the
bold solid and dashed lines display the theoretical predictions from
\citet{Vink2000} for objects with $T_{\rm eff} >$ 23\,000~K and 12\,500~K $<
T_{\rm eff} <$ 23\,000~K, respectively, where the difference is related to
an almost sudden change in the ionisation equilibrium of Fe around $T_{\rm
eff} \approx $ 23\,000~K (from Fe~IV to Fe~III), the so-called bi-stability
jump \citep[ and references therein]{Vink2000}. Due to this change and below
this threshold, the line acceleration is predicted to increase in the lower
and intermediate wind, because of the increased number of driving lines
being available.

Dashed-dotted, dashed-dotted-dotted and dotted lines refer to the findings 
from \citet{Kudritzki1999}, who derived these relations from {\it observed}
wind momenta of early B, mid B and A supergiants, respectively.

The wind momenta of our sample objects have been overplotted, by using 
averaged mass-loss rates (from our {\it two} \Ha\ profiles, see also
Sect.\,\ref{Mdot_variability}), for group I/II (large symbols) and IV
objects (small symbols). Objects with disk-like features are indicated by
``d'', and typical error bars are displayed in the lower right. Finally, we
have denoted objects with \Ha\ in absorption by additional crosses, to
indicate that the wind-momenta of these objects are subject to errors
somewhat larger than typical, of the order of 0.3 dex, due to problems in
deriving reliable values for the velocity field exponent, $\beta$ (cf.
\citealt{Puls96} and \citealt{Markova2004}).

The left and right figure allow to compare our findings with the displayed
predictions and observations, both for objects as a function of
spectral type (left) and as a function of their position with respect to
the predicted bi-stability jump (right).

At first, let us point out that differences in their behaviour as a function
of luminosity class are minor, and that there is no obvious difference
between group I/II and group IV objects, i.e., the wind-momenta for 
periodic pulsators and comparison objects behave similarly.

From the left figure then, we see the following: The position of the early
B-types (B0/1, filled circles) is consistent with both the theoretical
predictions and the findings from \citet{Kudritzki1999}, except for two
objects with uncertain positions and one object with disk-like features.
Late B-type supergiants\footnote{Note that this is the first investigation
with respect to this class of objects.} (B6 and later, triangles) nicely
follow the observed relation for A supergiants (dotted), but are located
below theoretical predictions (dashed). Only for mid B supergiants
(B2{\ldots}5, open circles), we find a difference to earlier results.
Whereas Kudritzki et al. have derived a very strict relation for {\it all}
B2/3 supergiants of their sample (dashed-dotted-dotted), located
considerably below the relation for early type objects (a finding which
still lacks theoretical explanation), our sample follows a non-unique trend.
Though high luminosity objects (with $\log L/L_{\odot} > 5$) behave similar
to the sample studied by Kudritzki et al. (with a somewhat smaller offset),
lower luminosity objects follow the (theoretical) hot star trend (bold),
but might also be consistent with the low $T_{\rm eff}$ relation
(dashed) when accounting for the larger errors in $D_{\rm mom}$ (\Ha\
absorption objects).

The right figure displays our sample objects as a function of $T_{\rm eff}$,
differentiating for objects with $T_{\rm eff} > $23\,000~K (filled circles) 
and objects with 12\,500~K $< T_{\rm eff} <$23\,000~K (open circles), i.e.,
with temperatures below and above the predicted bi-stability jump. Cooler
objects with $T_{\rm eff} <$12\,500~K (no predictions present so far) are
indicated by triangles. The situation is similar as above: Almost all hotter
objects follow the predicted trend (filled circles vs. bold line), and also
a large number of cooler objects follows this trend or the alternative one
(dashed). Additionally, however, there are a number of cool objects which
lie below both predictions, particularly at intermediate luminosities with
$4.7 < \log L/L_{\odot} < 5.4$.

Interestingly, there is only one ``real'' outlier, an \Ha\ absorption object
of group IV (HD\,157038), which is an evolved star being significantly
enriched in He and N. Note that the three objects with disk-like features
display a rather ``normal'' wind-momentum rate, though one of those
(HD\,157246) lies at the lower limit of the complete sample. Nevertheless,
we will omit the latter stars (as well as group III objects) for all further
analyses regarding mass-loss to obtain clean results.
 
Summarising the results from above, no obvious differences compared to
earlier findings could be identified within our complete sample (pulsators
{\it and} comparison objects), with the only exception that the {\it unique}
trend for mid B-type supergiants claimed by \citet{Kudritzki1999} could not
be confirmed by us: Part of these objects seem to behave ``normal'', whereas
another part shows indeed a momentum deficit. Note also that our findings
are in no contradiction to those from \citet{Crowther2006} (keeping in mind
that their sample only included B supergiants not later than B3).

\section{Position in the HR and log T$_{\rm eff}$-log g-Diagram}
\label{position}

After having derived the atmospheric parameters with satisfactory precision,
we are now in a position to tackle the major objective of this paper, namely
to try and clarify if the opacity mechanism could be held responsible for
the variability of our targets, as proposed by \citet{Waelkens98} and
\citet{Aerts00a}. In order to do so, we calculated the luminosities of the
targets from the effective temperatures and stellar radii derived from our
analysis. When comparing the relative difference in temperature and
luminosity between our values and those derived by \citet{Waelkens98}, we
notice that, in general, effective temperatures agree fairly well, though a
trend towards lower spectroscopic values is visible. The typical differences
are less than 5\% in logarithmic scale, with a maximal difference of 10\%.
On the other hand, the changes in luminosity are significant (up to 40\% in
the log, see Fig.\,\ref{relative_change}).

Since the temperatures agree so well, the luminosity difference is mainly
due to the difference in stellar radii. As discussed in
Sect.\,\ref{error_others}, the typical errors of our luminosities are of the
order of $\Delta \log L \approx 0.19{\ldots}0.22$, so that at least half of
the displayed difference should be attributed to problems with the values
provided by \citet{Waelkens98}.  This is not surprising, since these authors
used photometric calibrations for main-sequence stars because they had no
spectroscopic data available.  We are confident that our values are more
trustworthy, and that the estimated error bars reliable.

\begin{figure}
\centering
\includegraphics[angle=90,width=8.6cm]{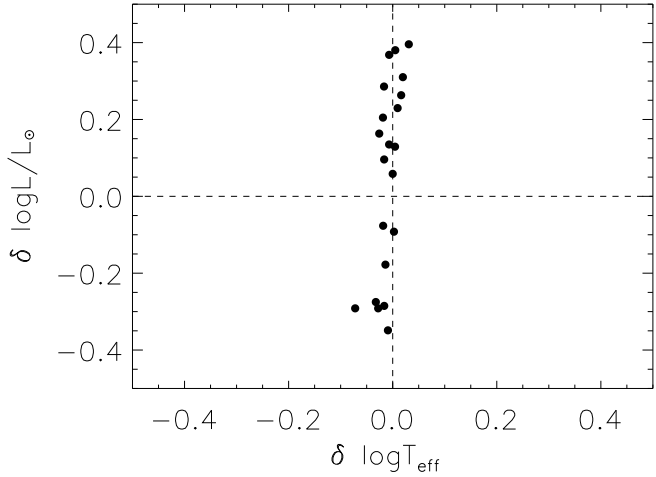}
\caption{The relative difference in luminosity against the relative
difference in effective temperature between photometrically
\citep{Waelkens98} and spectroscopically (this paper) determined stellar
parameters (with the relative difference defined as $\rm \delta x =
(x_{Lefever} - x_{Waelkens})/x_{Lefever}$).}
\label{relative_change}
\end{figure}

\begin{figure}
\centering
\includegraphics[angle=90,width=8.6cm]{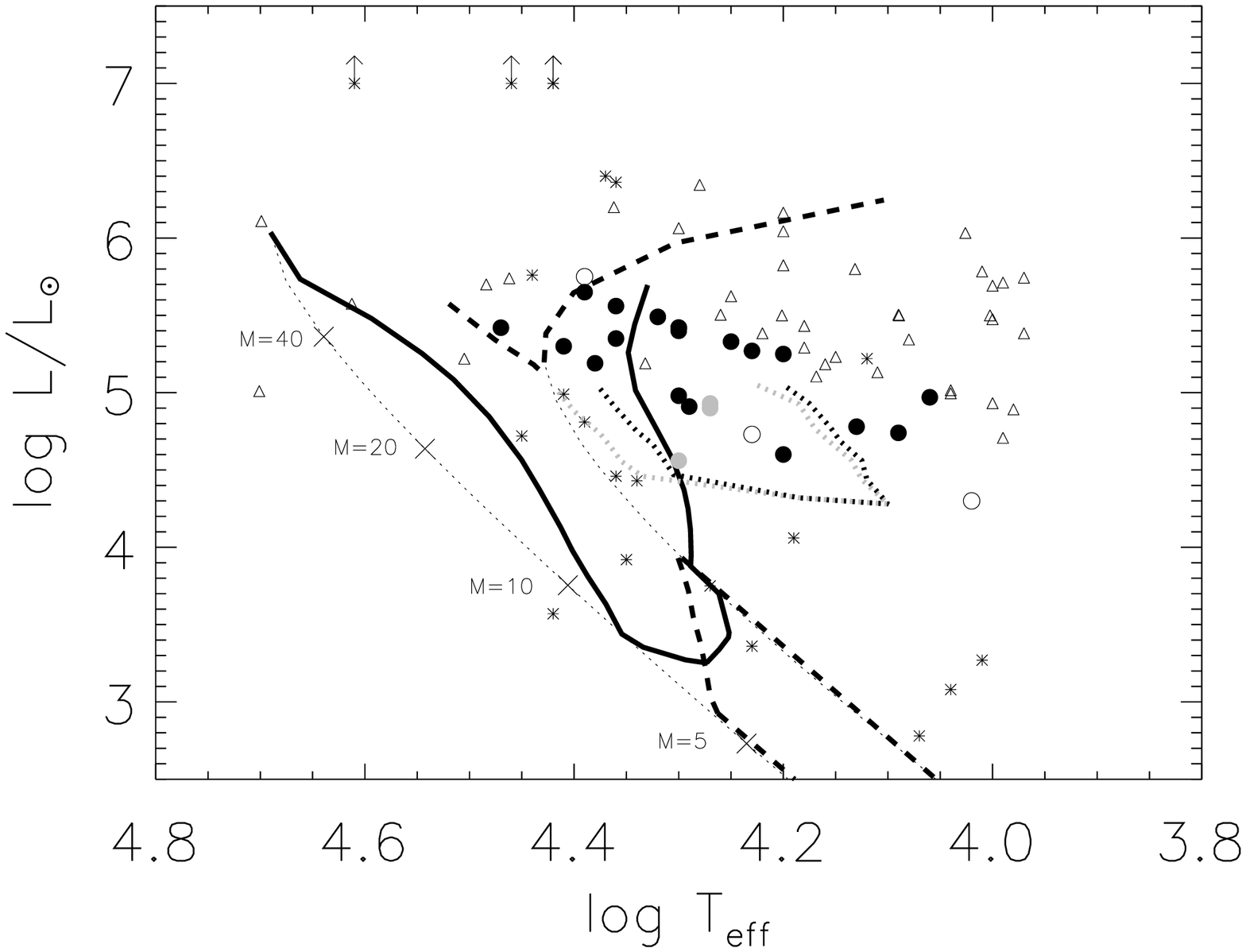}\\
\caption{Position of the sample stars in the HR diagram. Open triangles indicate
some well-studied periodically variable supergiants reported in the literature
\citep{Burki78, VanGenderen85, Lamers98, vanLeeuwen98}. The position of the
sample stars as derived by \citet{Waelkens98} is indicated by asterisks
(arrows indicate a lower limit).  The new position derived in this study 
is marked by circles (filled - group I; grey - group II;
open - group III). The dotted lines represent the ZAMS (four initial ZAMS
masses - in $M_{\odot}$ - are indicated) and TAMS. Theoretical instability domains
for the $\beta$ Cep (thick solid line) and the SPB stars (dashed lines) for
main-sequence models are shown \citep{Pamyatnykh99}, together with the instability
domains for post-TAMS models with $\ell$ = 1 (grey dotted) and $\ell$ = 2 
(black dotted) g-modes computed by \citet{Saio06}.}
\label{HRD}
\end{figure}

\begin{figure}
\centering
\includegraphics[angle=90,width=8.6cm]{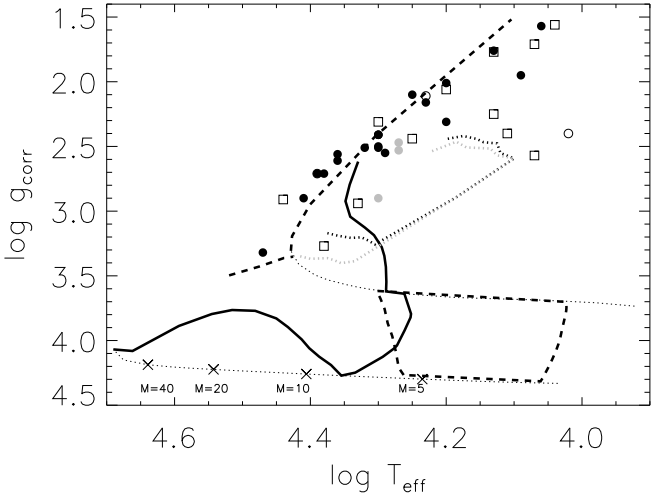}\\
\caption{Position of the sample stars and of the low-order p-mode ($\beta$ Cep -
thick solid line) and high-order g-mode (SPB-type - thick dashed line)
instability domains in the log$T_{\rm eff}$-log$g$ diagram
\citep{Pamyatnykh99} for main sequence stars. Post-TAMS model
predictions for $\ell$ = 1 (grey dotted) and $\ell$ = 2 (black dotted) g-modes are shown
for B stars with masses up to 20 M$_{\odot}$ \citep{Saio06}.
A few initial ZAMS masses are given in units of $M_{\odot}$.  Circles have the
same meaning as in Fig.\,\ref{HRD}.  Squares represent the group of comparison
stars (group IV).}
\label{logg_logTeff}
\end{figure}

In Fig.\,\ref{HRD} we compare the position of the variable B-type
supergiants in the HR diagram from spectroscopic and photometric derivations
and place them among known variable supergiants available in the literature.
It shows that the new position of the sample supergiants is quite different
from the one obtained by \citet{Waelkens98}. The stars lie now at the
low-luminosity region of the known $\alpha$~Cyg variables.  Also shown are
the theoretical instability strips of opacity-driven modes in $\beta$ Cep
and SPB stars, derived from main sequence (pre-TAMS) models
\citep{Pamyatnykh99}, and post-TAMS model predictions for $\ell$ = 1 and
2 for B stars up to 20 M$_{\odot}$ \citep{Saio06}. The former strips do not
encompass our targets at first sight, whereas the latter do, at least
for all stars hotter than 15\,000~K. A far more appropriate comparison between
the position of these strips and the sample stars is achieved from a $(\log
T_{\rm eff}$, $\log g$) plot. Indeed, the position of the targets in such a
diagram is much more reliable because it is free from uncertainties on the
radii and quantities derived thereof. Such a plot is provided in
Fig.\,\ref{logg_logTeff}. {\it It encapsulates the major result of our study.}
We see that all group I stars fall very close to the higher-gravity
limit of the predicted pre-TAMS instability strip and within
the post-TAMS instability strip predictions of gravity modes in evolved stars.
This implies that the observed periodic variability is indeed compatible with
non-radial gravity oscillations excited by the opacity mechanism.  The
photometric period(s) found for our sample stars, ranging from one to a few
weeks, and the high percentage of multiperiodic stars, fully support this
interpretation.

\begin{figure*}
\centering
\hspace{-0.97cm} 
\includegraphics[width=14.cm, height=9.cm]{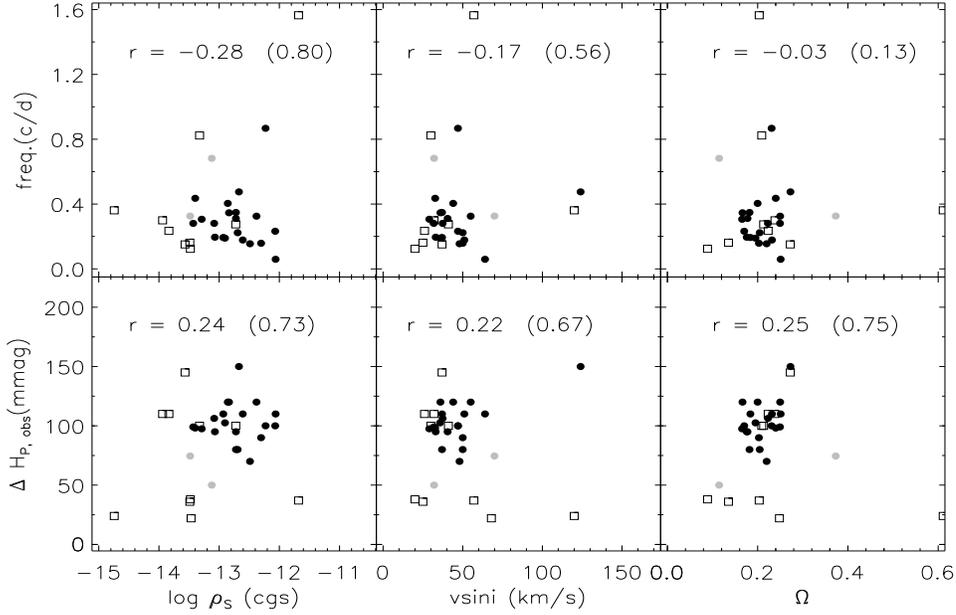}
\caption{Frequency and peak-to-peak variation as a function of the mean wind
density, $\log \rho_S$, at the sonic point, $v \sin i$ and $\Omega = v
\sin i/v_{\rm crit}$. Symbols have the
same meaning as in Fig\,\ref{logg_logTeff}.  Spearman rank correlation
coefficients, $r$, are calculated for the periodically variable B supergiants of
group I and II joined. The two-sided significance is given between braces.}
\label{photometry}
\end{figure*}

The position of the comparison stars in this (log $T_{\rm eff}$, log~$g$)
diagram is also interesting. In Section\,\ref{photometry_section} we argued
that five among these so-called {\it non-periodically\/} variable
supergiants, show amplitudes and periods comparable to the ones of the
sample stars.  We see that these five objects also lie at the same
higher-gravity limit of the instability strip. From the remaining seven
objects, with much lower amplitude variability, there are three that have
very high (projected) rotational velocities, which results in lower amplitudes,
\citep{DeCat02}, namely HD\,64760, HD\,157246 and HD\,165024.  Together
with HD\,149038, which nicely falls within the SPB instability domain, we
suggest these nine targets to be newly discovered oscillators in gravity
modes, unlike the remaining three objects HD\,46769, HD\,125288 and
HD\,86440 which lie, even when taking into account the error bar in $T_{\rm
eff}$, completely outside the predicted strip.

\section{Mass loss, Rotation and Photometric Variability
\label{Mdot_variability}}

\citet{Glatzel99} suggested the occurrence of pulsationally-driven mass-loss
in supergiants to be due to strange-mode instabilities on theoretical grounds.
While we found evidence for non-radial gravity modes in our sample of
``low-mass'' stars, it is worthwile to investigate if these modes play a
role in the mass-loss. If this is the case, one expects a correlation
between the frequency and/or amplitude of the oscillations and the mass-loss
rate. In view of the normal behaviour of our sample stars in terms of the
WLR relation, we can anticipate this relation to be weak, at best.  Also
rotation can introduce periodic variability and enhance mass-loss, either
alone or in combination with non-radial oscillations \citep[and references
therein]{Rivinius03}.

In order to investigate both possible mass-loss contributors, we plot the
photometric variability parameters as a function of (i) the wind density, $\log
\rho_S$, at the sonic point, (ii) the projected rotational velocity, $v \sin
i$, and (iii) the ratio $\Omega$ of $v \sin i$ to the critical
velocity\footnote{calculated here in the spherical approximation, since most
of our objects do not rotate too fast.}(see Fig.\,\ref{photometry}). We have
chosen to consider the wind density at the sonic point (i.e., $\log \rho_S$
= $\log (\dot{M} / (4 \pi R_*^2 c_S)$, with $c_S$ the isothermal sound
speed), because this is the lowermost place where mass-loss is initiated and
oscillations could have an effect.  The Spearman rank correlation
coefficients, $r$, computed for group I and II stars (indicated in each
panel in Fig.\,\ref{photometry}) reveal only two mild correlations related
to the mean wind density. Except for a similarly mild correlation
between $\Omega$ and the peak-to-peak variation, no other connection
between rotation and photometric variability can be detected. 

At first, a positive trend between the observed amplitude of variation
and the mean wind density is visible. At the same time, we notice a
tendency that higher mass-loss is present when the detected
frequencies are lower, i.e., the oscillation periods are longer. In
general, the oscillation periods scale as the mean density within the
star so this downward trend may suggest that the role of oscillations
to help in increasing the mass-loss is more evident in more evolved
stars. However, it concerns only weak correlations, which are 
not significant in a statistical sense. Moreover, the trends weaken if we
add the five comparison stars with significant amplitudes. 

The only seemingly ``true'' correlation occurs between $v \sin i$ and the mean
wind density, for which we find the Spearman rank correlation coefficient to
be 0.62 with a significance of 99\% (Fig.\,\ref{vsini_logQ}).  At first
sight, one may think that this is consistent with the well-known result that
rapid rotation helps in driving the mass-loss.  However, the correlation is
not significant if we consider $\Omega$ rather than $v \sin i$.  
This seems to point out that the observed correlation between $v \sin i$ and
log $\rho_S$ is a selection effect, in the sense that we have mostly very slow
rotators in our sample. Indeed, a clear correlation between rotation and the
mean wind density is only expected for objects rotating faster than $\sim$ 
70\% of their critical velocity, which is not at all the case for our
sample stars. 

\begin{figure}
\centering
\includegraphics[angle=90,width=8.6cm]{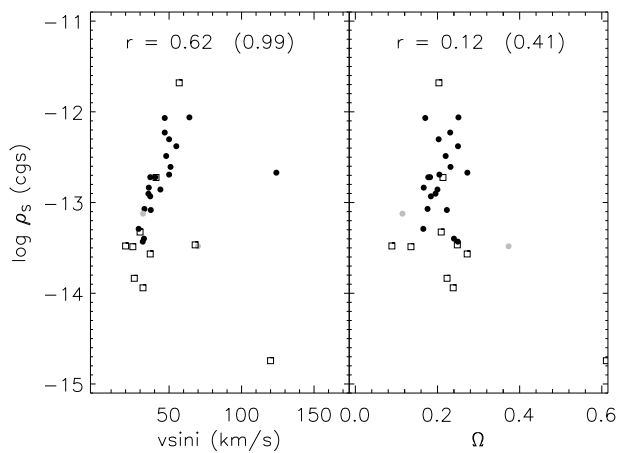}
\caption{Mean wind density, $\log \rho_S$, as a function of $v \sin i$ and
$\Omega$.
Spearman rank correlation coefficients, $r$, for the periodically variable B
supergiants are indicated (interpretation: see Fig.\,\ref{photometry}). Symbols
have the same meaning as in Fig\,\ref{logg_logTeff}.}
\label{vsini_logQ}
\end{figure}

\citet{Prinja1986} investigated the variability in UV P\,Cygni profiles of
early-type stars and concluded that variability in $\dot{M}$ occurs at the
10\% level, on time scales of a day or longer. Changes of a factor two or
larger were never observed. Only last year, \citet{Markova2005} investigated
the wind variability in 15 O type supergiants, using H$\alpha$ as a
signature. They found variations of the order of 4\% with respect to the
mean value of $\dot{M}$ for stars with strong winds and of $\pm$ 16\% for
stars with weak winds. The ratio of maximum to minimum mass-loss rate over
the time interval for their sample ranges from 1.08 to 1.47 (with a mean
ratio of 1.22), with a tendency that weaker winds show larger changes. Both
numbers are fully in agreement with \citet{Prinja1986}. For our sample
supergiants, we find that the ratio of maximum to minimum mass-loss
rate\footnote{but remember that we have only two observations at our
disposal, separated by typically one year.} ranges from 1.05 up to 1.88,
with a mean ratio for the whole sample of 1.31, which is still in agreement
with \citet{Prinja1986}, in the sense that these numbers do not exceed a
factor two. On the other hand, the maximum relative differences amount to
60\% , with an average of 22\% (without distinction between weak or strong
winds), which is somewhat higher than what is expected for ``normal'' O
supergiants. This might support the idea of a connection between the mass
loss and the variability in these stars.

At present we cannot conclude definitively to have found evidence for mass-loss
enhancement through oscillations. We merely find an {\it indication\/} that
higher photometric amplitudes seem to be accompanied by a higher wind density
and a higher relative change in the mass-loss rate. The validity of these
suggested connections would merit from further study through a more refined
time-resolved analysis of both the oscillation and wind properties of a few
selected targets.

\section{Discussion and Summary} 

We derived stellar and wind parameters of 40 B-type supergiants
using the NLTE atmosphere code FASTWIND and high-resolution, high
signal-to-noise spectroscopic data of selected H, He, Si lines. For the
majority of the objects (excluding group III), these parameters are
accurate within the discussed assumptions, and reliable error bars have
been estimated as well.

The primary aim of our study was to investigate if the origin of the
variability found in HIPPARCOS data of 28 of these stars could be
gravity-mode oscillations excited by the opacity mechanism which is also at
work in main sequence B stars. In order to assess this suggestion made by
\cite{Waelkens98}, we needed to achieve accurate values of the effective
temperature and gravity of the sample stars in order to compare them with
those of stellar models for which theoretical computations predict
oscillations.  The conclusion is clearcut: {\it all\/} the sample stars for
which we could derive reliable stellar parameters have $T_{\rm eff}$ and
$\log g$ values consistent with those of stellar models in which
gravity modes are predicted. They are all situated at the high $\log g$ limit of 
the instability strip of gravity modes in evolved stars, covering the entire
range in spectral type B. Due to the lack of a clear correlation between the observed 
photometric variability and the stellar rotation, and in view of the large
fraction of stars with multiperiodic behaviour, we suggest our sample stars
to be opacity-driven non-radial oscillators. 

Our study involved a sample of twelve comparison B supergiants which were
not selected to be periodically variable by the HIPPARCOS team.
Nevertheless, we found that nine of these behave similarly to our original
sample stars and we suggest them to be new $\alpha\,$Cyg variables with
gravity modes as well.  We thus end up with a sample of 37 non-radial
pulsators in the upper part of the HR diagram. The occurrence of
opacity-driven oscillations in that area was recently proven to be correct
for the B2 Ib/II star HD\,163899 from a 37-days ultra-precise MOST space-based
photometric lightcurve \citep{Saio06}. This star was found to oscillate in
48 frequencies with amplitudes below 4\,mmag.  Our findings and those by
\citet{Saio06} open up the area of asteroseismology to the upper-part of the
HR diagram and lead to excellent prospects for fine-tuning of evolutionary
models from the ZAMS towards the supernova stage from seismic sounding of B
supergiant interiors. In order to achieve this, one needs to obtain
long-term high-precision time series in photometry and spectroscopy to
achieve mode identification, and derive accurate stellar parameters in the
way achieved in this paper.

We find marginal evidence for a connection between the photometric
amplitudes and the wind density, in the sense that the oscillations may help
the line driving at the base of the wind. Firm conclusions in this direction
require further detailed study of a few sample stars though. Since our 
sample stars display similar wind momentum luminosity relations (WLR) as
found for ``normal'' B-supergiants (if there is something like ``normal'' at
all), the enhancement of the mass-loss cannot be very strong. Interestingly,
however, we also found one (important) difference compared to previous
evidence. Whereas the results from \citet{Kudritzki1999} 
strongly implied that {\it all} mid B-type supergiants display a
WLR which lies significantly below the theoretically predicted
one, our findings show
that there is no such unique (and problematic) correlation. Part of our
sample objects show this dilemma indeed (in particular those with $\log
L/L_{\odot} > 5$), whereas lower luminosity objects were found to be rather
consistent with theory. In this investigation we also derived, for
the first time, wind-momenta of late-type B-supergiants, which follow the
trend of A-supergiants surprisingly well.

While trying to achieve a high-precision estimate of the stellar and wind
parameters through line-profile fitting, we came across several interesting
features and results. The most important one is that we found ways to achieve
good estimates for the parameters, despite the limited number of available
lines, thanks to the high predictive power for different parameter estimates
of each of the available lines. We reached this conclusion because we could
compare our analyses with similar ones based on more spectral lines available in
the literature. This also led to a new effective temperature calibration in good
agreement with previous ones. We used this calibration to provide a practical
recipe to compute $T_{\rm eff}$ as a function of spectral type. 

We are currently automating the fitting procedure and will apply our methodology
to a large sample of early-type stars, comprising all massive stars in the 
archive with ground-based preparatory observations for the COROT space mission
\citep{Solano2005}, for which we have full spectra at our disposal.

\begin{acknowledgements}

We thank P.\ Crowther for providing us with the CTIO/JKT spectra for the stars
we had in common, and A.A.\ Pamyatnykh for providing us with the theoretical
instability domains B stars.  KL and CA are supported by the Research Council of
the Catholic University of Leuven under grant GOA/2003/04. We thank T.\
Vanneste, B.\ Nicolas and L.\ Eylenbosch for their contribution to this study in
the framework of their Master Thesis and Drs P. De Cat, L.\ Decin and J.\ De
Ridder for their contribution to the data gathering.
\end{acknowledgements}

\bibliographystyle{aa}

\clearpage
\tiny
\Ltable
\begin{landscape}
\begin{longtable}{rrlccccrccclrcccrrrc}
\caption{\label{finalparameters_SG} 
Stellar and wind parameters of all sample supergiants, subdivided by
group. Spectral types are taken from SIMBAD or recent literature. The
visual magnitudes, V, are from SIMBAD as well. Gravities as obtained
from the line fits are denoted by $g_{\rm eff}$, and those
approximately corrected for the centrifugal acceleration by $g_{\rm
corr}$ (see text). Absolute visual magnitudes M$_V$ from the
calibrations by Schmidt-Kaler (1982), used to derive the stellar radii
(in combination with the theoretical fluxes).\\
For those sample stars which have been observed in H$\alpha$ twice
(typically one year apart), we provide the derived wind parameters for
both observations (when both coincide, this is indicated by (2)).
Wind-strength parameter, $\log Q$, in the same units as in
Table\,\ref{comp_Crowther}, $\dot M$ in units of $M_{\odot}/yr$ and
modified wind momentum rate, $D_{\rm mom}$, in cgs.  All velocities
are given in km/s and have their usual meaning. Question marks for
v$_{\rm macro}$ denote those cases where the line profiles could be
fitted without any macroturbulent velocity. \\
In the last column, we indicate the type of \Ha\ emission: e/i/a
indicate emission/intermediate/absorption profiles, respectively.}
\\
\hline \hline
group & HD & SpT & V & T$_{\rm eff}$ & log $g_{\rm eff}$ & log $g_{\rm corr}$
 & M$_V$ & R$_*$/R$_{\odot}$ & log $L/L_{\odot}$ & log $Q$ & $\dot{M}$ & 
$v_{\infty}$ & $\beta$ & log D$_{\rm mom}$ & yhe & $v_{\rm micro}$ & $v_{\rm macro}$ &  $v \sin i$ & emission \\ 
\hline
\endfirsthead
\caption{continued.}\\
\hline \hline
group & HD & SpT & V & T$_{\rm eff}$ & log $g_{\rm eff}$ & log $g_{\rm corr}$
 & M$_V$ & R$_*$/R$_{\odot}$ & log $L/L_{\odot}$ & log $Q$ & $\dot{M}$ & 
$v_{\infty}$ & $\beta$ & log D$_{\rm mom}$ & yhe & $v_{\rm micro}$ & $v_{\rm macro}$ & $v \sin i$ & emission \\ 
\hline
\endhead
\hline
I & 168183  & O9.5Ib & 8.26 & 30000 & 3.30 & 3.32 & -5.86 & 19 & 5.42 & -13.58 & 0.15E-06 & 1700 & 1.3 & 27.84 & 0.10 & 10 & 60-90 & 124 & a \\ 
 &   &  &  &  &  &  &  &  &  & -13.46 & 0.20E-06 &  &  & 27.97 &  &  & &  &  \\ 
 & 89767  & B0Ia & 7.23 & 23000 & 2.55 & 2.56 & -6.4 & 30 & 5.35 & -13.09 & 0.85E-06 & 1600 & 2.5 & 28.67 & 0.20 & 15-20 & 75-80 & 47 & e \\ 
 &   &  &  &  &  &  &  &  & & -13.17 & 0.70E-06 & &  & 28.58 &  &  & &  & \\ 
 & 94909  & B0Ia & 7.35 & 25000 & 2.70 & 2.71 & -6.9 & 36 & 5.65 & -12.77 & 0.20E-05 & 1450 & 1.8 & 29.04 & 0.10 & 20 & 75 & 64 & e \\ 
 &   &  & &  &  &  &  & &  & -12.96 & 0.13E-05 &  &  & 28.85 &  &  &  & & \\ 
 & 93619  & B0.5Ib & 6.98 & 26000 & 2.90 & 2.90 & -5.95 & 22 & 5.30 & -12.88 & 0.75E-06 (2) & 1470 & 1.2 & 28.51 & 0.10 & 15 & 75 & 47 & i \\ 
 & 91943  & B0.7Ib & 6.69 & 24000 & 2.70 & 2.71 & -5.95 & 23 & 5.19 & -13.28 & 0.30E-06 & 1400 & 2.5 & 28.10 & 0.10 & 15 & 82 & 48 & i \\ 
 &   &  &  &  &  &  &  &  &  & -13.55 & 0.16E-06 & &  & 27.83 &  &  &  & & \\ 
 & 96880 & B1Ia & 7.62 & 20000 & 2.40 & 2.41 & -6.9 & 43 & 5.42 & -13.46 & 0.40E-06 (2) & 1200 & 2.2 & 28.29 & 0.10 & 15 & 65 & 44 & e \\ 
 & 115363  & B1Ia & 7.82 & 20000 & 2.40 & 2.41 & -6.9 & 43 & 5.42 & -12.98 & 0.12E-05 & 1200 & 3.0 & 28.77 & 0.10 & 15 & 50 & 55 & e \\ 
 &   &  &  &  &  &  &  & &  & -12.86 & 0.16E-05 &  &  & 28.89 &  &  &  & &  \\ 
 & 148688  & B1Ia & 5.33 & 21000 & 2.50 & 2.51 & -6.9 & 42 & 5.49 & -12.90 & 0.14E-05 & 1200 & 3.0 & 28.83 & 0.10 & 15 & 40 & 50 & e \\ 
 &   &  &  & &  & &  &  &  & -13.01 & 0.11E-05 &  &  & 28.73 &  &  & & &  \\ 
 & 170938  & B1Ia & 7.92 & 20000 & 2.40 & 2.41 & -6.9 & 43 & 5.42 & -13.21 & 0.71E-06 & 1200 & 3.0 & 28.54 & 0.10 & 15 & 55 & 51 & e \\ 
 &  &  &  &  &  & & &  &  & -13.26 & 0.63E-06 &  & & 28.49 & &  &  &  & \\ 
 & 109867  & B1Iab & 6.26 & 23000 & 2.60 & 2.61 & -6.9 & 38 & 5.56 & -13.38 & 0.50E-06 & 1400 & 2.5 & 28.43 & 0.10 & 15 & 70 & 50 & i \\
 &  &  &  &  &  & & &  &  & -13.36 & 0.53E-06 &  & & 28.46 & &  &  &  & \\ 
 & 154043  & B1Ib & 7.11 & 20000 & 2.50 & 2.51 & -5.8 & 26 & 4.98 & -13.49 & 0.20E-06 & 1300 & 2.0 & 27.92 & 0.10 & 13 & 65 & 37 & a \\ 
 & 106343  & B1.5Ia & 6.24 & 20000 & 2.50 & 2.50 & -6.9 & 42 & 5.40 & -13.07 & 0.52E-06 & 800 & 2.0 & 28.23 & 0.10 & 15 & 65 & 44 & i \\ 
 &   &  &  &  & &  &  &  & & -13.04 & 0.55E-06 &  &  & 28.25 & &  &  & &   \\ 
 & 111990  & B1/B2Ib & 6.77 & 19500 & 2.55 & 2.55 & -5.7 & 25 & 4.91 & -13.26 & 0.14E-06 & 750 & 1.8 & 27.52 & 0.10 & 15 & 62 & 36 & i \\ 
 & 92964  & B2.5Iae & 5.40 & 18000 & 2.10 & 2.10 & -6.95 & 48 & 5.33 & -13.14 & 0.28E-06 & 520 & 3.0 & 27.80 & 0.10 & 15 & 50 & 31 & e \\ 
 &   &  &  &  &  &  &  &  & & -13.19 & 0.25E-06 &  & & 27.75 & &  & & &  \\ 
 & 53138  & B3Ia & 3.00 & 17000 & 2.15 & 2.16 & -7.0 & 50 & 5.27 & -13.09 & 0.31E-06 & 490 & 2.5 & 27.83 & 0.20 & 15 & 45 & 38 & i \\ 
 &   &  &  &  &  &  &  &  &  & -13.26 & 0.21E-06 &  &  & 27.66 &  &  & &  & \\ 
 & 102997  & B5Ia & 6.55 & 16000 & 2.00 & 2.01 & -7.0 & 55 & 5.25 & -13.01 & 0.23E-06 & 325 & 1.5 & 27.54 & 0.10 & 12 & 50 & 39 & e \\ 
 &   &  &  &  &  &  &  & &  & -12.88 & 0.31E-06 &  &  & 27.67 &  &  &  & & \\ 
 & 108659  & B5Ib & 7.31 & 16000 & 2.30 & 2.31 & -5.4 & 26 & 4.60 & -13.36 & 0.58E-07 (2) & 470 & 1.0 & 26.94 & 0.10 & 10 & 40 & 29 & a \\ 
 & 80558  & B6Iab & 5.91 & 13500 & 1.75 & 1.76 & -6.2 & 45 & 4.78 & -13.37 & 0.50E-07 & 250 & 1.5 & 26.72 & 0.10 & 11 & 45 & 28 & i \\ 
 &   & &  &  &  &  & &  & & -13.20 & 0.75E-07 &  &  & 26.89 &  &  &  & &  \\ 
 & 91024  & B7Iab & 7.62 & 12500 & 1.95 & 1.95 & -6.2 & 50 & 4.74 & -13.09 & 0.95E-07 & 225 & 1.5 & 26.97 & 0.10 & 8 & 30 & 25 & i \\ 
 &   & & &  &  &  &  &  & & -13.14 & 0.85E-07 & &  & 26.93 & &  & & &  \\ 
 & 94367  & B9Ia & 5.27 & 11500 & 1.55 & 1.57 & -7.1 & 77 & 4.97 & -12.81 & 0.24E-06 & 175 & 1.0 & 27.36 & 0.10 & 10 & 28 & 31 & i \\ 
 &   &  & &  &  &  &  & & & -12.93 & 0.18E-06 & &  & 27.24 &  &  &  &  & \\ 
\hline
II  
 & 47240  & B1Ib & 6.18 & 19000 & 2.40 & 2.48 & -5.8 & 27 & 4.93 & -13.41 & 0.17E-06 & 1000 & 1.5 & 27.74 & 0.15 & 15 & 55 & 94 & i \\ 
 &   &  & &  &  &  &  &  &  & -13.26 & 0.24E-06 &  &  & 27.89 &  &  &  &  & \\ 
 & 54764  & B1Ib/II & 6.06 & 19000 & 2.45 & 2.53 & -5.8 & 26 & 4.90 & -14.07 & 0.30E-07 (2) & 900 & 1.5 & 26.93 & 0.10 & 10 & 70 & 108 & a \\ 
 & 141318  & B2II & 5.77 & 20000 & 2.90 & 2.90 & -4.8 & 16 & 4.56 & -13.76 & 0.30E-07 & 900 & 1.5 & 26.83 & 0.10 & 10 & 50 & 32 & a \\ 
\hline
III & 105056  & B0Iabpe & 7.34 & 25000 & 2.70 & 2.71 & -6.9 & 40 & 5.75 & -12.60 & 0.40E-05 & 1600 & 3.0 & 29.40 & 0.10 & 15 & ? & 61 & e \\ 
 &   &  &  &  &  &  &  &  &  & -12.50 & 0.51E-05 &  &  & 29.51 &  &  &  & &  \\ 
 & 98410  & B2.5Ib/II & 8.83 & 17000 & 2.10 & 2.11 & -5.6 & 27 & 4.73 & -12.71 & 0.30E-06 & 500 & 2.0 & 27.69 & 0.15 & 15 & 70 & 31 & e \\ 
 &   &  &  &  &  &  &  &  &  & -12.58 & 0.41E-06 & &  & 27.82 &  &  & & &  \\ 
 & 68161  & B8Ib/II? & 5.66 & 10500 & 2.40 & 2.40 & -5.6 & 43 & 4.30 & -14.42 & 0.30E-08 (2) & 200 & 0.9 & 25.39 & 0.10 & 10 & ? & 17 & a \\ 
\hline
IV & 149038  & O9.7Iab & 4.91 & 28000 & 2.90 & 2.91 & -6.45 & 25 & 5.53 & -12.58 & 0.23E-05 & 1700 & 1.3 & 29.09 & 0.20 & 15 & 75 & 57 & i \\ 
 & 64760  & B0.5Ib & 4.23 & 24000 & 3.20 & 3.27 & -5.95 & 24 & 5.23 & -13.16 & 0.42E-06 & 1400 & 0.8 & 28.25 & 0.10 & 15 & ? & 230 & i \\ 
 & 157246  & B1Ib & 3.31 & 21500 & 2.65 & 2.94 & -5.8 & 25 & 5.08 & -14.11 & 0.30E-07 & 1000 & 1.5 & 26.97 & 0.10 & 15 & ? & 275 & i \\ 
 & 157038  & B1/B2IaN & 6.41 & 20000 & 2.30 & 2.31 & -6.9 & 43 & 5.42 & -13.25 & 0.50E-06 & 1000 & 3.0 & 28.31 & 0.20 & 15 & 40 & 41 & e \\ 
 & 165024  & B2Ib & 3.66 & 18000 & 2.30 & 2.39 & -5.7 & 26 & 4.80 & -15.25 & 0.18E-08 & 840 & 1.5 & 25.68 & 0.10 & 15 & 85-50 & 95-120 & a \\ 
 & 75149  & B3Ia & 5.47 & 16000 & 2.05 & 2.06 & -7.0 & 39 & 4.95 & -13.43 & 0.10E-06 & 500 & 2.5 & 27.29 & 0.10 & 15 & 60 & 30 & i \\ 
 & 58350  & B5Ia & 2.40 & 13500 & 1.75 & 1.77 & -7.0 & 65 & 5.10 & -13.17 & 0.14E-06 & 250 & 2.5 & 27.25 & 0.15 & 12 & 40 & 37 & i \\ 
 & 86440  & B5Ib & 3.50 & 13500 & 2.25 & 2.25 & -5.4 & 31 & 4.45 & -13.64 & 0.40E-07 & 470 & 0.9 & 26.81 & 0.10 & 10 & ? & 20 & a \\ 
 & 125288  & B6Ib & 4.35 & 13000 & 2.40 & 2.40 & -5.35 & 31 & 4.39 & -13.23 & 0.40E-07 & 250 & 0.9 & 26.54 & 0.10 & 10 & 30 & 25 & a \\ 
 & 106068  & B8Iab & 5.95 & 12000 & 1.70 & 1.72 & -6.2 & 50 & 4.66 & -13.35 & 0.44E-07 & 200 & 2.0 & 26.59 & 0.10 & 8 & 25 & 26 & i \\ 
 & 46769  & B8Ib & 5.79 & 12000 & 2.55 & 2.57 & -5.2 & 30 & 4.22 & -13.09 & 0.37E-07 & 200 & 1.0 & 26.40 & 0.10 & 12 & 35 & 68 & a \\ 
 & 111904  & B9Ia & 5.80 & 11100 & 1.55 & 1.56 & -7.38 & 95 & 5.09 & -13.25 & 0.12E-06 & 175 & 3.0 & 27.11 & 0.10 & 10 & ? & 32 & i \\ 
\hline
\end{longtable}
\end{landscape}
\normalsize 

\Online
\appendix
\onecolumn
\section{Spectral Line Fits of All Sample and Comparison Stars}

\begin{figure*}[ht]
\label{fits_groupI_1}
\caption{Spectral line fits for the periodically variable
  B-type supergiants with reliable parameters: group I.} 
\vspace{0.5cm}\hspace{1.7cm} H$\alpha$ 1  \hspace{2.1cm} H$\alpha$ 2 
\hspace{2.4cm} H$\gamma$ 
\hspace{1.8cm} He~I 4471  \hspace{1.6cm} He~I 6678  \hspace{1.5cm} Si~III \\ 
\rotatebox[origin=l]{90}{\hspace{0.4cm} \object{HD 168183} (O9.5 Ib)}
\hspace{0.2cm} 
\includegraphics[width=3.5cm,height=18.0cm,angle=90]{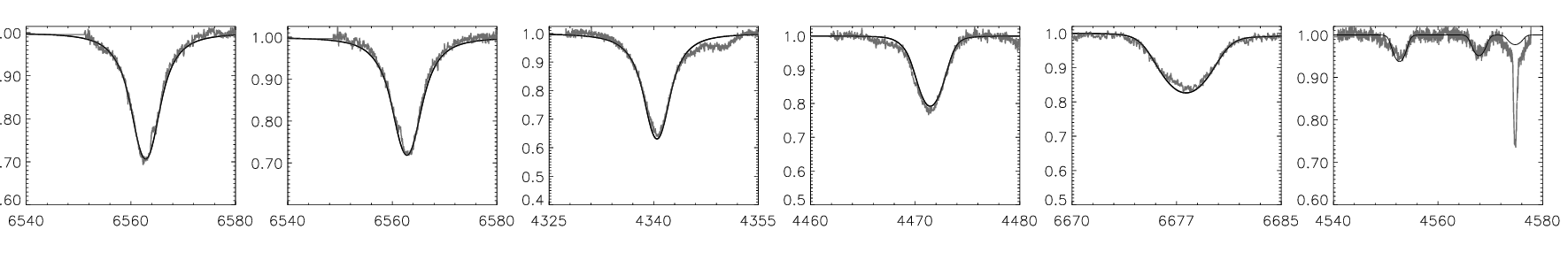}\\
\rotatebox[origin=l]{90}{\hspace{0.4cm} \object{HD 89767} (B0 Iab)}
\hspace{0.2cm}
\includegraphics[width=3.2cm,height=18.0cm,angle=90]{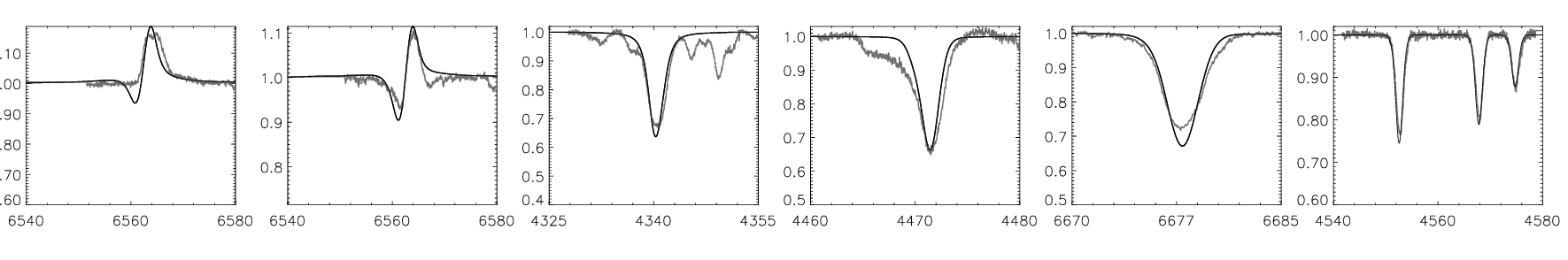}\\
\rotatebox[origin=l]{90}{\hspace{0.4cm} \object{HD 94909} (B0 Ia)}
\hspace{0.2cm}
\includegraphics[width=3.2cm,height=18.0cm,angle=90]{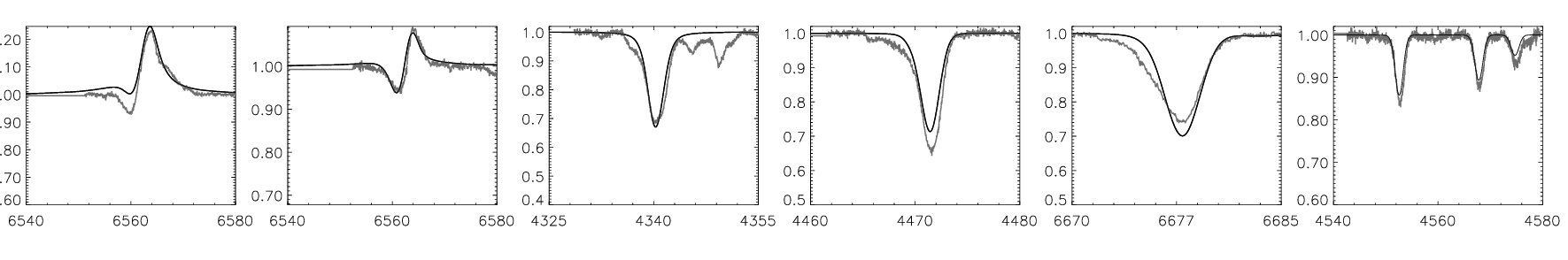}\\
\rotatebox[origin=l]{90}{\hspace{0.4cm} \object{HD 93619} (B0.5 Ib)}
\hspace{0.2cm}
\includegraphics[width=3.2cm,height=18.0cm,angle=90]{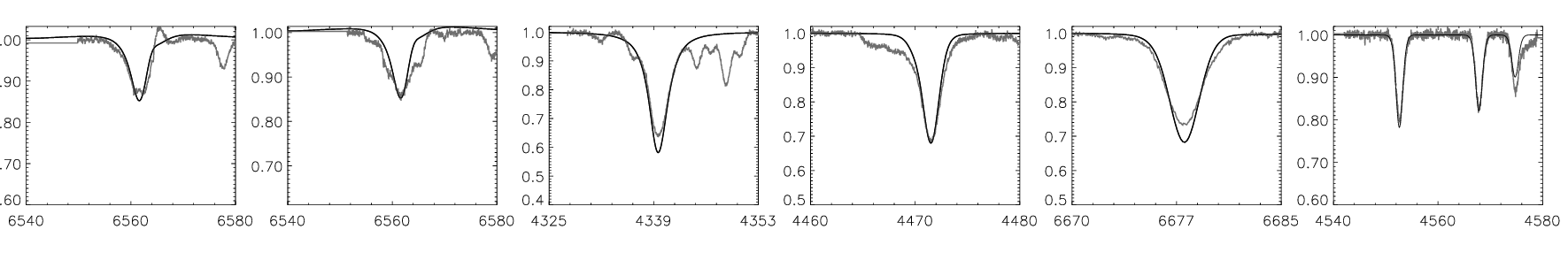}\\
\rotatebox[origin=l]{90}{\hspace{0.4cm} \object{HD 91943} (B0.7 Ib)}
\hspace{0.2cm}
\includegraphics[width=3.2cm,height=18.0cm,angle=90]{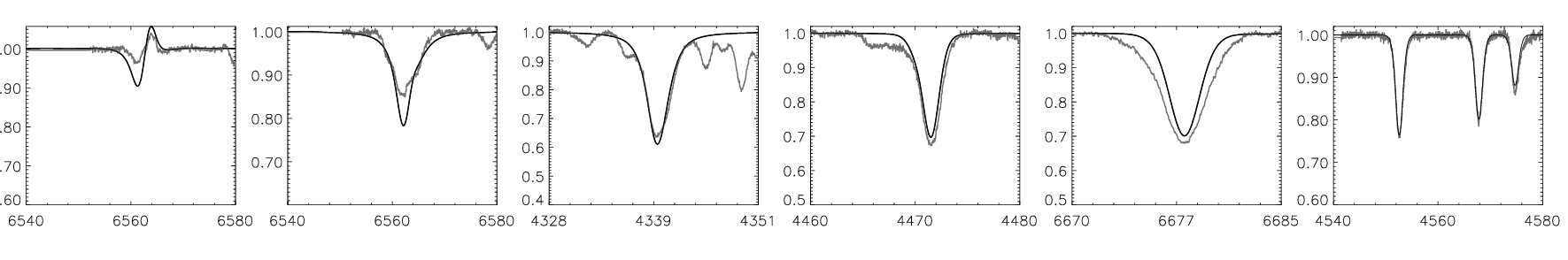}\\
\rotatebox[origin=l]{90}{\hspace{0.4cm} \object{HD 96880} (B1 Ia)}
\hspace{0.2cm} 
\includegraphics[width=3.5cm,height=18.0cm,angle=90]{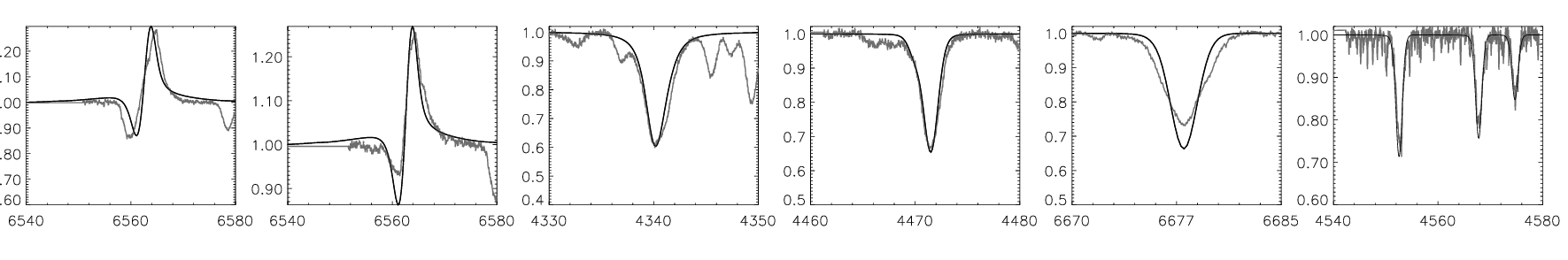}\\
\end{figure*}

\begin{figure*}[ht]
\label{fits_groupI_2}
\caption{Spectral line fits for the periodically variable
  B-type supergiants with reliable parameters: group I (continued).} 
\vspace{0.5cm}\hspace{1.7cm} H$\alpha$ 1  \hspace{2.1cm} H$\alpha$ 2 
\hspace{2.4cm} H$\gamma$ 
\hspace{1.8cm} He~I 4471  \hspace{1.6cm} He~I 6678  \hspace{1.5cm} Si~III \\ 
\rotatebox[origin=l]{90}{\hspace{0.4cm} \object{HD 115363} (B1 Ia)}
\hspace{0.2cm}
\includegraphics[width=3.5cm,height=18.0cm,angle=90]{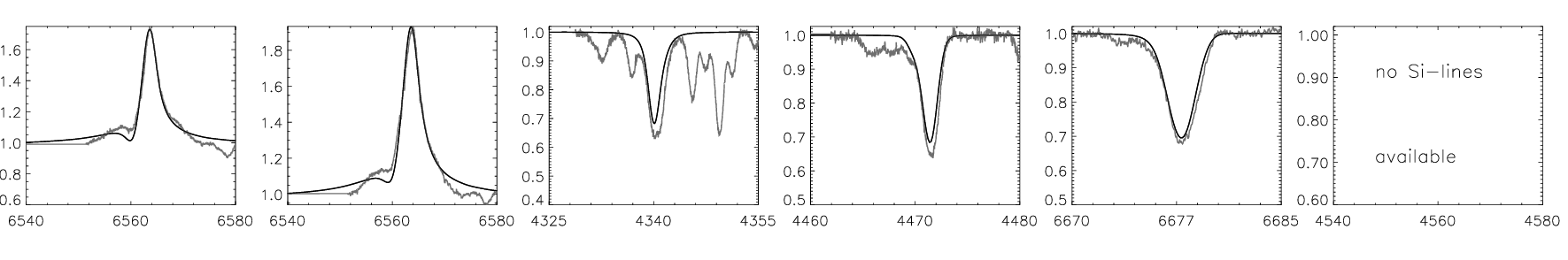}\\
\rotatebox[origin=l]{90}{\hspace{0.4cm} \object{HD 148688} (B1 Ia)}
\hspace{0.2cm}
\includegraphics[width=3.2cm,height=18.0cm,angle=90]{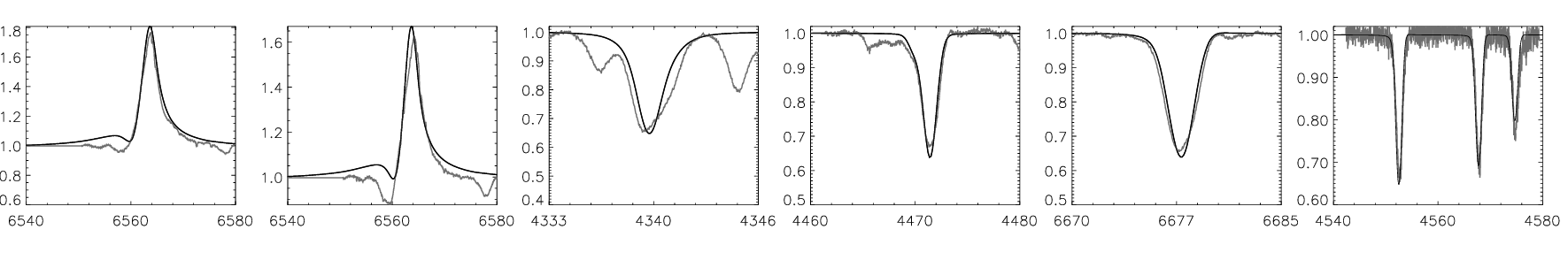}\\
\rotatebox[origin=l]{90}{\hspace{0.4cm} \object{HD 170938} (B1 Ia)}
\hspace{0.2cm}
\includegraphics[width=3.5cm,height=18.0cm,angle=90]{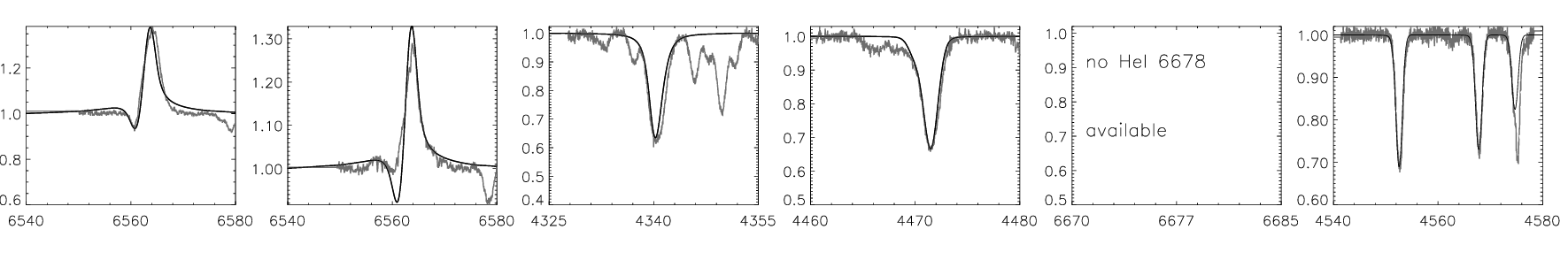}\\
\rotatebox[origin=l]{90}{\hspace{0.4cm} \object{HD 109867} (B1 Iab)}
\hspace{0.2cm}
\includegraphics[width=3.2cm,height=18.0cm,angle=90]{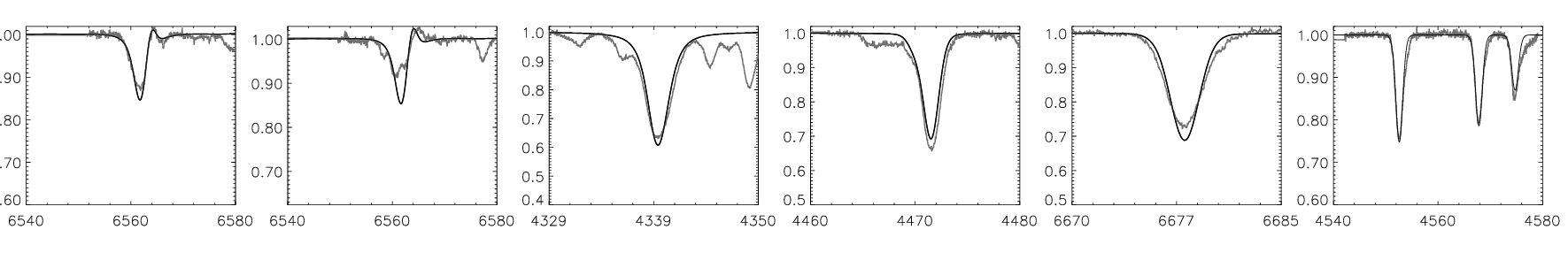}\\
\rotatebox[origin=l]{90}{\hspace{0.4cm} \object{HD 154043} (B1 Ib)}
\hspace{0.2cm}
\includegraphics[width=3.2cm,height=18.0cm,angle=90]{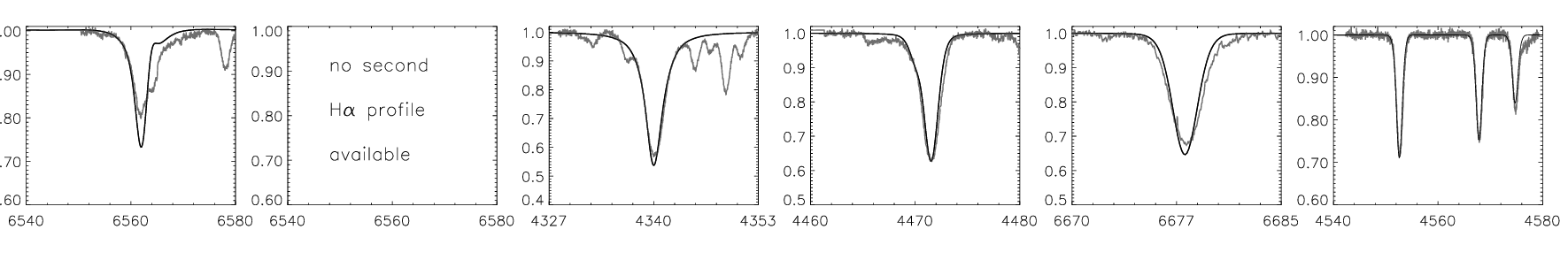} \\
\rotatebox[origin=l]{90}{\hspace{0.4cm} \object{HD 106343} (B1.5 Ia)}
\hspace{0.2cm}
\includegraphics[width=3.5cm,height=18.0cm,angle=90]{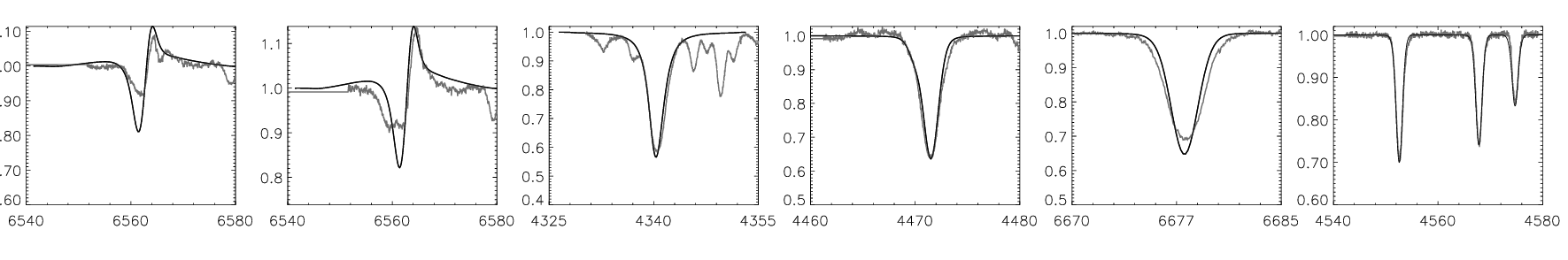}\\
\end{figure*}

\begin{figure*}[ht]
\label{fits_groupI_3}
\caption{Spectral line fits for the periodically variable
  B-type supergiants with reliable parameters: group I (continued).} 
\vspace{0.5cm}\hspace{1.7cm} H$\alpha$ 1  \hspace{2.1cm} H$\alpha$ 2 
\hspace{2.4cm} H$\gamma$ 
\hspace{1.8cm} He~I 4471  \hspace{1.6cm} He~I 6678  \hspace{1.5cm} Si~II \\ 
\rotatebox[origin=l]{90}{\hspace{0.4cm} \object{HD 111990} (B2 Ib)}
\hspace{0.2cm}
\includegraphics[width=3.5cm,height=18.0cm,angle=90]{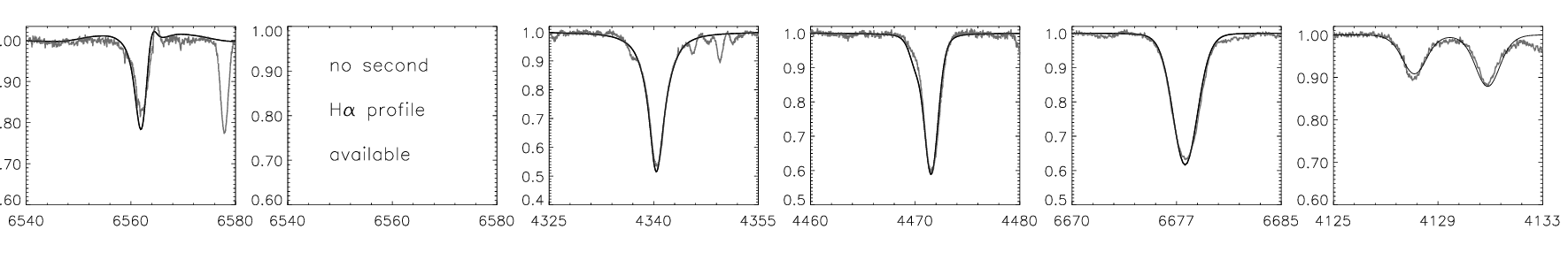}\\
\rotatebox[origin=l]{90}{\hspace{0.4cm} \object{HD 92964} (B2.5 Iae)}
\hspace{0.2cm}
\includegraphics[width=3.5cm,height=18.0cm,angle=90]{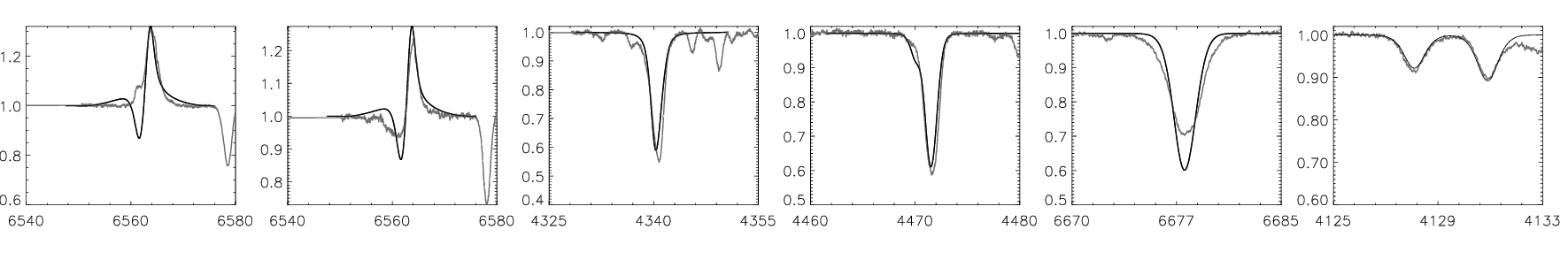}\\
\rotatebox[origin=l]{90}{\hspace{0.4cm} \object{HD 53138} (B3 Ia)}
\hspace{0.2cm}
\includegraphics[width=3.5cm,height=18.0cm,angle=90]{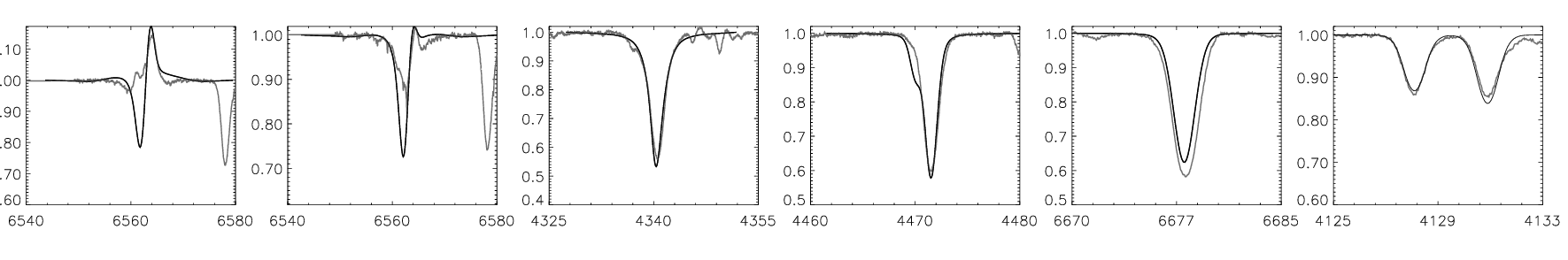}\\
\rotatebox[origin=l]{90}{\hspace{0.4cm} \object{HD 102997} (B5 Ia)}
\hspace{0.2cm}
\includegraphics[width=3.2cm,height=18.0cm,angle=90]{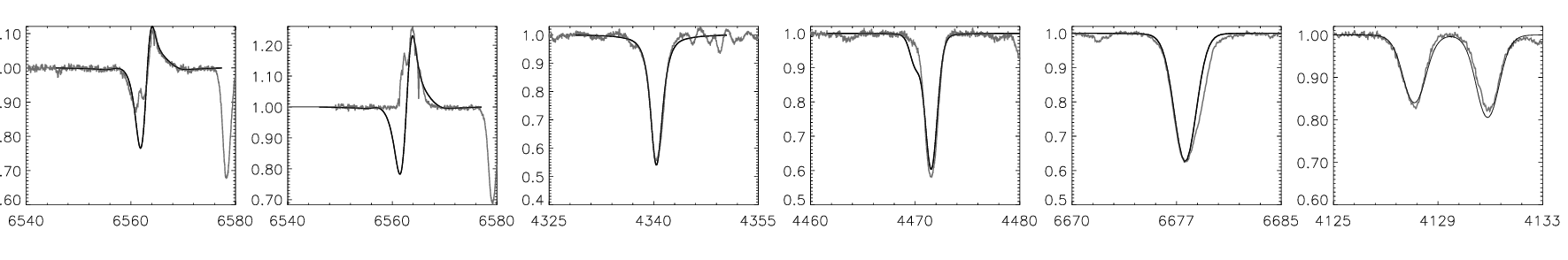}\\
\rotatebox[origin=l]{90}{\hspace{0.4cm} \object{HD 108659} (B5 Ib)}
\hspace{0.2cm}
\includegraphics[width=3.2cm,height=18.0cm,angle=90]{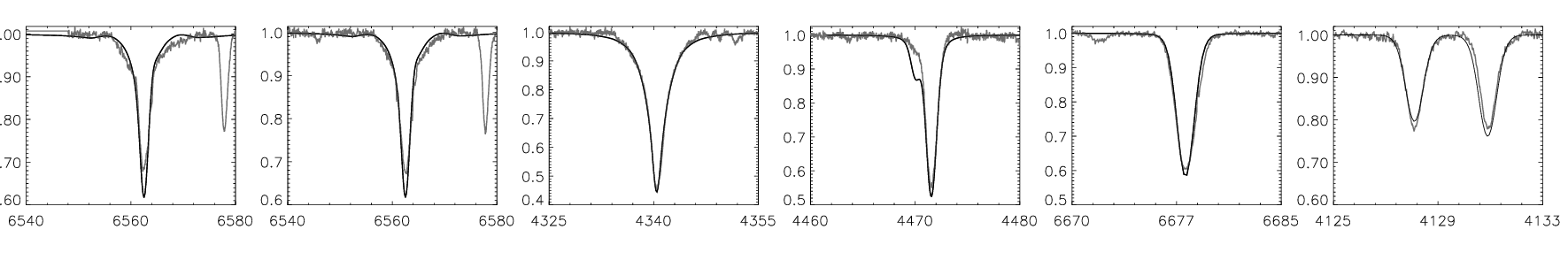}\\
\rotatebox[origin=l]{90}{\hspace{0.4cm} \object{HD 80558} (B6 Iab)}
\hspace{0.2cm}
\includegraphics[width=3.2cm,height=18.0cm,angle=90]{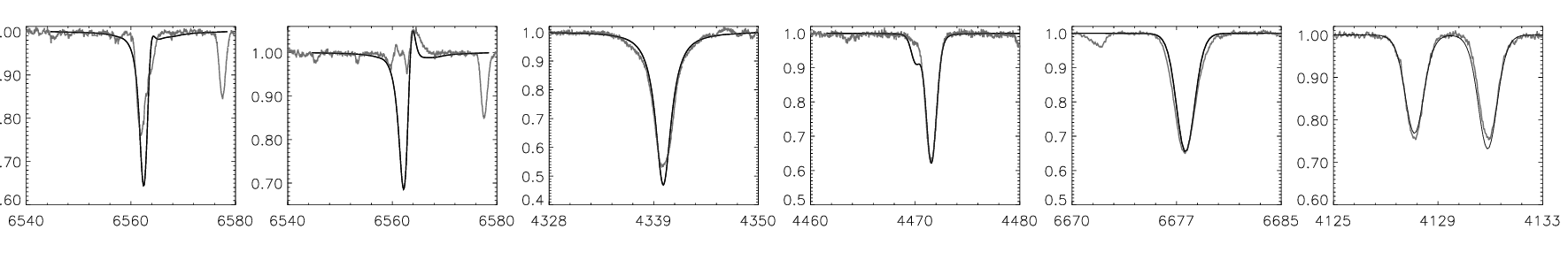}\\
\end{figure*}

\begin{figure*}[ht]
\label{fits_groupI_4}
\caption{Spectral line fits for the periodically variable
  B-type supergiants with reliable parameters: group I (continued).} 
\vspace{0.5cm}\hspace{1.7cm} H$\alpha$ 1  \hspace{2.1cm} H$\alpha$ 2 
\hspace{2.4cm} H$\gamma$ 
\hspace{1.8cm} He~I 4471  \hspace{1.6cm} He~I 6678  \hspace{1.5cm} Si~II\\ 
\rotatebox[origin=l]{90}{\hspace{0.4cm} \object{HD 91024} (B7 Iab)}
\hspace{0.2cm}
\includegraphics[width=3.2cm,height=18.0cm,angle=90]{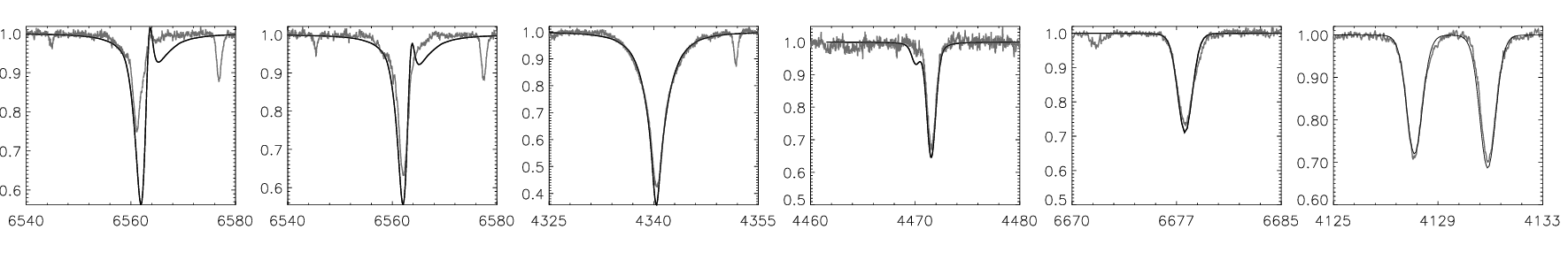}\\
\rotatebox[origin=l]{90}{\hspace{0.4cm} \object{HD 94367} (B9 Ia)}
\hspace{0.2cm}
\includegraphics[width=3.2cm,height=18.0cm,angle=90]{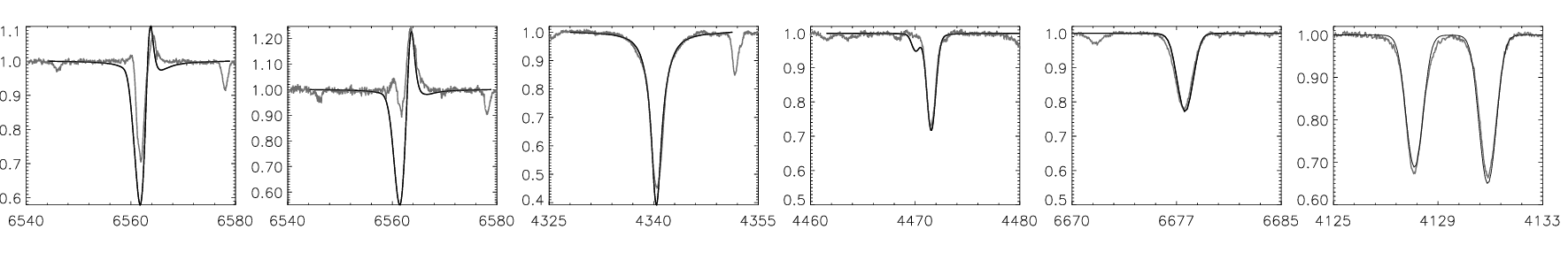}\\
\end{figure*}

\begin{figure*}[ht]
\label{fits_groupII}
\caption{Spectral line fits for the periodically variable B-type
supergiants with a potential $T_{\rm eff}$ dichotomy: group II. Only our
favoured solution is displayed (see text).} 
\vspace{0.5cm}\hspace{1.7cm} H$\alpha$ 1  \hspace{2.1cm} H$\alpha$ 2 
\hspace{2.4cm} H$\gamma$ 
\hspace{1.8cm} He~I 4471  \hspace{1.6cm} He~I 6678  \hspace{1.5cm} Si~III \\
\rotatebox[origin=l]{90}{\hspace{0.4cm} \object{HD 54764} (B1 Ib/II)}
\hspace{0.2cm}
\includegraphics[width=3.2cm,height=18.0cm,angle=90]{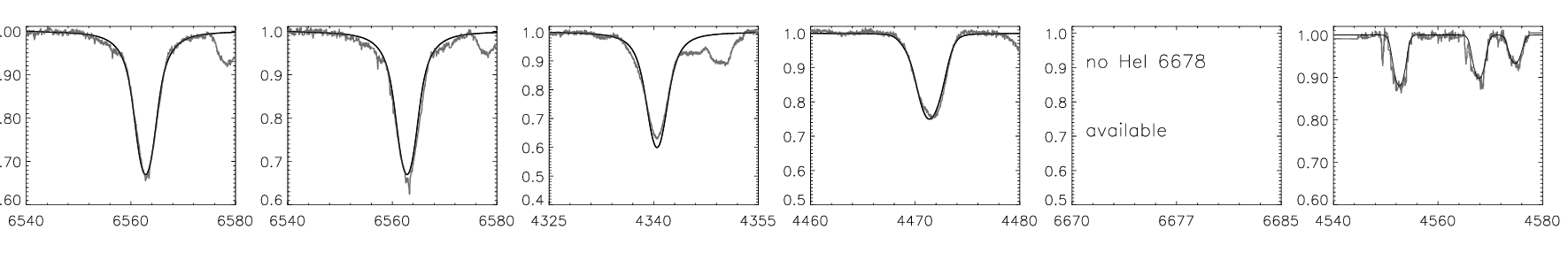}
\\
\rotatebox[origin=l]{90}{\hspace{0.4cm} \object{HD 47240} (B1 Ib)}
\hspace{0.2cm}
\includegraphics[width=3.5cm,height=18.0cm,angle=90]{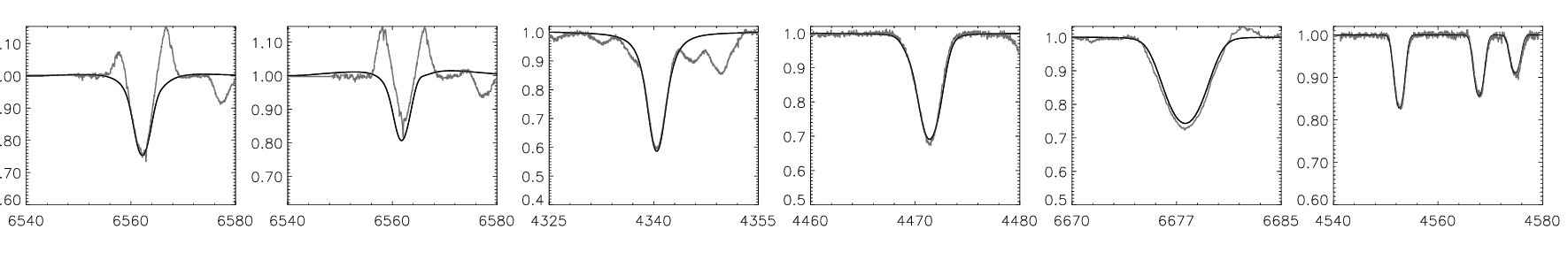}
\\
\rotatebox[origin=l]{90}{\hspace{0.4cm} \object{HD 141318} (B2 II)}
\hspace{0.2cm}
\includegraphics[width=3.5cm,height=18.0cm,angle=90]{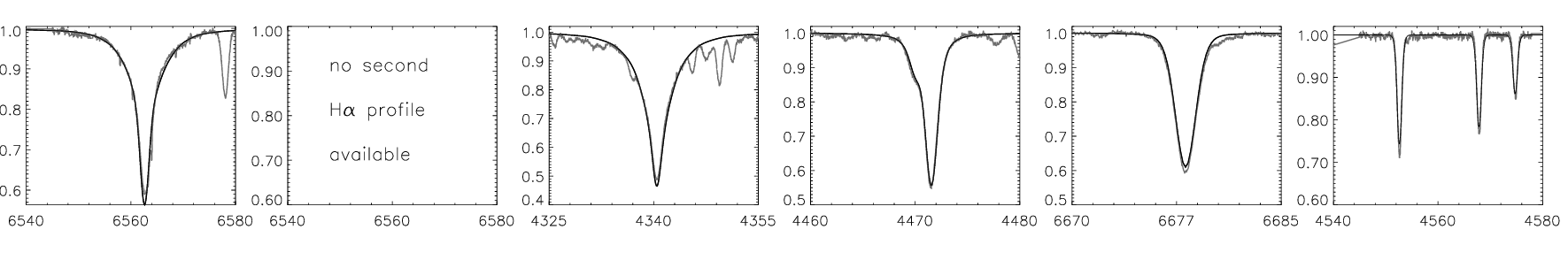}\\
\end{figure*}

\begin{figure*}[ht]
\label{fits_groupIII}
\caption{Spectral line fits for the periodically variable B-type
supergiants with parameters to be considered as indicative numbers only: group
III. } 
\vspace{0.5cm}\hspace{1.7cm} H$\alpha$ 1  \hspace{2.1cm} H$\alpha$ 2 
\hspace{2.4cm} H$\gamma$  \hspace{1.8cm} He~I 4471
 \hspace{1.6cm} He~I 6678  \hspace{2.0cm} Si~II/III \\
\rotatebox[origin=l]{90}{\hspace{0.4cm} \object{HD 105056} (B0 Iabpe)}
\hspace{0.2cm}
\includegraphics[width=3.2cm,height=18.0cm,angle=90]{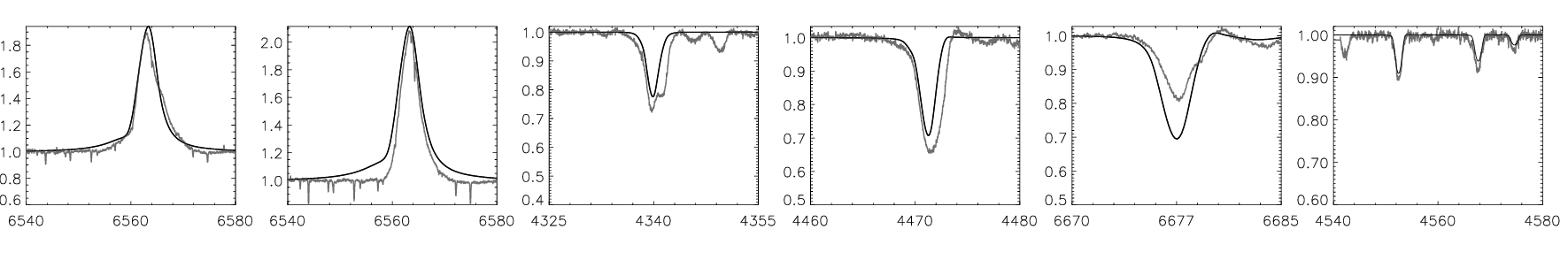}\\
\rotatebox[origin=l]{90}{\hspace{0.4cm} \object{HD 98410} (B2.5 Ib/II)}
\hspace{0.2cm}
\includegraphics[width=3.2cm,height=18.0cm,angle=90]{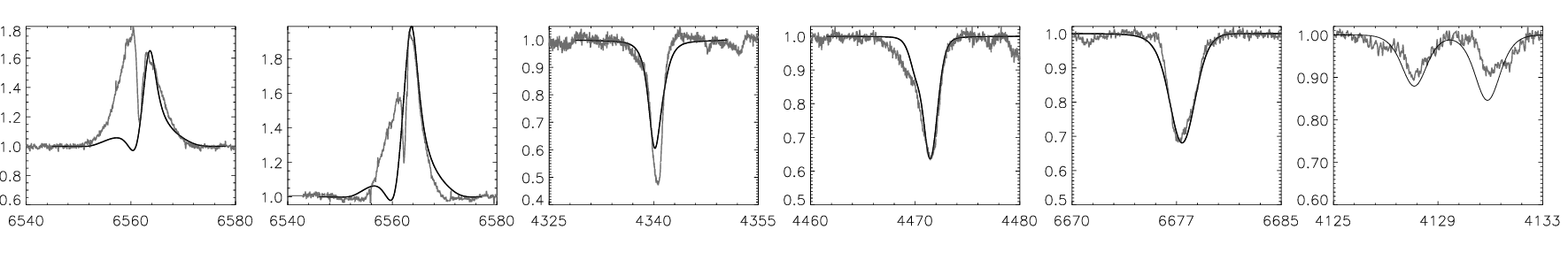}\\
\rotatebox[origin=l]{90}{\hspace{0.4cm} \object{HD 68161} (B8 Ib/II?)}
\hspace{0.2cm}
\includegraphics[width=3.2cm,height=18.0cm,angle=90]{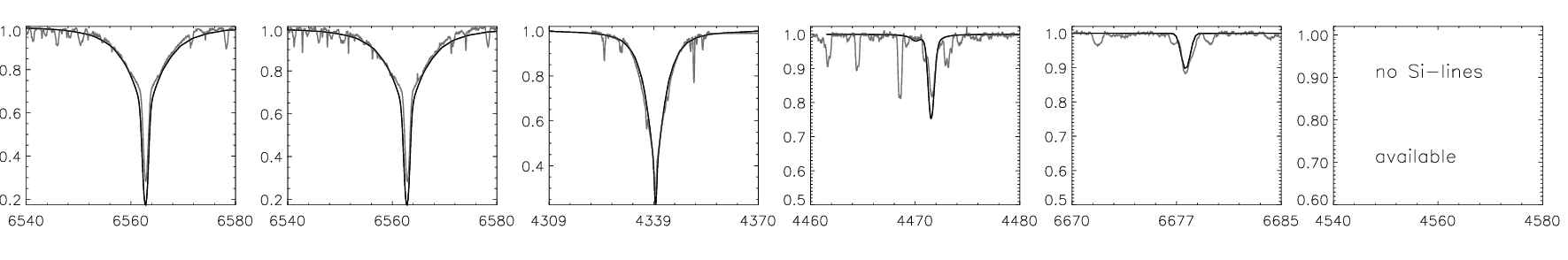}\\
\end{figure*}
  
\begin{figure*}[ht]
\label{fits_groupIV}
\caption{Spectral line fits for the comparison stars (previously not known 
to exhibit any periodic variability). No Si availabe.} 
\vspace{0.5cm}\hspace{1.7cm} H$\alpha$  \hspace{2.1cm} H$\gamma$ \hspace{1.8cm}
He~I 4471 \hspace{2.3cm} H$\alpha$ 
\hspace{2.0cm} H$\gamma$  \hspace{1.5cm} He~I 4471 \\
\rotatebox[origin=l]{90}{\hspace{0.4cm} \object{HD 149038} (O9.7 Iab)}
\hspace{0.2cm}
\includegraphics[width=3.2cm,height=8.0cm,angle=90]{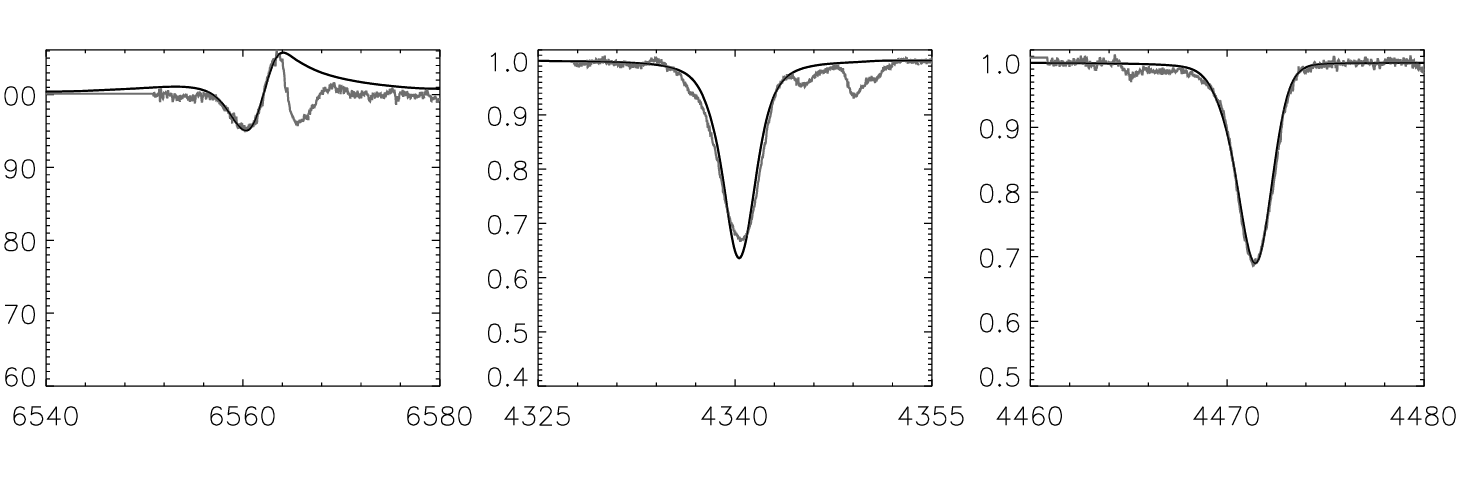}
\rotatebox[origin=l]{90}{\hspace{0.4cm} \object{HD 64760} (B0.5 Ib)}
\hspace{0.2cm}
\includegraphics[width=3.2cm,height=8.0cm,angle=90]{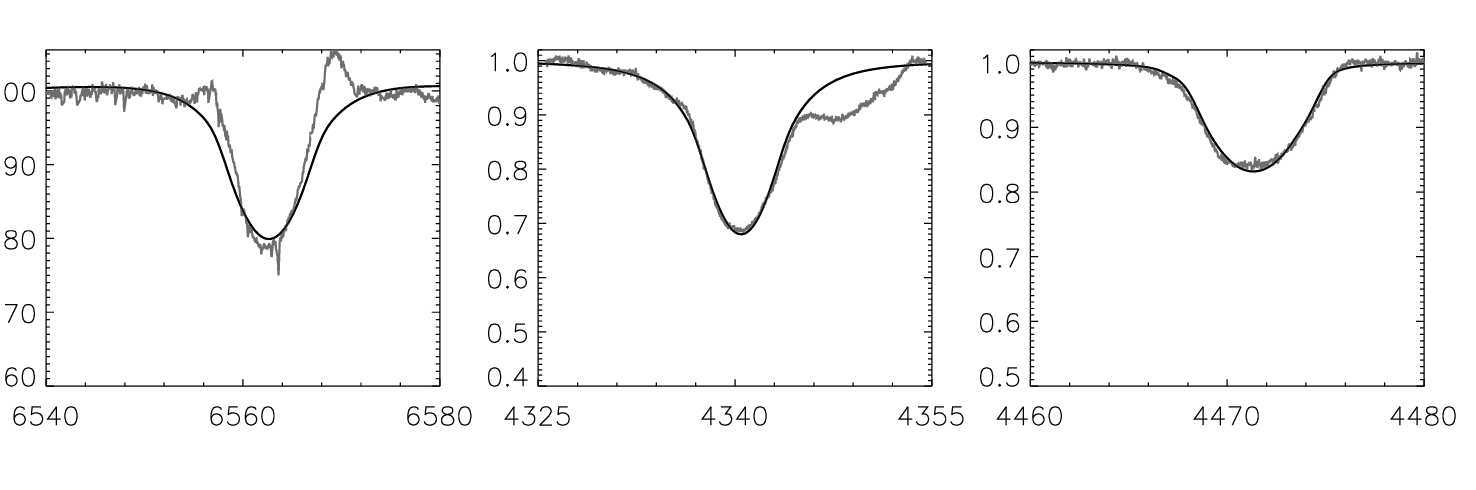}\\
\rotatebox[origin=l]{90}{\hspace{0.4cm} \object{HD 157246} (B1 Ib)}
\hspace{0.2cm}
\includegraphics[width=3.2cm,height=8.0cm,angle=90]{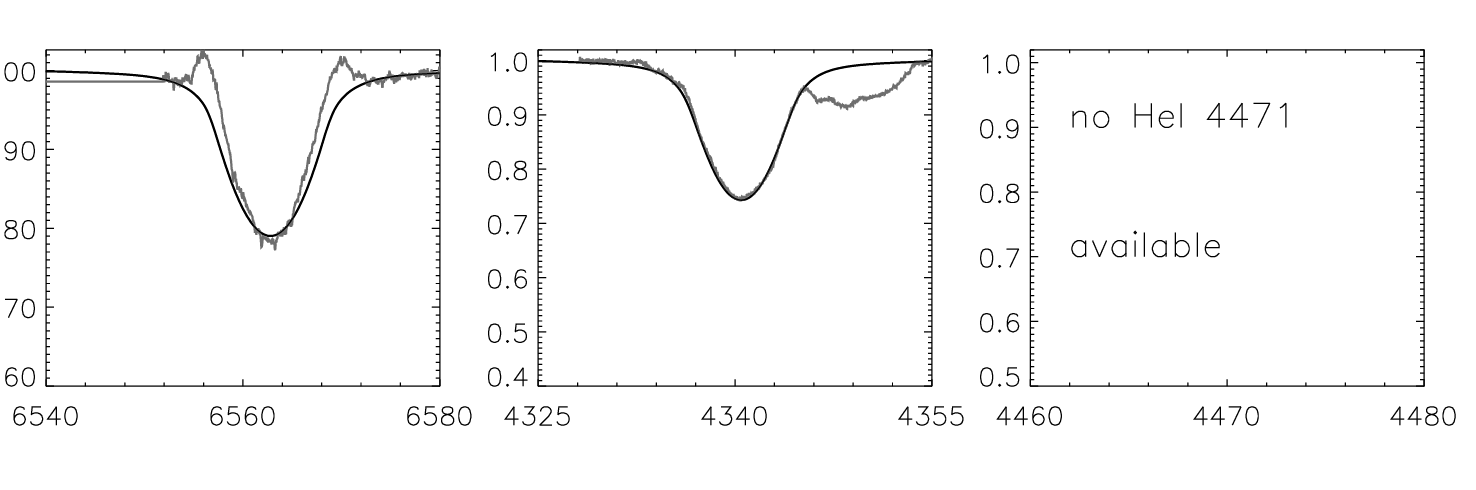}
\rotatebox[origin=l]{90}{\hspace{0.4cm} \object{HD 157038} (B1/2 IaN)}
\hspace{0.2cm}
\includegraphics[width=3.2cm,height=8.0cm,angle=90]{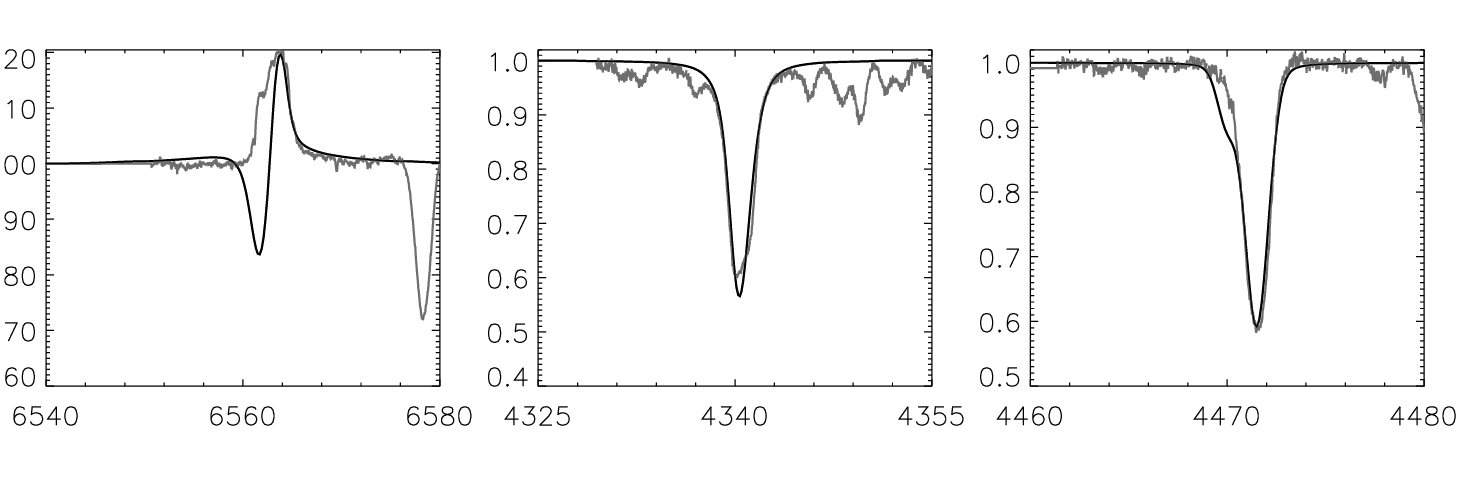}\\
\rotatebox[origin=l]{90}{\hspace{0.4cm} \object{HD 165024} (B2 Ib)}
\hspace{0.2cm}
\includegraphics[width=3.2cm,height=8.0cm,angle=90]{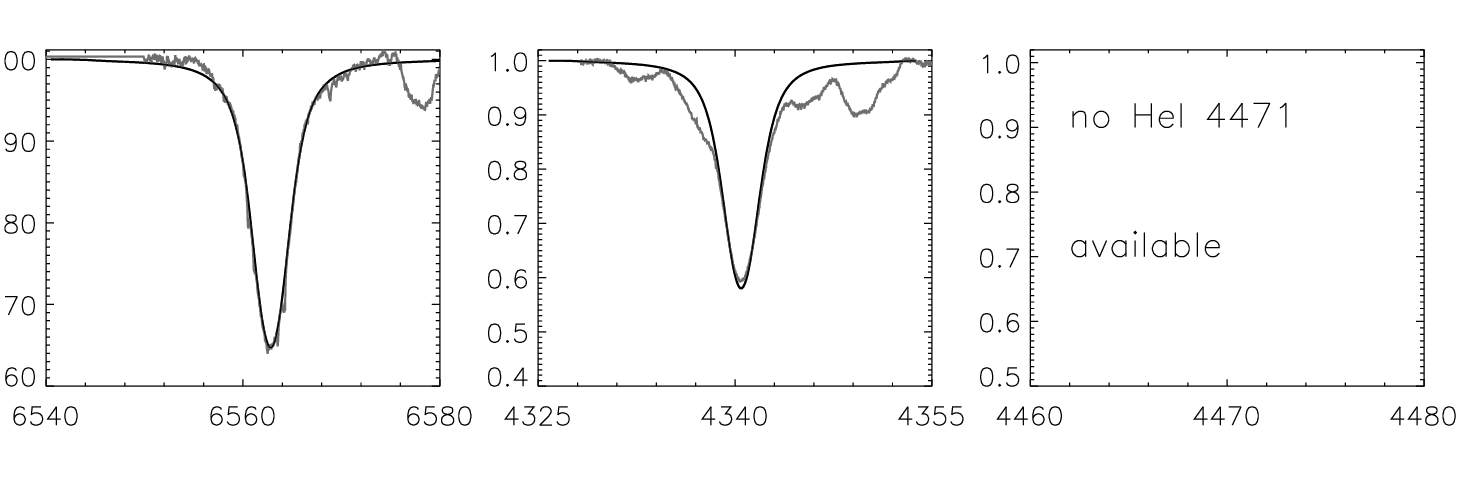}
\rotatebox[origin=l]{90}{\hspace{0.4cm} \object{HD 75149} (B3 Ia)}
\hspace{0.2cm}
\includegraphics[width=3.2cm,height=8.0cm,angle=90]{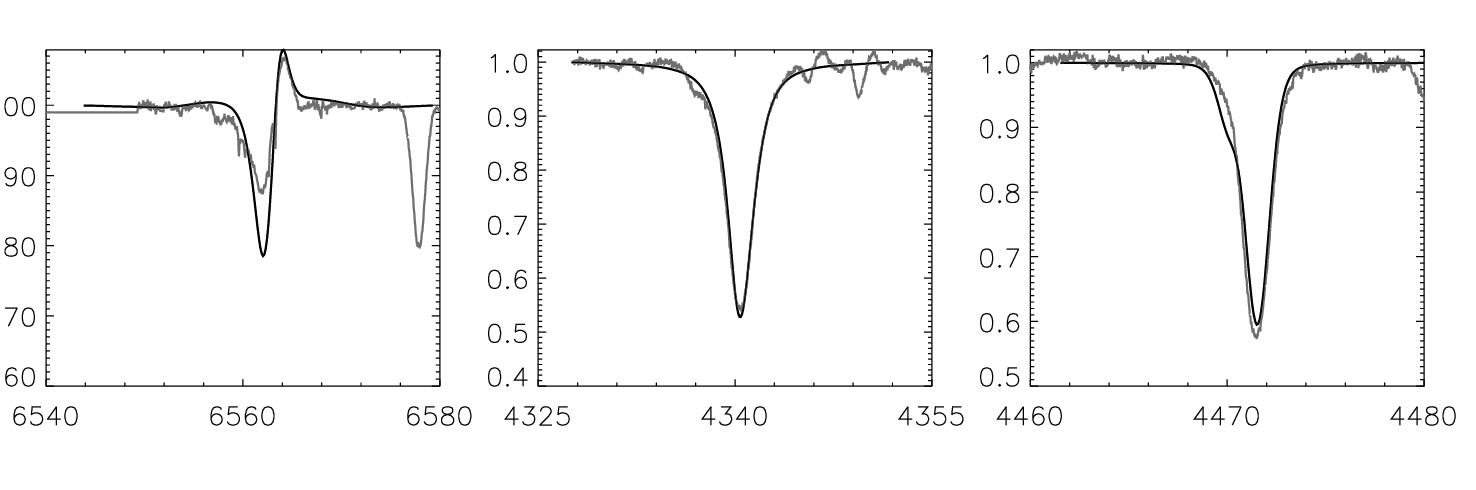}\\
\rotatebox[origin=l]{90}{\hspace{0.4cm} \object{HD 58350} (B5 Ia)}
\hspace{0.2cm} 
\includegraphics[width=3.2cm,height=8.0cm,angle=90]{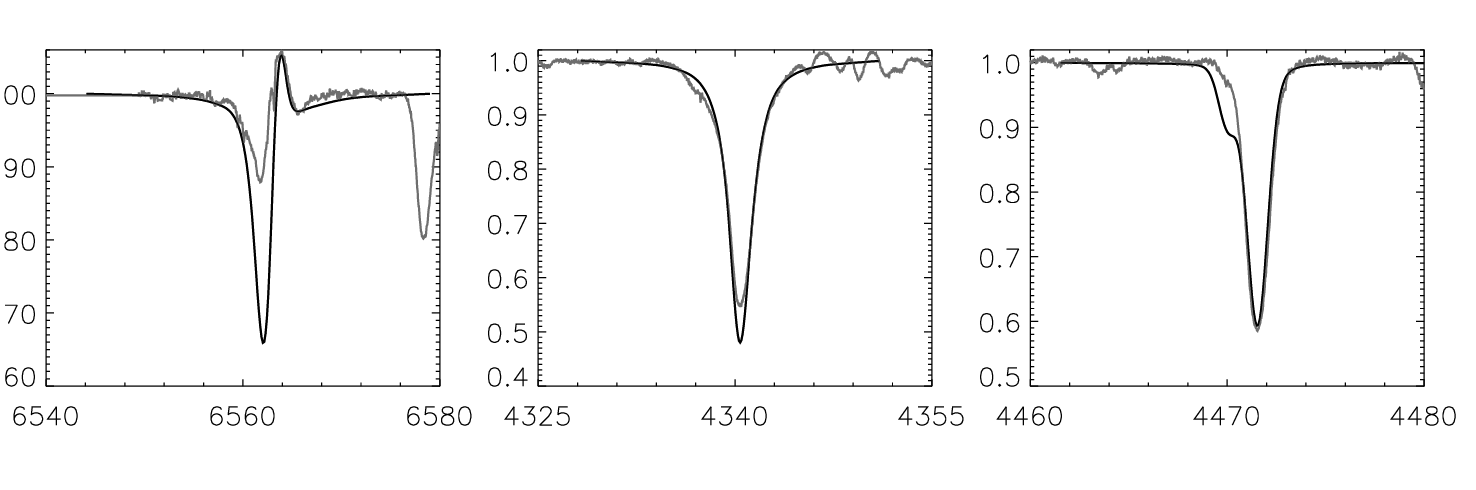}
\rotatebox[origin=l]{90}{\hspace{0.4cm} \object{HD 86440} (B5 Ib)}
\hspace{0.2cm} 
\includegraphics[width=3.2cm,height=8.0cm,angle=90]{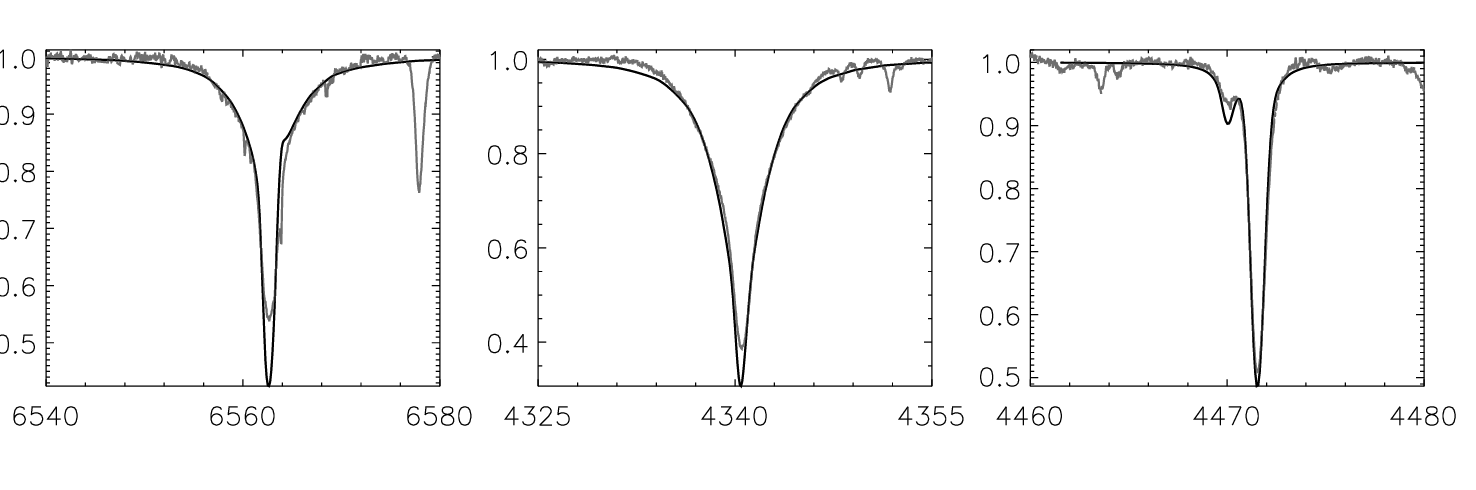}\\
\rotatebox[origin=l]{90}{\hspace{0.4cm} \object{HD 125288} (B6 Ib)}
\hspace{0.2cm}
\includegraphics[width=3.2cm,height=8.0cm,angle=90]{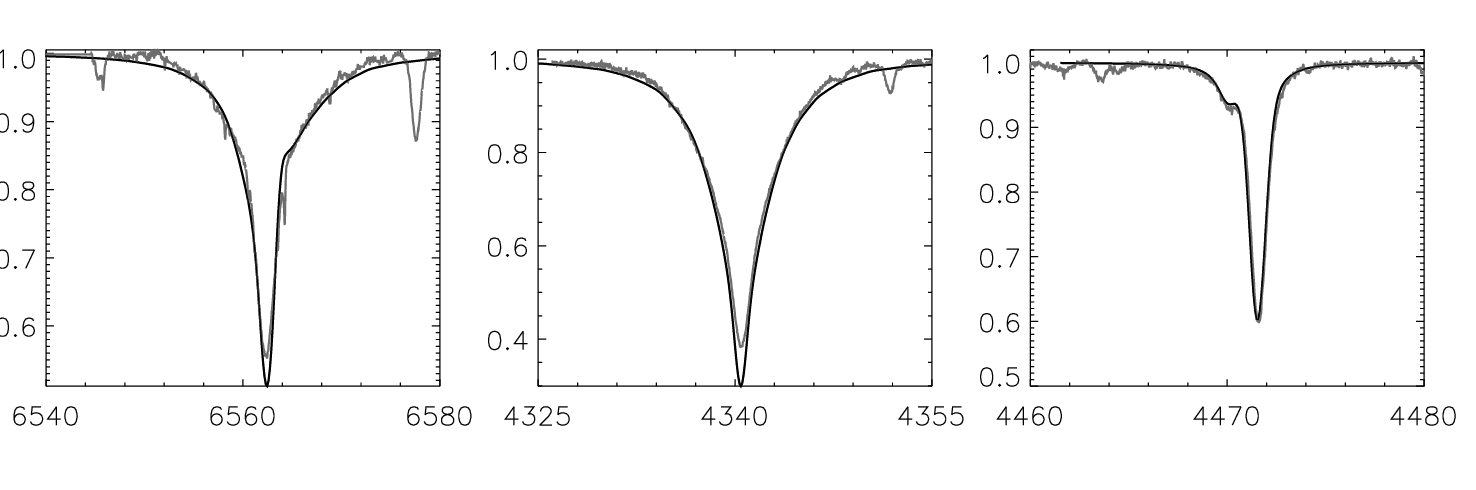}
\rotatebox[origin=l]{90}{\hspace{0.4cm} \object{HD 106068} (B8 Iab)}
\hspace{0.2cm}
\includegraphics[width=3.2cm,height=8.0cm,angle=90]{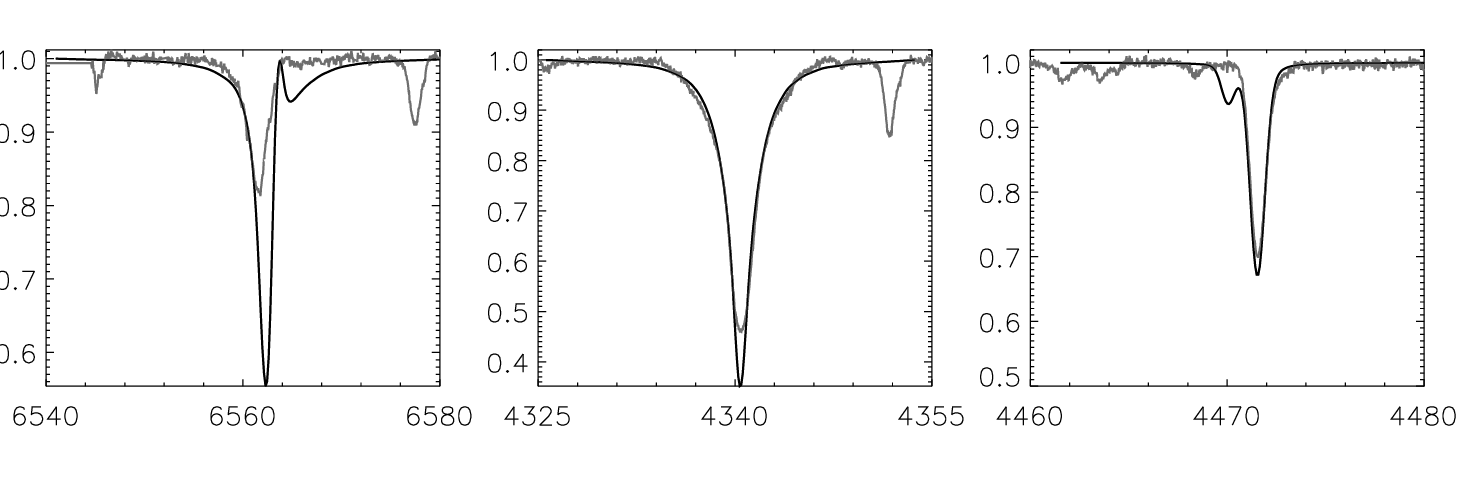}\\
\rotatebox[origin=l]{90}{\hspace{0.4cm} \object{HD 46769} (B8 Ib)}
\hspace{0.2cm}
\includegraphics[width=3.2cm,height=8.0cm,angle=90]{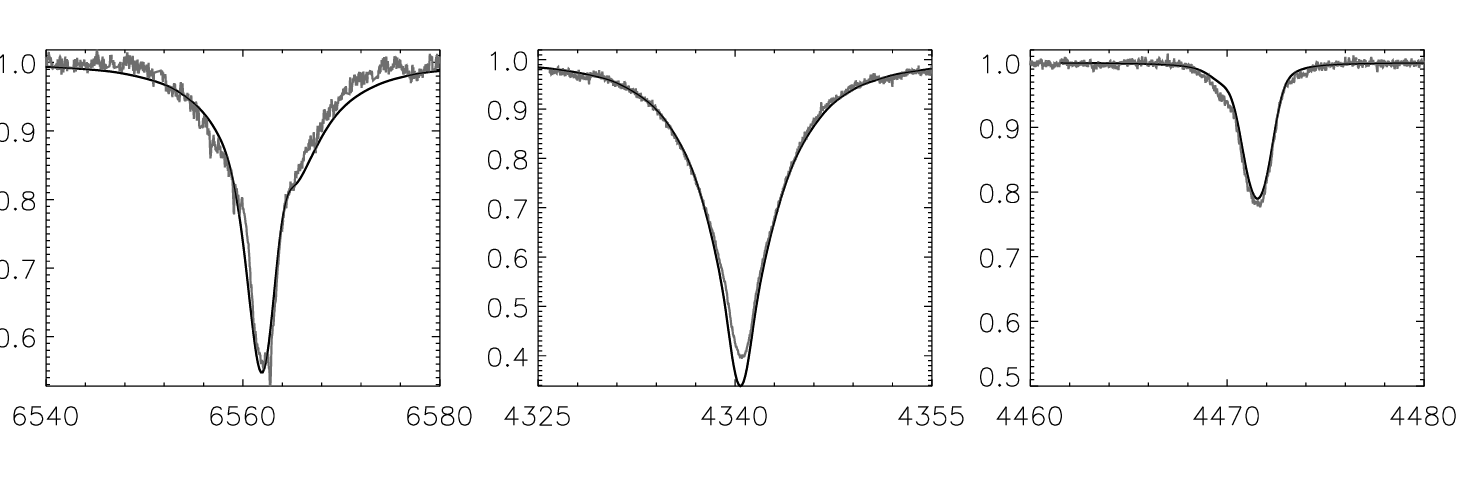}
\rotatebox[origin=l]{90}{\hspace{0.4cm} \object{HD 111904} (B9 Ia)}
\hspace{0.2cm}
\includegraphics[width=3.2cm,height=8.0cm,angle=90]{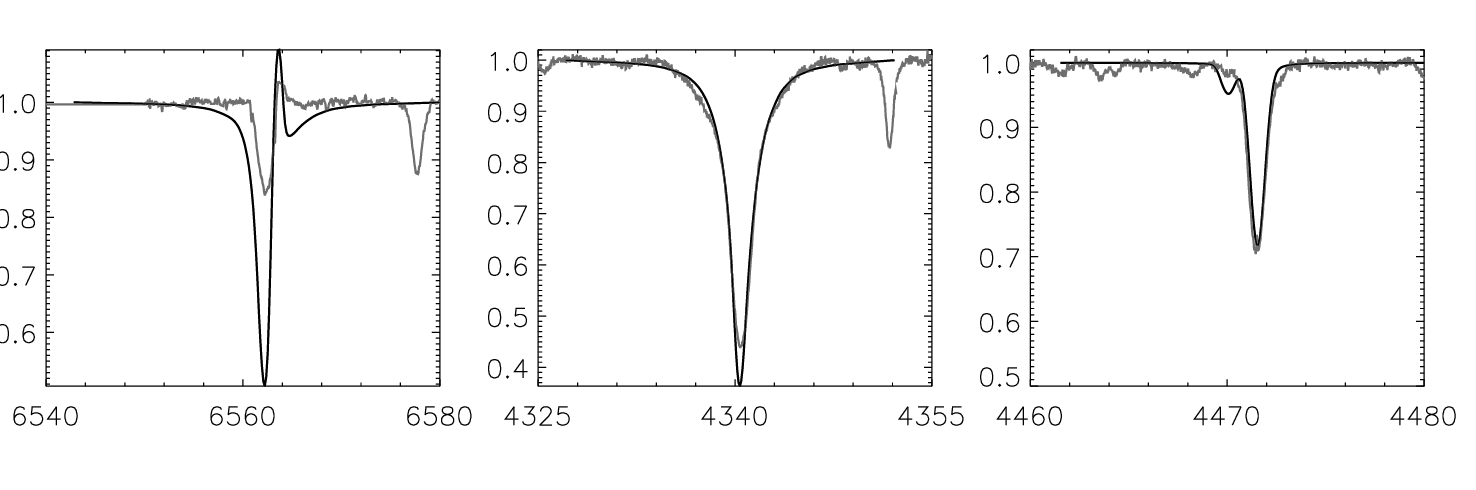}\\
\end{figure*}
\clearpage

\twocolumn
\section{Individual Discussion of All Sample and Comparison Stars}

In the following, we discuss all sample and comparison stars one by one
and group per group. The order of description is based on the spectral type
of each object {\it as adopted in this paper} (see discussion of individual
objects below and also Sect.\,\ref{calib}). Within each spectral type,
stars are ordered following their HD number. This order refers to the order
of the objects in Figs.~A.1. to A.7. as well.

\subsection{Stars Discarded from the Sample}

We have good reasons to remove \textbf{\object{HD\,51110}} (B9 Ib/II) from
our sample.  From our spectra, it immediately turned out that HD\,51110 is
very He-weak (all He lines almost absent). We ascribe this chemical
abundance anomaly to the gravitational settling of helium. This is only
expected to show up in the stable, non-convective atmospheres of the higher
gravity stars (white dwarfs and sub-dwarfs), as hypothesized by
\citet{Michaud83} and confirmed by \cite{Fabbian05} for hot HB stars in
NGC1904. The extreme strength of our observed H$\gamma$ and H$\alpha$
profiles are consistent with this hypothesis. Thus, we consider this star as
being misclassified and hence unimportant for our study.

For \textbf{\object{HD\,147670}} (B8/B9 II) only H$\alpha$ was measured.
Since we cannot derive stellar parameters from one line, we are forced to
exclude it from our sample as well.\\

\subsection{Group I Objects \label{comments_groupI}}

The first group of periodically variable B-type supergiants consists of
those objects for which we were able to derive accurate results, on the
basis of our assumptions, as discussed in Sects.\,\ref{linefitting} and
\ref{erroranalysis}. They are the following:\\

\noindent \textbf{\object{HD\,168183} (O9.5 Ib)}
was already earlier found to be a short-period (single-lined) binary,
exhibiting large variations in radial velocity between consecutive nights (from
$-52$ up to $92$ km/s) and having a period of about 4 days \citep{Bosch1999}.
From our data we confirm both the short period {\it and} the huge radial
velocity changes, going from $-60$ km/s to $80$ km/s. 
The fact that HD\,168183 is a member of the open cluster NGC 6611 allows 
to derive its radius as 19 R$_{\odot}$ (from d = 2.14 $\pm$ 0.10
-- \citealt{Belikov1999}; V = 8.18, A$_V$ = 2.39 -- \citealt{Hillenbrand1993}). 

Several different spectral types are mentioned for this star: B1 Ib/II (SIMBAD),
B0 III \citep{Evans2005}, O9.5 I (\citet{Hillenbrand1993}, confirmed by
\citet{Bosch1999}). We accept O9.5 I as the ``true'' spectral type,
relying on our spectroscopically derived temperature of 30\,000~K, which is in
agreement with the observational $T_{\rm eff}$ scale for O type supergiants
provided by \citet{Martins2005}. From the H$\alpha$ profile, we conclude that
this object is of luminosity class Ib. 

The peculiar feature seen in the Si~III 4574 line is a known instrumental
defect, which is also present in the spectrum of HD\,170938.\\

\noindent \textbf{\object{HD\,89767} (B0 Ia)}
We obtain a reasonable fit for this object and no further comments are required.\\

\noindent \textbf{\object{HD\,94909} (B0 Ia)} 
This object was intensively discussed in the text (see
Section\,\ref{Comparison}).\\

\noindent \textbf{\object{HD\,93619} (B0.5 Ib)}
From the fundamental parameters derived from the profile fits, this appears to
be a typical B0.5 Ib star. The small emission bump in the red wing of
H$\alpha$ seems to show up and vanish from one observation to the other.\\

\noindent \textbf{\object{HD\,91943} (B0.7 Ib)}
For the discussion of this object, we refer to Section\,\ref{Comparison}.\\

\noindent  \textbf{\object{HD\,96880} (B1 Ia)} 
Also for this star the fit quality is good except for He~I~6678 which
would require a larger macroturbulence), and no further comments are required.\\

\noindent \textbf{\object{HD\,115363} (B1 Ia)}
The observations of all targets discussed in this paper were done at the end
of the lifetime of the CAT telescope.  Unfortunately, at the time when it
was completely closed down, in September 1998, we did not have Si measured
for this target yet. He~I~4471, which is a reasonable temperature estimator
in the temperature range (at least if we adopt a solar
Helium content), gives us a perfect fit at a temperature of 20\,000~K. Even
more, HD\,115363 is rather similar to HD\,170938, which has the same
spectral type and for which we do have Si~III measured. For the latter
object, we find exactly the same temperature and gravity. Taken together, we
feel confident that the derived parameters are credible. Note also the
perfect fit of the H$\alpha$ emission.\\

\noindent \textbf{\object{HD\,148688} (B1 Ia)}
has been fully discussed in the paper (see Section\,\ref{Comparison}).\\

\noindent \textbf{\object{HD\,170938} (B1 Ia)}
Although the emission peak observed in the H$\alpha$ profile of HD\,170938
is lower than the one observed in HD\,115363, marking a difference in wind
properties (with mass-loss rates differing within a factor of about two to
four), all other stellar parameters are identical in both B1 Ia supergiants.
Note that also in this star we detect the known instrumental defect in the
Si~III~4574 line.\\

\noindent \textbf{\object{HD\,109867} (B1 Iab)}
is remarkable because of the very large changes in radial velocity, 
extending from -70 to 20 km/s.\\

\noindent \textbf{\object{HD\,154043} (B1 Ib)}
For this poorly known supergiant we are able to obtain a convincing fit. 
This is one of the few stars that does not show any radial velocity changes
between the several line measurements. The radial velocity amounts to -20
km/s, and Si~III appears to be asymmetric (see Section 4.1). \\ 

\noindent \textbf{\object{HD\,106343} (B1.5 Ia)}
Apart from the fact that the blue part of H$\alpha$ is not fitting
perfectly (both in the absorption trough and in the wings), we can be quite
sure about the stellar parameters. The derived parameters confirm
previous results from \citet{Lamers1995} and \citet{Bianchi1994} obtained by
means of UV spectroscopy.\\

\noindent \textbf{\object{HD\,111990} (B2 Ib)}
For this supergiant, part of a double system, only one
H$\alpha$ profile has been observed, which can be fitted very nicely. The
P~Cygni profile shows only a very small emission peak, which points to a
moderate wind density.  Note that from this star on we switch from the
analysis of Si~III to the analysis of Si~II.\\

\noindent \textbf{\object{HD\,92964} (B2.5 Iae)}
This object is one of the few sample stars that exhibits a clear
asymmetry in the line profile, especially visible in H$\gamma$.  Since we
are able to obtain an acceptable fit without further assumptions, we ascribe
this asymmetric behaviour to the strong wind which is affecting the
photospheric lines. As already mentioned, does the He~I~6678 line require a
macro-turbulent velocity twice as large as the one we derive from the Si
lines.\\

\noindent \textbf{\object{HD\,53138} (B3 Ia)}
has been discussed in the paper (see Section\,\ref{Comparison}). Note that
the fits and the parameters provided in Table\,\ref{finalparameters_SG}
refer to the model at 17\,000~K, which has been discussed as a compromise
between the results obtained from our spectra and those obtained from the
complete JKT spectrum with additional ionisation stages from Si and
He.\\

\noindent \textbf{\object{HD\,102997} (B5 Ia)}
The wind strength of the first measurement is lower than in the second
observation, turning the P~Cygni profile with a partly refilled absorption
wing into a pure emission profile. Unfortunately we are unable to correctly
reproduce this pure emission. This is the first object (out of two) for
which we find an effective temperature larger than predicted from our
calibration (see discussion at the end of Section\,\ref{calib}).\\

\noindent \textbf{\object{HD\,108659} (B5 Ib)}
Quite unknown among the supergiants, there is only one previous temperature
estimate for this object, derived by \citet{Waelkens98} using photometric
measurements ($T_{\rm eff}$ = 11\,750~K). Our spectral fits indicate a much
higher temperature, at least by 4\,000~K. Though it seems that the derived
mass-loss rate is too low (synthetic cores too deep), a further increase in
$\dot M$ would result in a small red emission peak which is not observed.
Also for this second B5 target, we find a temperature larger than expected
for its spectral type (see Section\,\ref{calib})).\\

\noindent \textbf{\object{HD\,80558} (B6 Iab)}
As indicated especially by the second H$\alpha$ profile, the wind of
HD\,80558 might show a non-spherical distribution, if one
regards its shape as the beginning of a double-peaked structure. In so far, the
results of our wind analysis can be considered only as a very rough
estimate.\\

\noindent \textbf{\object{HD\,91024} (B7 Iab)}
From this spectral type on, He~I~4471 becomes strongly sensitive to changes
in effective temperature (see Fig.\,\ref{isocontour_heI}), on the expense of
its reaction to changes in surface gravity. Thus, $T_{\rm eff}$ is easily
fixed at 12,500~K, with a gravity of $\log g = 1.95$ following from the
H$\gamma$ line wings. Note that for this object the {\it observed} Si~II
components are of different strength (as predicted), and thus could be
fitted without any compromise.

HD\,91024 has a moderate wind, which is slightly refilling the wings of
H$\alpha$.  By inspection of the first H$\alpha$ profile, we see an almost
flat red wing (at continuum level), with a very steep decline into absorption.
This also occurs for HD\,106068, and cannot be represented by our 
synthetic profiles, at least at the inferred gravities.\\

\noindent  \textbf{\object{HD\,94367} (B9 Ia)}
has a moderately strong wind, displayed by the (strongly variable)
emission peak of H$\alpha$. Again, the overall shape of the profile cannot
be modelled, and only the emission peak has been fitted.\\

\subsection{Group II Objects \label{comments_groupII}}

We recall from Section\,\ref{linefitting} that, for some stars, the lack of
Si~II and/or Si~IV prohibits to choose between two equally well fitting
models at different positions in parameter space.  In order to break this
dichotomy, we proceed as follows. First of all, we aim at a good fit of
He~I~4471 which is a reasonable temperature indicator, though best-suited
only at latest spectral types (below 15\,000~K, see
Fig\,\ref{isocontour_heI}). Still, we have to keep in mind that this line
depends on the (unknown) helium abundance. By combining the He~I~4471
diagnostics (using solar Helium abundances) with knowledge about the spectral
type of the star, we are then able to obtain some clue about which
temperature is the correct one.  

We take HD\,54764 as a prototypic example to illustrate the dichotomy
problem in some detail.\\

\noindent \textbf{\object{HD\,54764} (B1 Ib/II)}
For HD\,54764, relying on both the Si~III~triplet and the He~I~4471 line, we
are able to find two very different models that both give a reasonable match
with the observations: $T_{\rm eff}$ = 19\,000~K, $\log g$ = 2.45 and $T_{\rm
eff}$ = 26\,000~K, $\log g$ = 2.9, respectively. By comparing the
corresponding synthetic lines, one can hardly see any difference (see
Fig\,\ref{fits_HD54764}), though the He~I lines do favour the cooler
solution.

\begin{figure*}[ht]
\vspace{0.5cm}\hspace{2.2cm} H$\alpha$ 1  \hspace{2.5cm} H$\alpha$ 2 
\hspace{2.5cm} H$\gamma$ 
\hspace{2.5cm} He~I~4471  \hspace{1.2cm}  Si~III~4552-4567-4574\\ 
\rotatebox[origin=l]{90}{\hspace{0.4cm} HD\,54764 (B1 Ib/II)} \hspace{0.2cm} 
\includegraphics[width=3.2cm,height=17.0cm,angle=90]{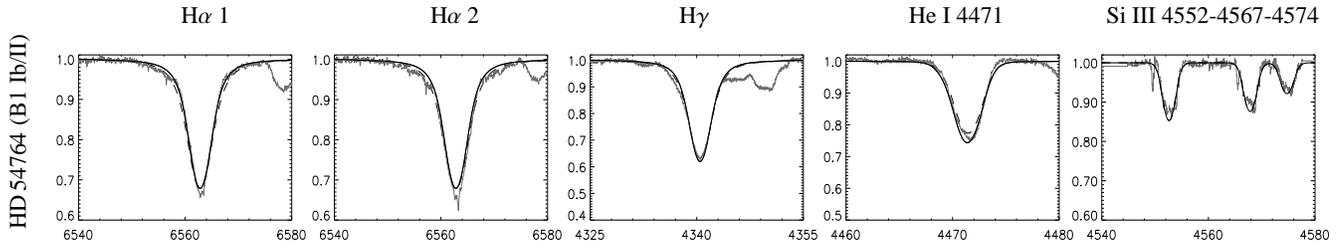}
\caption{Spectral line fits for HD\,54764 - a prototype for
supergiants belonging to group II: two models, located in completely different
parameter domains (regarding $T_{\rm eff}$ and log $g$), produce
similar line profiles. Bold: cool model ($T_{\rm eff}$ = 19\,000K,
log $g$ = 2.4); dashed: hot model ($T_{\rm eff}$ = 26\,000K, log $g$ =
2.9).} 
\label{fits_HD54764}
\end{figure*}

\begin{figure*}[ht]
\hspace{1.0cm}\includegraphics[width=10.cm,height=17.0cm,angle=90]{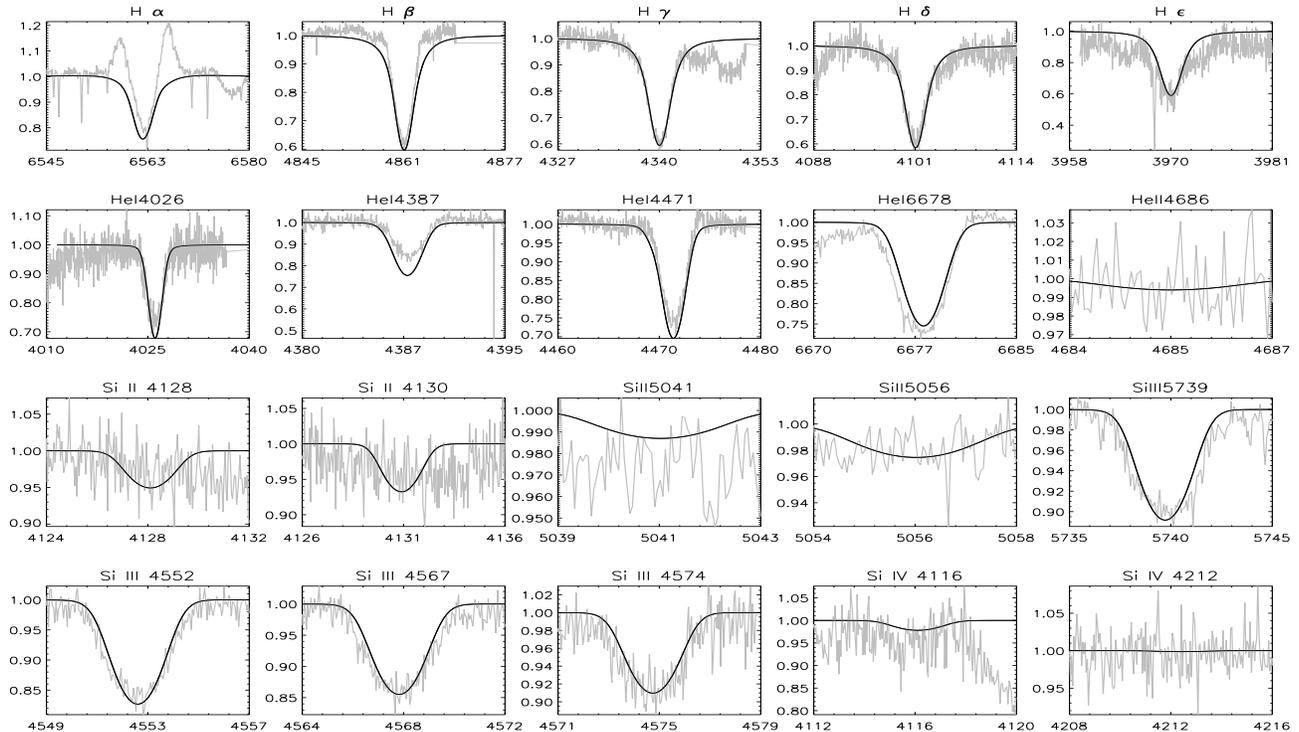}
\caption{FEROS spectrum of HD\,47240: 
synthetic line profiles at $T_{\rm eff} = 19\,000$ K and log $g$ = 2.4.} 
\label{fits_HD47240}
\end{figure*}

\citet{Lyubimkov2004} derived values for $T_{\rm eff}$, log $g$ and A$_V$
for 102 stars, based on their colour indices. Although photometric methods
rapidly lose their predictive capacity in the massive star domain, it is
quite interesting to have a look at their results for HD\,54764, which are
$T_{\rm eff}$ = 25\,500 $\pm$ 800 K and $\log g$ = 3.56 $\pm$ 0.17 (note the
rather optimistic error bars). Whereas this effective temperature would
perfectly match our hotter model, the gravity is certainly way too high,
which is immediately reflected in the Balmer line wings. Unfortunately we
could not detect any He~II or Si~IV lines in the observed spectral range, to
unambiguously decide on $T_{\rm eff}$.  By concentrating on the secondary
diagnostics mentioned above (spectral type), the cooler model is the more
plausible one, if HD\,54764 is actually of spectral type B1. Note that we
find similar temperatures for all other B1 supergiants, being in agreement
also with the former $T_{\rm eff}$ calibration by \citet{Lennon1993}. 

HD\,54764 is one of the three supergiants within our sample that
exhibits clear asymmetries in its line profile, in this case most likely 
because of an optical companion \citep{Abt1984}.\\

\noindent \textbf{\object{HD\,47240} (B1 Ib)}
is another example of finding two models with completely different stellar
parameters, both fitting the Si~III lines very well: one model with $T_{\rm
eff} = 19\,000$ K and log $g$ = 2.4, and a second one with $T_{\rm eff} =
24\,000$ K and log $g$ = 2.8. Again, also the He~I lines are too similar to
allow for a clear distinction.

In contrast to the first group~II object, for HD\,47240 we do have a
complete FEROS spectrum at our disposal, provided in GAUDI, the preparatory
archive of ground-based observations in the framework of COROT
\citep{Solano2005}. The drawback of this spectrum is its very low S/N, so
that the Si~II and Si~IV lines, although being very different in strength,
still lie in the noise level. At least from He~II~4686, on the other hand,
the hotter solution can be ruled out with high confidence, since in this
case the line is predicted to be much stronger than it is observed. Again,
also for this object its spectral type points towards the cooler solution.
The complete comparison with the FEROS observations is given in
Fig.\,\ref{fits_HD47240}.  

It is worth mentioning that this star might be a binary and that the
H$\alpha$ morphology (double-peaked structure) is typical for a fast rotator
observed almost equator-on \citep{Morel2004}. Note that we do observe this
structure also for the fast rotators HD\,64760 and HD\,157246.\\

\noindent \textbf{\object{HD\,141318} (B2 II)}
For this object, the differences between the parameters of the cool and the
hot solution are not as large as for the previous cases, namely ($T_{\rm
eff}$, log $g$) = (20\,000~K, 2.9) and (22\,000~K, 3.2), respectively. 
Although the forbidden component of He~I~4471 is fitting slightly better for
the hotter model, the Si~III triplet is better represented for the cooler
one. Note that an intermediate model at 21\,000~K does {\it not} give a
good fit. From its spectral type then, we prefer the cooler solution.

\subsection{Group III Objects \label{comments_groupIII}}

Group III constitutes of those (three) stars for which we cannot claim a
similar accuracy as obtained for the previous two groups, either because
they are somewhat extreme, either because of their peculiar spectrum that
complicates a reliable fit, or a combination of both. All three objects
belong to the group of 10 stars that were added to the sample of
\citet{Waelkens98}, and are possible chemically peculiar stars. We have
classified these stars as ``unreliable'', and the derived parameters have to
be considered with caution.\\

\noindent \textbf{\object{HD\,105056} (B0 Iabpe)}
is a nitrogen-rich, carbon-deficient supergiant, exhibiting very strong
emission lines, up to twice the continuum level, and has been classified as
an ON9.7 Iae supergiant in several studies. Its peculiar nature and the
extremely dense wind hamper a correct modelling of the photospheric lines
(e.g., H$\gamma$), since these are severely contaminated by the wind. Though
nitrogen enrichment usually goes along with the enrichment of helium, we
used a compromise solution for He~I~4471 and 6678 at normal abundance,
guided by the temperature we found from fitting the Si~III triplet.\\

\noindent \textbf{\object{HD\,98410} (B2.5 Ib/II)}
Similar to the last object, also HD\,98410 is an extreme supergiant, with a
very strong wind, H$\alpha$ completely in emission and strongly refilled
photospheric lines.

Because of the restricted wavelength range around each line, the
normalisation of the H$\alpha$ profile became more problematic than typical,
thus increasing the uncertainty of the derived mass-loss rate. 

The difficulty to fit H$\alpha$ and H$\gamma$ in parallel might point to the
presence of strong clumping in the wind (e.g., \citealt{Repolust2004}), 
and the actual mass-loss rate might be considerable lower than implied by
H$\alpha$. Taken together with the peculiar shape of one of these profiles
(which might be explained by an equatorially enhanced wind), we ``classify''
our analysis as unreliable.\\

\noindent \textbf{\object{HD\,68161} (B8 Ib/II?)}
Although the spectroscopically derived spectral type is B8 Ib/II,
\citet{Eggen1986} mentioned that photometry indicates a different luminosity
class, namely B8 IV. HD\,68161 has been considered by \citet{Paunzen1998} as
a probable, variable chemically peculiar star, in particular a variable star
of the $\alpha^2$ CVn type. If so, this star should be a main sequence star
of spectral type later than B8p (consistent with the photometrically derived
spectral type), exhibiting a strong magnetic field and abnormally strong
lines of Si among other elements. Since for this star we could not observe
the Si lines, we cannot confirm this conjecture. Note, however, that the
derived gravity is not so different from HD\,46769, which is a ``normal" B8
Ib supergiant, and that it is also in agreement with typical gravities for
these objects. Due to the discussed uncertainties we add this object
to our list of unreliable cases.

\subsection{Group IV Objects - the Comparison Stars \label{comments_groupIV}}

Group IV is the group of 12 bright comparison stars, selected from the
Bright Star Catalogue, previously not known to exhibit any periodic
variability. For these objects mostly three lines have been observed: He~I
4471, H$\gamma$ and H$\alpha$, which, in combination with our $T_{\rm eff}$
calibration for B-supergiants (see Section\,\ref{calib} and
Table\,\ref{calibration_table}), will be used to estimate the required
stellar and wind parameters. 

\noindent \textbf{\object{HD\,149038} (O9.7 Iab)} 
While SIMBAD lists this star as a B0 supergiant, a spectral type of O9.7 Iab
has been suggested by \citet{Walborn1996}, \citet{Lamers1999},
\citet{Maiz2004} and \citet{Fullerton2006}. Recent revisions of stellar
parameter calibrations in the O type regime \citep{Martins2005} predict
$T_{\rm eff} \approx 30500$ K, $\log g \approx$ 3.2, R $\approx$ 22.1
$R_{\odot}$ and log $L/L_{\odot} \approx 5.57$ for an O9.5 supergiant, where
these values have been used by \citet{Fullerton2006}. The latter authors
additionally revised the distance of this star, from 1.3 kpc
\citep{Georgelin1996} to 1.0 kpc. The radio mass-loss rate, $\log
\dot{M}_{\rm radio}$ (at time of observation), could be constrained as being
less than -5.51 $\pm 0.18~M_{\odot}$/yr, with a terminal velocity of
$v_{\infty} = 1750$ km/s and $\beta$ = 1.0. Our calibration predicts
somewhat lower effective temperatures at O9.7, but the general
agreement between our values and those stated above is convincing.\\

\noindent \textbf{\object{HD\,64760} (B0.5 Ib)} is a very fast rotating
supergiant. With a projected rotational velocity, $v \sin i$, of more than
220 km/s it is rather likely being observed at a very high inclination,
i.e., almost equator-on. Due to its particularly interesting wind structure,
HD\,64760 is amongst the best studied early B type supergiants \citep[being
the most recent investigations]{Prinja2002, Kaufer2002, Kaufer2006}. Its
richness in spectral features led to conclusive evidence for the existence
of a corotating two and four armed spiral structure, suggested to be
originating from stream collisions at the surface and perturbations in the
photoshere of the star \citep{Kaufer2002, Kaufer2006}. H$\alpha$ consists of
a double-peaked structure, with a blueward and a somewhat stronger, redward
shifted emission peak around the central absorption (as in Fig.~2 of
\citealt{Kaufer2002}). This star is clearly a Ib supergiant, with a moderate
mass-loss rate refilling the photospheric absorption. Of course, we are not
able to reproduce this double-peaked profile, and only the gravity can be
{\it derived} (together with some crude estimate for $\dot M$) for the
adopted effective temperature, $T_{\rm eff}$ = 24\,000~K, which is $\log g$
= 3.2. From the M$_V$ calibration, we finally have $R_{\ast} = 24
R_{\odot}$, where all these parameters are similar to those as reported by
\citet{Howarth1998}.\\ 

\noindent \textbf{\object{HD\,157246} (B1 Ib)}
This is another example for a rotationally modulated wind, similar to
HD\,64760. We find a projected rotational velocity of 275~km/s. The
H$\alpha$ profile shows the typical blue- and redward shifted emission
peaks, which are about equal in height. They suggest that the wind is
equatorially compressed, see also \citet{Prinja1997}.  With only two lines
at our disposal (i.e., H$\alpha$ and H$\gamma$), we can only probe that the
stellar parameters are consistent with those reported until now
\citep{Prinja2002}.\\

\noindent \textbf{\object{HD\,157038} (B1/2 IaN)}  
is enriched in nitrogen and helium (n(He)/n(H) $\approx$
0.2 according to our analysis), and the observed lines can be fitted 
reasonably well with a model of $T_{\rm eff}$ = 20\,000~K and
log $g$ = 2.3.\\ 

\noindent \textbf{\object{HD\,165024} (B2 Ib)}
As for HD\,157246, we have only two hydrogen lines at our disposal, and we
can derive only $\log g$ and an estimate of the wind properties. As we
have no direct means to derive the rotational velocity, we can only give a
range for v$_{\rm macro}$, depending on the adopted $v$ sin $i$ from the literature.
The most reasonable solution is a combination of $v$ sin $i$ = 120 km/s
and v$_{\rm macro}$ = 50 km/s (considering the macrotubulence derived
for objects of similar spectral type).\\ 

\noindent \textbf{\object{HD\,75149}  (B3 Ia)} 
is a poorly studied early B supergiant, with a dense wind. All lines
can be fitted at an effective temperature of 16\,000 K.\\

\noindent \textbf{\object{HD\,58350}  (B5 Ia)}
HD\,58350 or $\eta$ CMa is a well-studied B-type supergiant. By means of the
NLTE atmosphere code TLUSTY \citep{Hubeny2000}, \citet{McErlean1999} has
also analysed this star (cf. Sect.~\ref{compmcerlean}). Their best fitting
model has a temperature of 16\,000~K, whereas previous temperature estimates
were lower, between 13\,000 and 14\,000~K. From our calibrations
(Table\,\ref{calibration_table}), we find a typical value of 13\,500~K at
B5, which is just consistent with these lower values and gives also an
acceptable fit. On the other hand, the JKT spectrum (from the online LDF
atlas) shows that the effective temperature might be actually higher by
1\,000~K (still within the quoted error bars), since Si~II is too strong and
Si~III too weak (see Fig.\,\ref{fits_HD58350}).\\

\begin{figure}[ht]
\hspace{0.2cm}\includegraphics[angle=90,width=8.6cm]{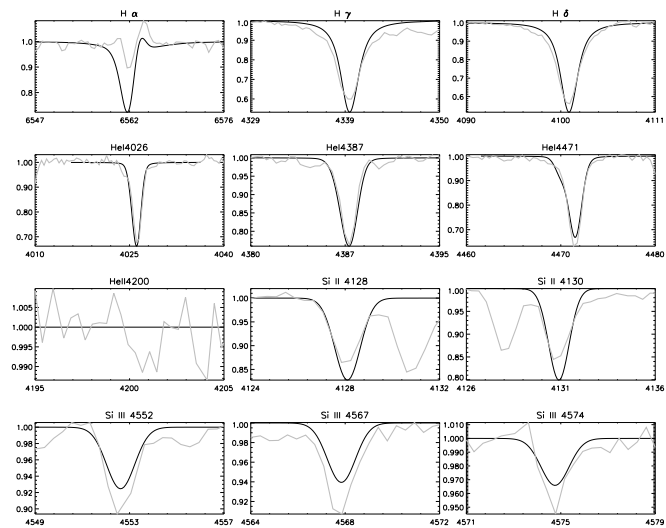}
\caption
{Spectral line fits for the JKT spectrum of HD\,58350. Gray: observed JKT
spectrum \citep{Lennon1993}, Black: predictions at $T_{\rm eff} =
13\,500$~K and $\log g$ = 1.75.} 
\label{fits_HD58350}
\end{figure}

\noindent \textbf{\object{HD\,86440}  (B5 Ib)}
No comment necessary, except that also the forbidden component of He~I~4471
fits very well, at an effective temperature of 13\,500~K.\\

\noindent \textbf{\object{HD\,125288} (B6 Ib)}
This is one other object for which we find a perfect fit of the
forbidden He~I~4471 component.\\

\noindent \textbf{\object{HD\,106068} (B8 Iab)}
This rarely studied bright B supergiant shows exactly the same feature 
in H$\alpha$ as we have found for HD\,91024 (group~I): a very flat red wing,
with a sudden steep decrease into the (blue-shifted) absorption. Of course, we cannot fit such a
profile.\\

\noindent \textbf{\object{HD\,46769}  (B8 Ib)}
For B8 objects, our calibration gives an effective temperature of 12,100~K,
which is also required to fit the He~I~4471 line (remember its sensitivity to
$T_{\rm eff}$ in this temperature domain). Models with a temperature increased
by 1\,000~K, compared to our calibration, would give a too strong He~I~4471 line,
and vice versa for lower temperatures.\\

\noindent \textbf{\object{HD\,111904} (B9 Ia)} 
is another poorly known object from the BSC. Contrary to
HD\,106068, the flat wing is this time in the blue part of the line and not
on the red side, i.e., there is some (unknown) refilling mechanism.  The
star, being a member of the open star cluster NGC 4755 in Crux, allows to
derive an M$_V$ of -7.38 $\pm$ 0.15 \citep{Slowik1995} (consistent with our
calibration within the adoted errors), and hence a radius of 95
R$_{\odot}$.\\
\end{document}